\documentclass[structabstract]{aa}  
\usepackage{graphicx}
\usepackage[]{natbib}
\usepackage{txfonts}
\begin{document}
   \title{Gamma-ray binaries beyond one-zone models: An application to LS~5039} 

   \titlerunning{One-zone models for gamma-ray binaries}
   \authorrunning{S. del Palacio, V. Bosch-Ramon, G. E. Romero}


   \author{Santiago del Palacio
          \inst{1,2} \thanks{Fellow of CONICET}, Valent\'i Bosch-Ramon \inst{3}
          \and
          Gustavo E. Romero\inst{1,2} \thanks{Member of CONICET}
          }

   \institute{Instituto Argentino de Radioastronom\'{\i}a (CCT La Plata, CONICET), C.C.5, (1894) Villa Elisa, Buenos Aires, Argentina.\\
                 \email{[sdelpalacio,romero]@iar-conicet.gov.ar}
                 \and
Facultad de Ciencias Astron\'omicas y Geof\'{\i}sicas, Universidad Nacional de La Plata, Paseo del Bosque, B1900FWA La Plata, Argentina.
         \and
             Departament d'Astronomia i Meteorologia, Institut de Ci\`encies del Cosmos (ICC), Universitat de Barcelona 
             (IEEC-UB), Mart\'i i
Franqu\`es 1, 08028 Barcelona, Catalonia, Spain\\
             \email{vbosch@am.ub.es}
             }

   \date{Received ; accepted }

 
  \abstract
   {Several binary systems hosting massive stars present gamma-ray emission. In most of these systems, despite detailed observational 
   information being available, the nature and the structure of the emitter are still poorly known.}
   {We investigate the validity of the so-called one-zone approximation for the high-energy emitter in 
   binary systems hosting a massive star. In particular, the case of LS~5039 is considered.}
   {Assuming a point-like emitter at rest, the presence of a nearby massive star, and the observed MeV and 
   GeV fluxes as a reference, a non-thermal leptonic model is systematically applied for different locations,
   magnetic fields, and non-radiative losses. This allows us to identify both the emitter configurations that are most
   compatible with observations and inconsistencies between model predictions and the available data.}
   {In the case of LS~5039, the best parameter 
   combination is fast non-radiative cooling and a low magnetic field. However, discrepancies appear when comparing the model 
   results at the MeV and GeV energy ranges with the 
   observed fluxes. Predictions fail when the orbital motion is included in the analysis, because emitters and energy
   budgets that are too large are required. Values of X-ray and TeV fluxes that are too high 
   are predicted in such a case, along half of the orbit.}  
   {We show that the radiation in LS~5039 does not come from only one electron population, and the emitter is likely extended and
   inhomogeneous with a low magnetic field. We suggest that the emitter moves at relativistic velocities with Doppler boosting 
   playing a significant role.}
   \maketitle
%

\section{Introduction}

Some galactic gamma-ray sources are high-mass binary systems in which one of the components is an early-type star of spectral 
type OB. Some of these binary systems have been detected 
from radio to high energies (HE; $E > 100$ MeV) and/or very high energies (VHE; $E > 100$ GeV) \citep{paredes2013,dubus2013}.
Depending on the nature of the companion (Cn), the systems can be classified as a compact binary (either a microquasar or 
a binary hosting a young pulsar) or a massive star binary. In a microquasar, the Cn is a stellar-mass  
black hole (BH) or a neutron star (NS) with a weak magnetic field, which is capable of accreting material coming from the star and 
generating relativistic jets \citep[e.g.,][]{mirabel1999, boschramon2006, massi2008, boschramon2009}. 
In a binary with pulsar, the Cn is a young NS with a strong magnetic field that powers an
intense relativistic wind \citep[e.g.,][]{maraschi1981,tavani1997,khangulyan2007,romero2007}. Finally, 
in a massive star binary, the Cn is another massive star with a strong stellar wind \citep[e.g.,][]{eichler1993,benaglia2003,reimer2006}.
The non-thermal gamma-ray emission from all these types of systems present the signature of the Cn orbital motion around the massive 
star in the form of modulation
and correlation of the radiation at different energy bands. This has led to the conclusion that the
massive star plays a crucial role in determining the high-energy phenomenology \citep[e.g.,][]{bednarek2007,khangulyan2008,dubus2008}.

The non-thermal emission from high-mass binaries is generated by ultra-relativistic particles, likely accelerated in strong 
shock-waves in plasma flows. The non-thermal energy could be supplied by accretion and transported by jets in microquasars, 
or carried by supersonic winds of massive stars or the relativistic wind of a pulsar.
Most of the accelerated particles cool down locally through interactions with ambient matter, magnetic fields, and radiation fields. 
The result at high energies of these interactions depends strongly on the
massive star, as it provides targets (mostly ultraviolet -UV- photons) for Inverse Compton (IC) scattering and baryons for proton-proton 
($pp$) collisions among other radiation processes. We note, however, that leptons cool down and radiate more efficiently 
than hadrons under typical conditions. Additionally, the VHE radiation coming from the inner regions of a high-mass binary 
is likely to undergo absorption due to
 pair creation in the stellar radiation field \citep[see, e.g.,][for an assessment of the importance of the different processes]
 {boschramon2009}.  

The study of gamma-ray binaries allows us to gain knowledge on physical processes occurring in extreme astrophysical
environments. However, there are still many unknown features about the particle acceleration mechanism, the structure 
of the emitter, and even the nature of the Cn in several cases. These uncertainties manifest themselves in the simplicity of the
models adopted and in the departure of their predictions from an accurate representation of the phenomenology of the sources. 
In this work, we take advantage of the few assumptions required by a simple, one-zone model to carry out a robust exploration of the 
model validity, which can be useful to sketch physical properties of the objects whose treatment is formally beyond this kind of models.

The structure of this paper is as follows: In Sect.~\ref{lim}, we present a simple and robust tool based on an one-zone model to
 thoroughly investigate the effects of the emitter-star-observer geometry and different energy losses on the resulting 
radiation. In Sect.~\ref{res}, we apply this tool to the system LS~5039 to perform a more detailed analysis, and finally,
 in Sect.~\ref{disc}, we
discuss our results in the context of the current observational data of LS~5039 and summarize the main conclusions of this work.

\section{Limitations of one-zone models: Informative analysis}\label{lim}

\subsection{Description of the approach}\label{descr}

A one-zone model describes an emitting region in which particles are injected homogeneously and evolve under homogeneous conditions.
This is a very simple model, yet capable of incorporating the most relevant physical processes of a given system and of 
reproducing the main features of its observable quantities. Regardless of its simplicity, 
the one-zone leptonic model has proven to be a robust tool 
for studying the high-energy phenomenology in high-mass binary systems \citep[e.g.][]
{kaufman2002,boschramon2006, khangulyan2008,dubus2008,takahashi2009,araudo2009,zabalza2011a}. 

\begin{figure}
    \centering
    \includegraphics[width=0.4\textwidth, angle=0]{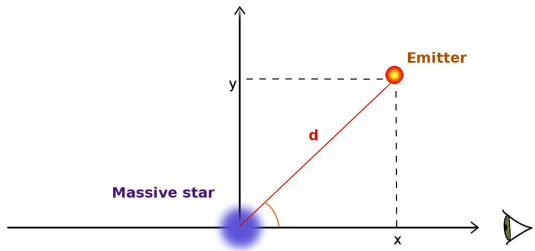}
    \caption[]{Sketch of the one-zone model considered in this work.}
    \label{fig:model}
  \end{figure}

High-mass binary systems 
can be characterized by the presence of a massive star and an accelerator of relativistic particles, which is 
a shock between colliding winds or a mechanism associated with a microquasar jet.
We consider both the massive star and the accelerator, regardless of its nature,
as point-like and, thus, homogeneous objects here. In addition, the accelerator and the emitter are assumed steady
and co-spatial, as electrons cannot travel long distances while 
emitting because of the short-cooling timescales. A sketch of the model is shown in Fig.~\ref{fig:model}.

The injection of relativistic electrons in the emitter is taken to follow an energy distribution 
$Q(E) \propto E^{-2}\exp{(-E/E_{\mathrm{max}})}$ for energies above 1 MeV, which is consistent with a Fermi~I acceleration 
process and also compatible with the 
observational features of the X-ray emission \citep[e.g.,][]{takahashi2009,zabalza2011a}. The relativistic 
electrons interact with the emitter magnetic and ambient stellar photon field producing a broad radiation spectrum.
The electron maximum energy (the cutoff energy $E_{\mathrm{max}}$ above) can be obtained by equating the cooling time to their
 acceleration time plus the constraint
derived from comparing the accelerator/emitter size and the particle gyroradius: $R>r_{\rm g}=E_{\rm max}/qB$. For simplicity,
 however, we take here $E_{\mathrm{max}}=30$~TeV, as
this is the expected maximum electron energy in LS~5039, the source studied below \citep[][such an energy requires a highly
 efficient accelerator]{khangulyan2008}.

Particles can lose energy through non-radiative losses, which can be through adiabatic cooling or particle escape. Their characteristic 
timescale is characterized as $t_{\mathrm{ad}}=d/v$ \citep[see, e.g.][]{takahashi2009}, where $d$ is the distance from the emitter to
 the star (a {\it loose} upper limit for the emitter size) and $v$ the velocity of the emitting flow.
The radiative processes dominant here are IC scattering and synchrotron emission, whereas VHE gamma-ray absorption takes place through 
pair production in the stellar photon field \citep{gould1967,blumenthal1970,aharonian1981}. We notice that the IC scattering 
takes place in the Klein-Nishina (KN) regime at such high energies and, therefore, has to be computed under such formalism.

We have not considered radiation reprocessing, although an electromagnetic (EM) IC cascade can develop for weak enough magnetic fields, 
increasing the effective transparency to VHE photons, while the secondary pair radiation can 
overcome the X-rays from the primary electron distribution in the emitter for stronger magnetic fields \citep{boschramon2008a}. 
We have also assumed that the emitting flow is at most mildly relativistic, as it would be the case for a standing shock in a
jet or a wind-colliding region, and, thus, we have not accounted
for Doppler boosting, which would significantly increase the model geometrical parameters. A thorough, albeit qualitative, discussion
of the impact of these assumptions is worthy and is presented in Sect.~\ref{disc}.

With all these considerations, only two parameters remain free in our model: the escape velocity, $v$, and the magnetic field 
to stellar photon energy density ratio, $\xi = u_{\mathrm{mag}}/u_{\mathrm{rad}}$. The value of $v$ is expected to be in the 
range $\sim 10^8$ -- $3\times10^{10}$~cm~s$^{-1}$, as it seems reasonable that the flow speed is between the stellar wind velocity 
and close to the speed of light ($c$); in particular, we adopt the values $v=10^8$~cm~s$^{-1}$ and $v=c$. As for $\xi$, a
reasonable range to study is $\xi=10^{-4}$ -- 1, as it goes from virtually no synchrotron cooling to a case when it becomes dominant. 
We have explored the parameter values $\xi=10^{-4}$, $10^{-2}$ and $1$.

\subsubsection{SEDs and maps: Fluxes, spectra and other emitter properties}\label{sedmap}

The spectral energy distribution (SED) is a measure of the amount of energy emitted per time and area units in a certain energy 
region. As both the IC scattering emission and the pair-production absorption strongly depend on the interaction angle, 
the emitter-star-observer geometry 
plays a crucial role in the resulting SEDs. Nonetheless, the study of individual SEDs may not be clear enough to explore these 
geometrical aspects when the emitter structure and location within the system are not known. Alternatively, one can make use of 
maps to display all the emitter spatial possibilities at once. First, one computes the particle population and the (absorbed) 
emission from an emitter placed in all the possible locations of the star-emitter-observer plane. Once this is done, one can 
extract any relevant quantity for each location and display it in the form of a two-dimensional map \citep[see, e.g.,][for similar
maps of gamma-ray absorption and particle acceleration]{dubus2006,khangulyan2008,boschramon2008b,romero2010}. 

We focus here on the total energy flux in the ranges of 0.3--10~keV ($F_{\rm X}$), 1--30~MeV ($F_{\rm MeV}$), 0.1--10~GeV 
($F_{\rm GeV}$), and 0.1--10~TeV ($F_{\rm TeV}$); the luminosity injected in the relativistic particles ($L_{\rm inj}$); 
and the minimum emitter radius
normalized to the stellar distance ($R/d$). We derive $R$ assuming a balance between the ram pressure of the stellar 
wind and the non-thermal electron pressure, which gives a lower limit for the emitter pressure. 

The energy fluxes, even without accounting for spectral features, already inform us if the model reproduces the observations by
comparing the predicted and the observed values at different bands. This comparison may suggest the presence of different 
populations of particles, or the occurrence (presence) of more processes (emitting sites) than those just assumed, for a particular 
source. Furthermore, the non-detection of sources with certain flux levels, or different-band flux combinations, can rule out the 
existence of objects with certain properties. 

The quantities $L_{\rm inj}$ and $R/d$ hint indirectly at major flaws of the model because either the luminosity budget is too 
high when we account for our background knowledge on the sources, or the point-like assumption ($R\gtrsim 0.5 \,d$) is violated. 
The former may suggest radiation beaming as a form of relaxing the energetic constraints, and the latter is probably pointing at a 
structured and extended emitter, although parameters/locations yielding $R\gg d$ are probably ruled out.

We have chosen $F_{\rm MeV}$ and $F_{\rm GeV}$ as the most relevant quantities for LS~5039, the source studied below. The observed 
values of $F_{\rm MeV}$ and $F_{\rm GeV}$ are very high, and trying to reconcile them with a broadband simple model, or studying 
their energetic and size requirements, might be very informative. Therefore, in what follows, the model fluxes in the MeV and 
GeV bands have been fixed to the observed values (although not simultaneously).

\subsection{An application to LS~5039}

The system LS~5039 is a well-studied high-mass binary located at 2.9~kpc \citep{moldon2012a}, which is proposed as a gamma-ray source 
by \cite{paredes2000} and confirmed by \cite{aharonian2005}. The effective temperature of the star is $T=3.9\times 10^4$~K; 
its radius, $R_* = 9.3\,R_{\odot}$ \citep{casares2005}; and the wind velocity, $v_\mathrm{w}(d) = 2400\,(1 - R_*/d)$~km~s$^{-1}$ 
\citep{kudritzki2000}. 
The Cn nature is still unclear; its mass is estimated to be $M = 3.7\,M_\odot$, but this value is highly dependent on the poorly 
known inclination of the orbit (for discussions on the nature of the Cn, see \citealt{casares2005}; and 
\citealt{boschramon2009,dubus2013} and references therein)\footnote{In addition, note that the radio structures detected
 in LS~5039 might be consistent with the presence of a non-accreting pulsar \citep{moldon2012b}.}. The orbit of LS~5039 
is mildly eccentric, $e=0.24-0.35$ \citep{casares2005,aragona2009,sarty2011}, with a semi-major axis $a\approx 3.5\,R_*$. We adopt
the value of $a$ as the binary spatial scale here. 

As mentioned, we normalize the results of the calculations to the detected MeV and GeV fluxes. In particular, the injection 
luminosity is set to reproduce the observed MeV and GeV energy fluxes: $F_{\rm GeV}=2.8 \times 10^{-10}$~erg~cm$^{-2}$~s$^{-1}$ 
in the range 0.1--10~GeV \citep{hadasch2012} and $F_{\rm MeV}=2.6\times 10^{-9}$~erg~cm$^{-2}$~s$^{-1}$
in the range 1--30~MeV \citep{collmar2014}. We note that the fluxes in these energy bands 
in LS~5039 vary along one orbit by a factor of a few, so we have taken intermediate values. The observed fluxes
in the X-ray and TeV energy bands are: $F_{\rm X}\approx (0.5-1.3) \times 10^{-11}$~erg~cm$^{-2}$~s$^{-1}$ \citep{takahashi2009} and 
$F_{\rm TeV}\approx (1.9-7.4)\times 10^{-12}$~erg~cm$^{-2}$~s$^{-1}$ \citep{aharonian2006}, depending on the orbital phase.

Accounting for the system and stellar parameters of LS~5039, the GeV and MeV fluxes, and the magnetic field and emitter 
velocity ranges introduced in Sect.~\ref{descr}, we have computed several broadband SEDs and maps of the quantities 
listed in Sect.~\ref{sedmap}.  

We adopt an upper-limit for the non-thermal injection luminosity of $\sim 10^{37}$~erg~s$^{-1}$, which is roughly what 
is required to explain the inferred MeV luminosities \citep{collmar2014}. Higher values would be in conflict with the
 two possible scenarios of LS~5039: For the accretion-jet scenario, such a powerful non-thermal emitter is in 
conflict with the lack of accretion features in the X-ray spectrum and present models of jet formation
\citep{boschramon2007,rea2011,barkov2012}. For the pulsar scenario, the lack of thermal X-ray evidence points
to spin-down luminosities below $10^{37}$~erg~s$^{-1}$ \citep{zabalza2011b}.

\section{Results}\label{res}

Several SEDs and a set of maps have been obtained for LS~5039. The results presented in this section are just a subset 
of those obtained and are chosen as the most useful to provide an insight into the physics of LS~5039. They also illustrate 
the approach described in the previous section. In the Appendix, maps for the extreme parameter cases are shown to provide 
a wider context to the maps presented here.

\subsection{Spectral energy distributions}\label{sed}

\begin{figure}
\centering
\includegraphics[width=0.262\textwidth, angle=270]{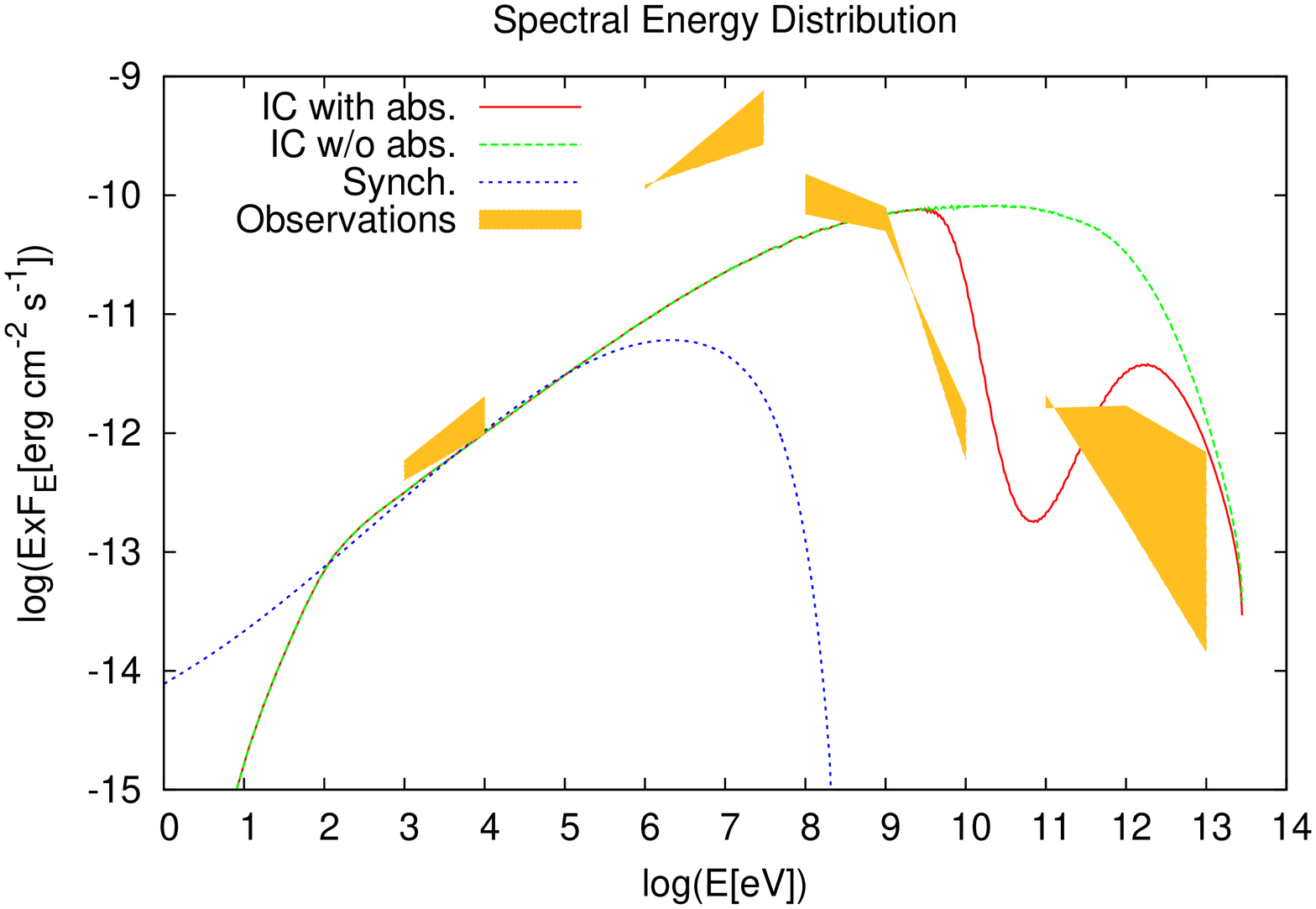}
\caption{Spectral energy distribution for an emitter with fast non-radiative losses and a weak magnetic field
located at $x=-a$ and $y=a$. The massive star is at (0,0), and the observer in the positive $x$-axis direction. The normalization was set 
  to reproduce an energy flux in the $0.1$--$10$ GeV range equal to $2.8 \times 10^{-10}$~erg~cm$^{-2}$~s$^{-1}$. Observational 
  constraints in X-rays, MeV, GeV, and TeV energies are also presented.}
\label{fig:10sed1}

\centering
\includegraphics[width=0.262\textwidth, angle=270]{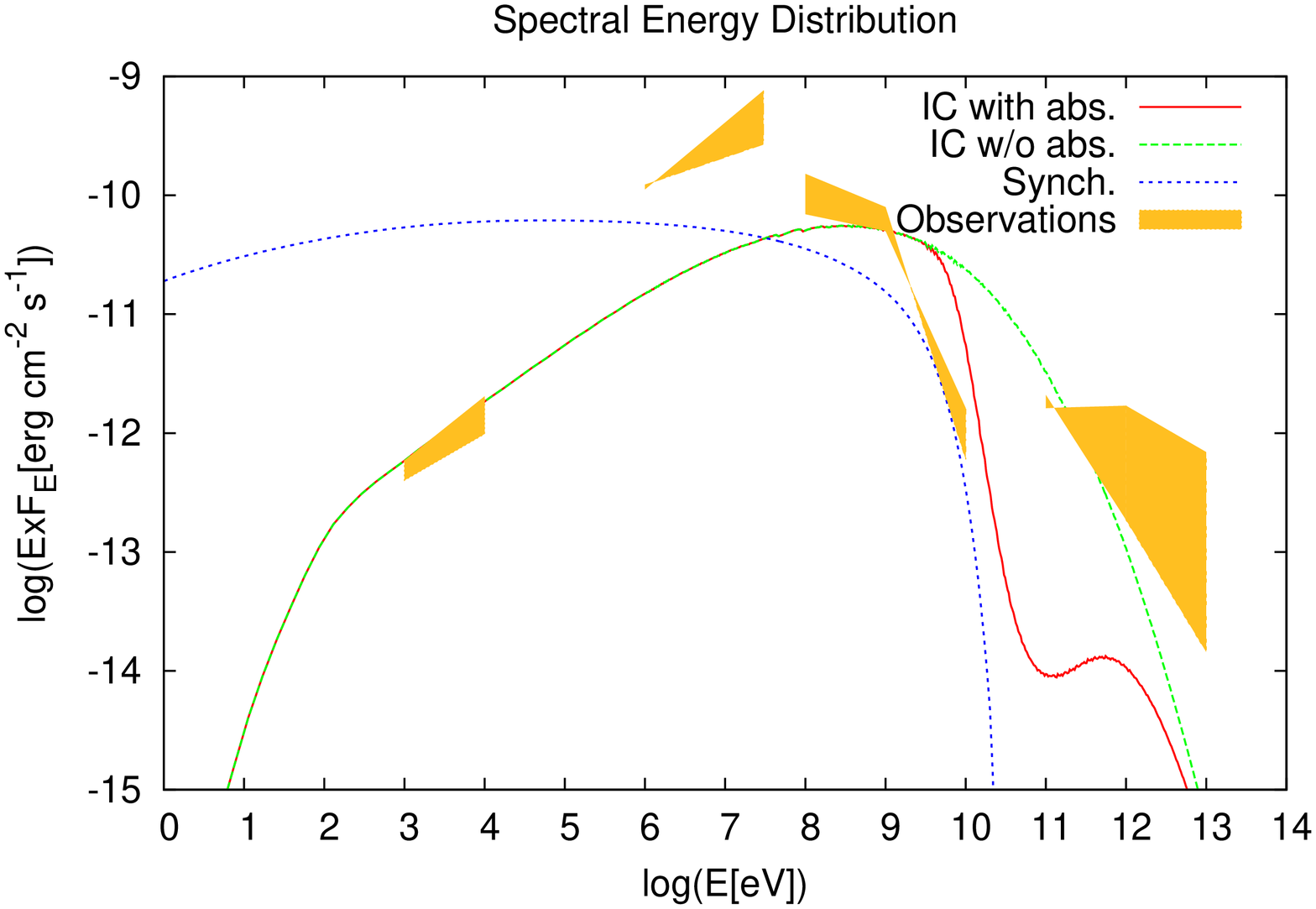}
\caption{As in Fig.~\ref{fig:10sed1} but for an emitter with fast non-radiative losses and a strong magnetic field.}
\label{fig:11sed1}
  
\centering
\includegraphics[width=0.262\textwidth, angle=270]{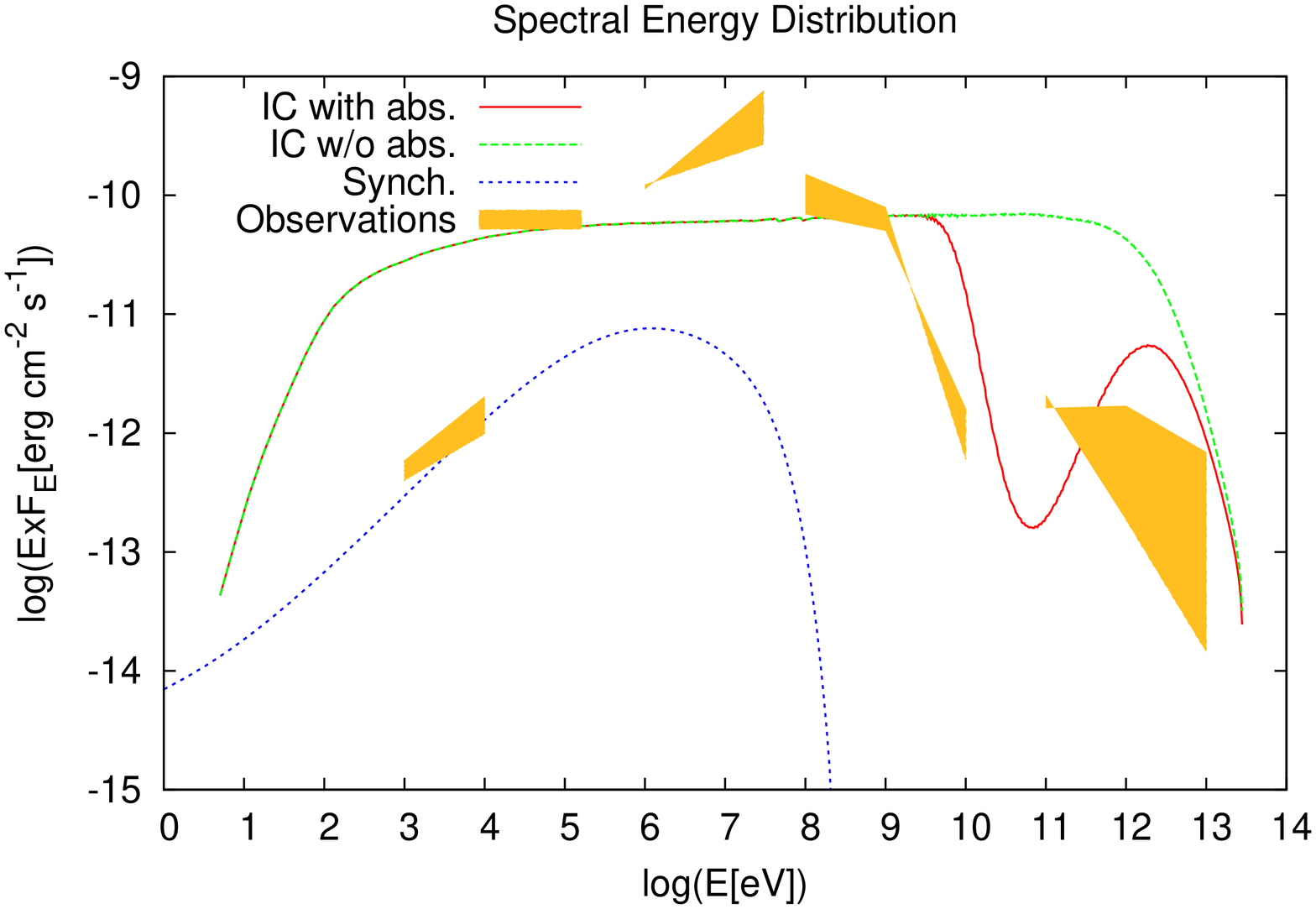}
\caption{As in Fig.~\ref{fig:10sed1} but for an emitter with slow non-radiative losses and a weak magnetic field.}
\label{fig:00sed1} 
  
\centering
\includegraphics[width=0.262\textwidth, angle=270]{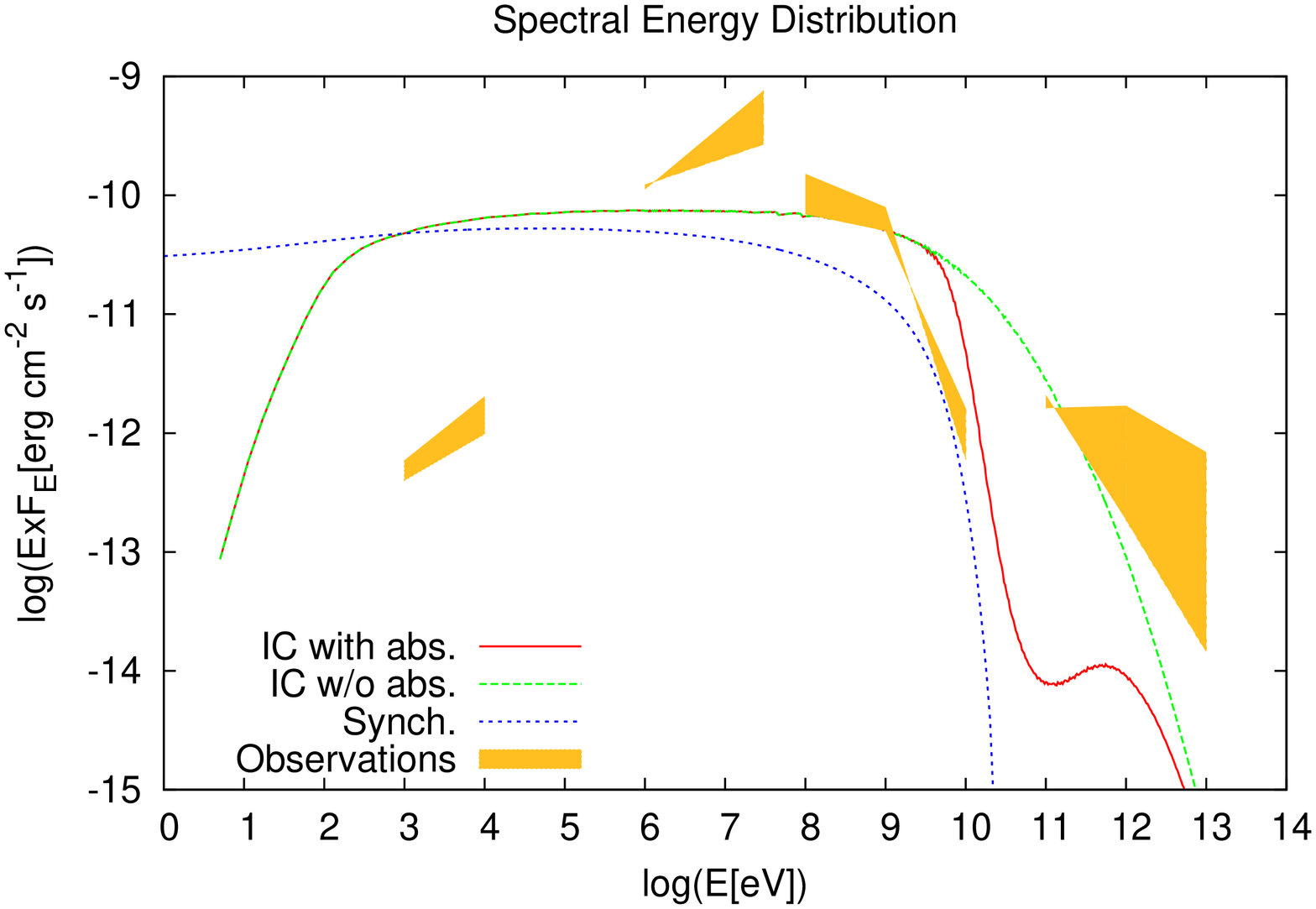}
\caption{As in Fig.~\ref{fig:10sed1} but for an emitter with slow non-radiative losses and a strong magnetic field.}
\label{fig:01sed1}
  
\end{figure}

\begin{figure}

\centering
\includegraphics[width=0.262\textwidth, angle=270]{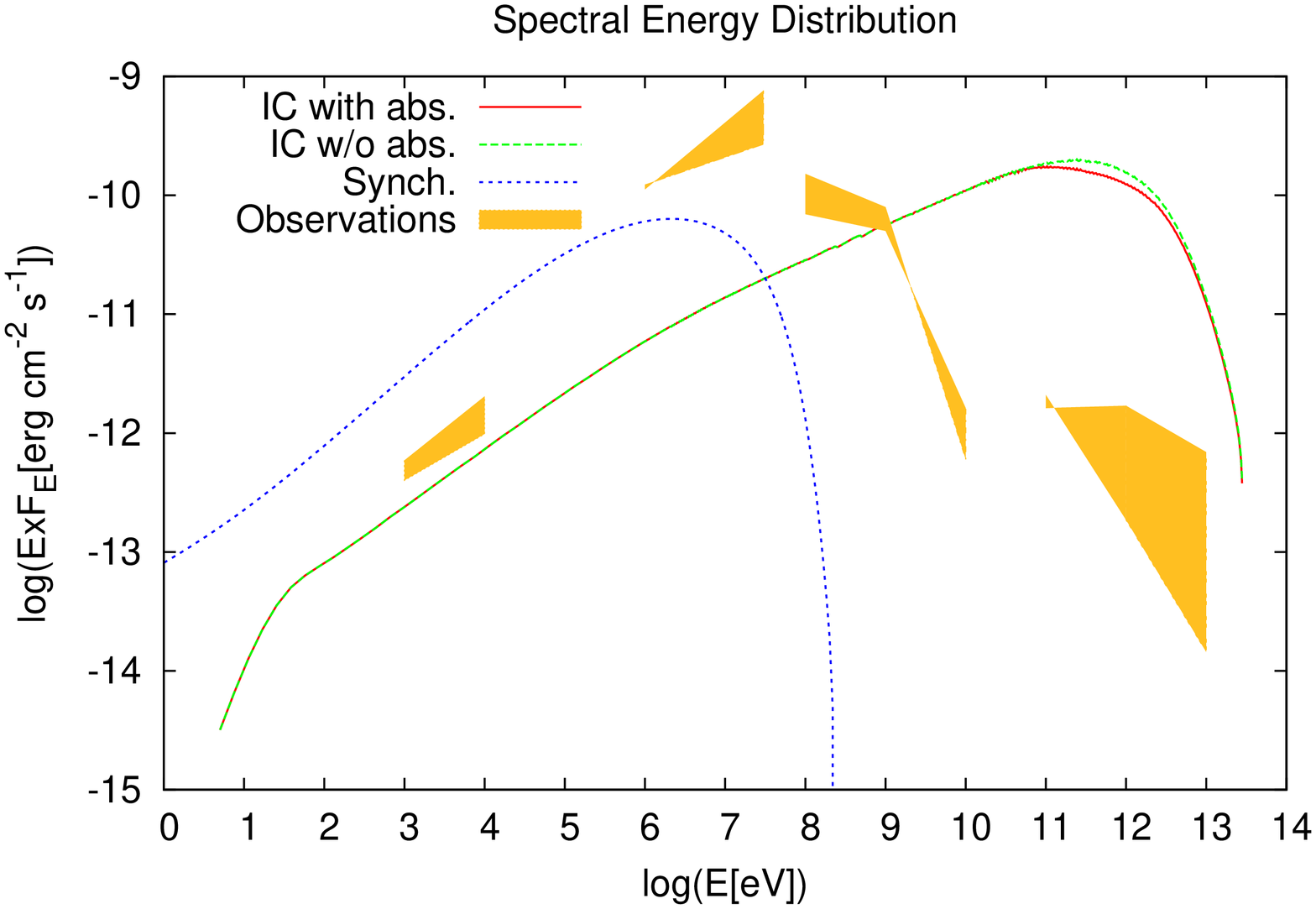}
\caption{Spectral energy distribution for an emitter with fast non-radiative losses and a weak magnetic field
located at $x=a$ and $y=a$. The massive star is at (0,0), and the observer in the positive $x$-axis direction. The normalization was set 
  to reproduce an energy flux in the $0.1$--$10$ GeV range equal to $2.8 \times 10^{-10}$~erg~cm$^{-2}$~s$^{-1}$. 
  Observational constraints in 
 X-rays, MeV, GeV, and TeV energies are also presented.} 
\label{fig:10sed2}

\centering
\includegraphics[width=0.262\textwidth, angle=270]{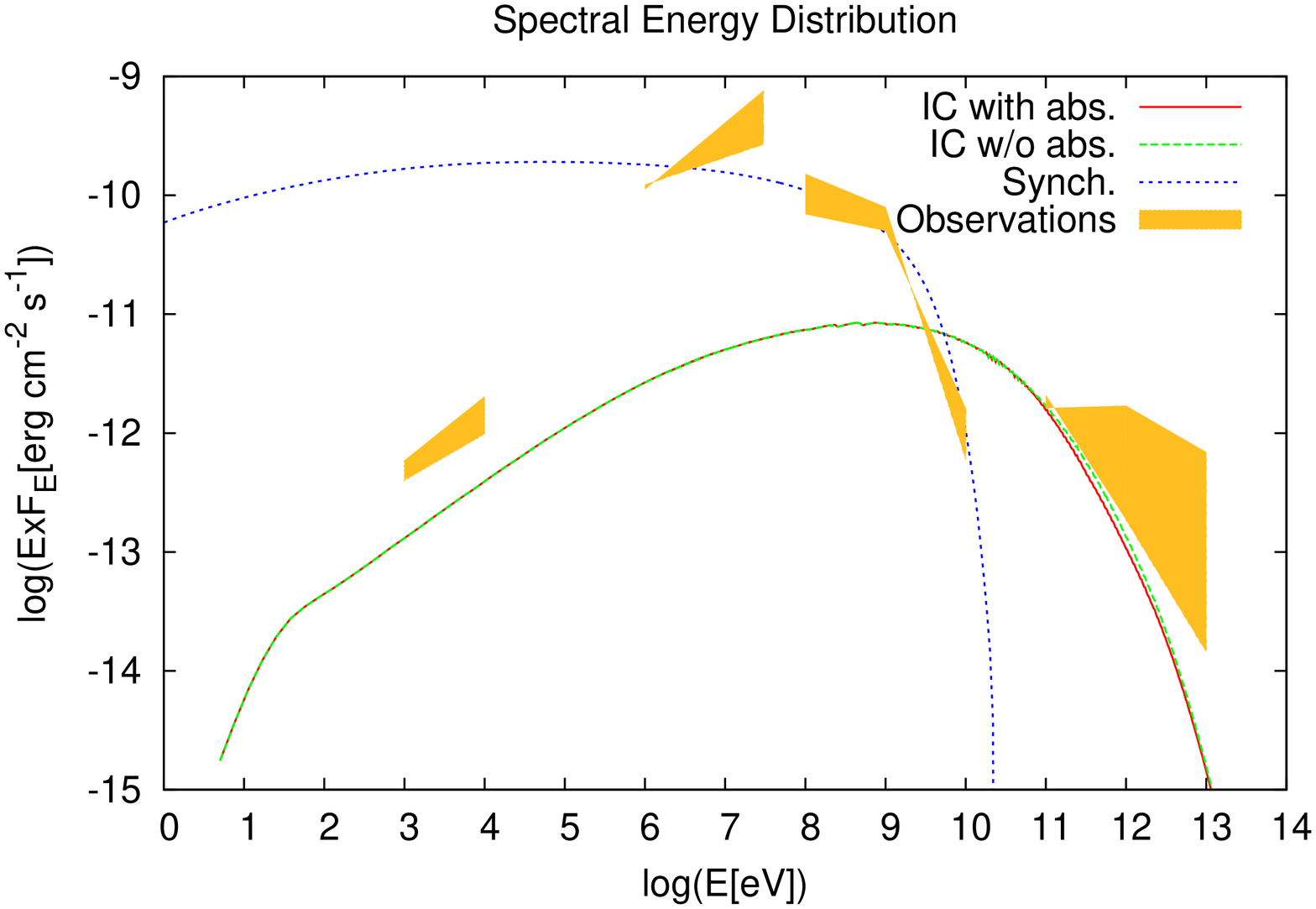}
\caption{As in Fig.~\ref{fig:10sed2} but for an emitter with fast non-radiative losses and a strong magnetic field.}
\label{fig:11sed2}
  
\centering
\includegraphics[width=0.262\textwidth, angle=270]{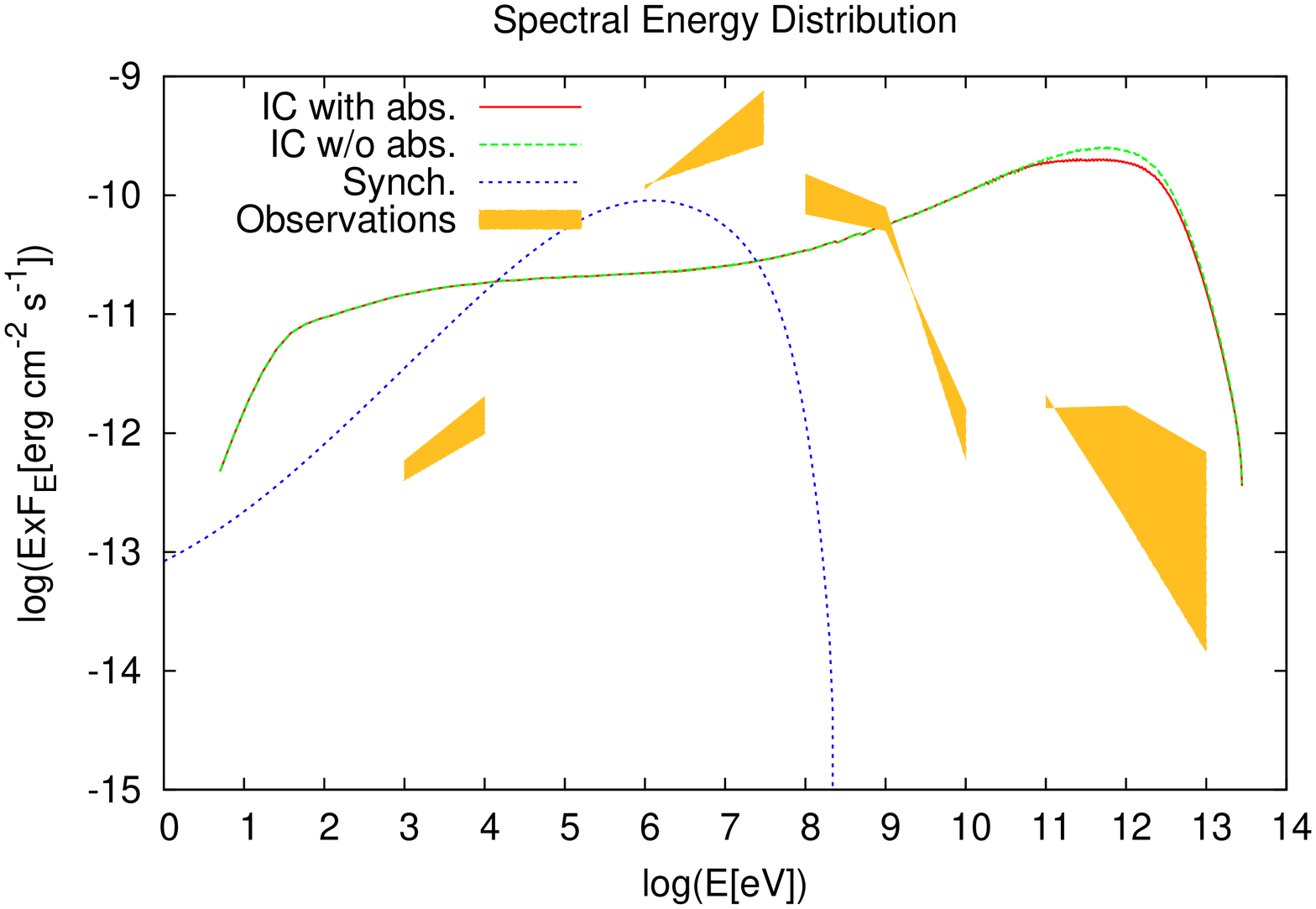}
\caption{As in Fig.~\ref{fig:10sed2} but for an emitter with slow non-radiative losses and a weak magnetic field.}
\label{fig:00sed2} 
  
\centering
\includegraphics[width=0.262\textwidth, angle=270]{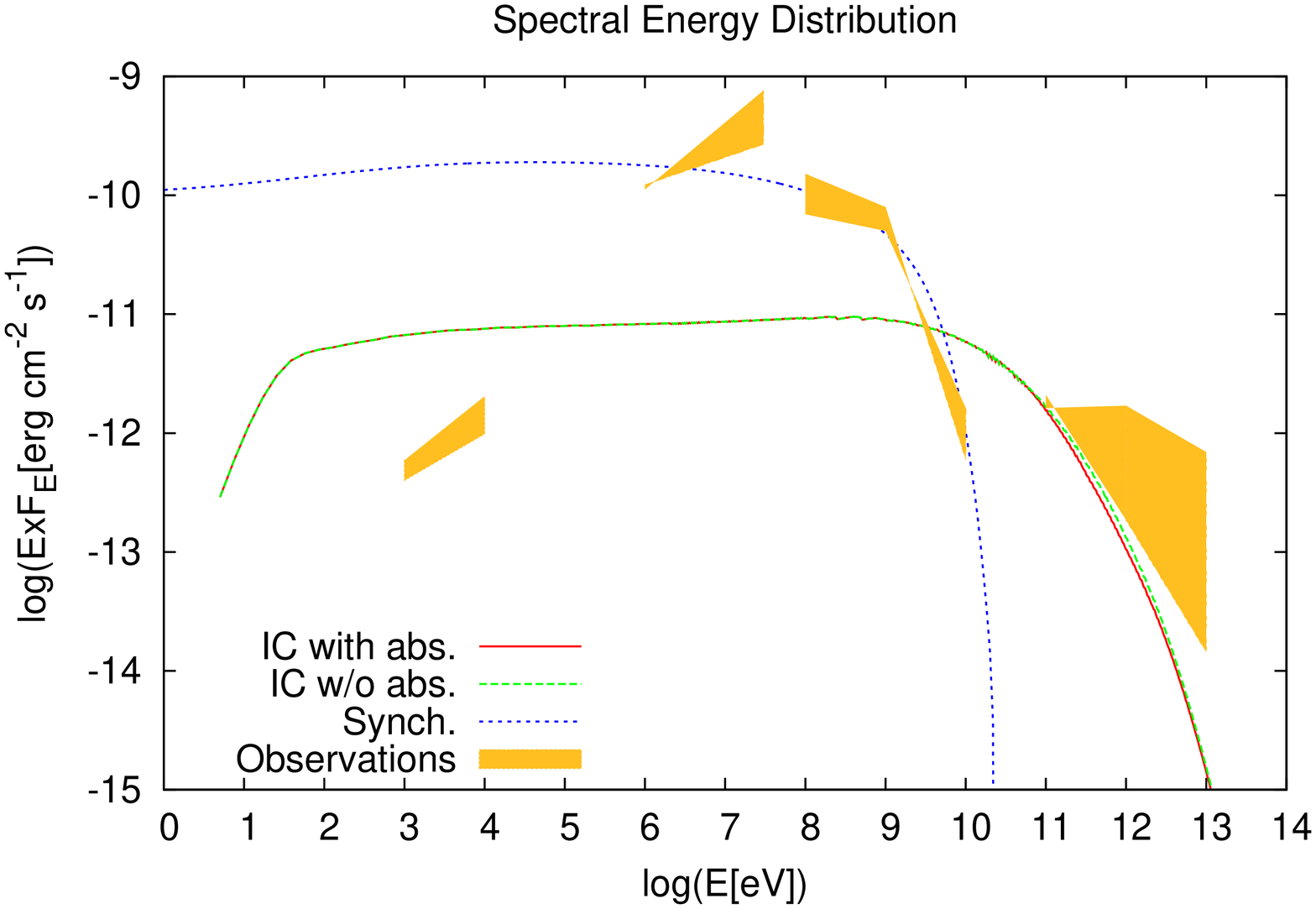}
\caption{As in Fig.~\ref{fig:10sed2} but for an emitter with slow non-radiative losses and a strong magnetic field.}
\label{fig:01sed2}
  
\end{figure}

Just to give a qualitative idea of how the different factors considered here (magnetic field intensity, importance of non-radiative losses 
and emitter location) affect the resulting spectra, we show SEDs for emitters with different locations and properties in 
Figs.~\ref{fig:10sed1}-\ref{fig:01sed2}. We present some extreme cases of fast/slow non-radiative losses ($v=c$ and $v=10^8$ cm s$^{-1}$, 
respectively), high/low magnetic fields ($\xi=1$ and $\xi=10^{-4}$, respectively, with $\xi = u_{\mathrm{mag}}/u_{\mathrm{rad}}$, 
which translates into $B$-fields of 
$\sim 10$ -- $10^2$ G and $\sim 0.1$ -- $1$ G, respectively, depending on location), and less extreme configurations of 
the emitter-star-observer. We choose the emitter positions $(-a,a)$ and $(a,a)$, which correspond to an emitter located 
roughly behind/in front of the star with respect to the observer and at a distance on the order of the system size. 
In all cases, the normalization was set to $F_{\rm GeV}=2.8 \times 10^{-10}$~erg~cm$^{-2}$~s$^{-1}$. We also show the various 
levels of observed 
emission in the four energy bands considered (X-rays, MeV, GeV, TeV), so the reader can easily judge the quality of the match.

These figures show the typical behavior for a synchrotron/IC one-zone emitter
with different IC and absorption geometries under different cooling regimes
\cite[see, e.g.][]{khangulyan2008,dubus2008,takahashi2009,zabalza2011a}. 

Regardless of the dominant cooling mechanism, gamma-ray absorption is weaker (stronger) and its maximum occurs at higher (lower)
energies when the emitter is in {\it front} of ({\it behind}) the star. The IC component becomes harder (softer) and, as well as the
absorption due to pair creation,
also weaker (stronger) when the emitter is in {\it front} of ({\it behind}) the star. Furthermore, the pair-creation threshold shifts
to higher energies when the emitter is in {\it front} of the star. All these
variations are related to the different electron-photon and photon-photon
interaction angles as seen from the observer such that smaller (larger) angles correspond to an emitter
{\it in front} of ({\it behind}) the star.  On the other hand, synchrotron
radiation is not affected by changes in the geometry (at the same distance to the star, which translates in the same value of the 
magnetic field).
This is clear in the figures, as no spectral changes are seen in the synchrotron component
between both geometries for the same cooling setup. 

Concerning the dominant cooling process, both synchrotron and IC emission under IC dominance become harder at the
energies in which the KN effect is relevant, as the KN IC cross section strongly drops
with energy. As this drop is actually faster than in synchrotron, it
implies that synchrotron cooling easily overcomes IC as the main cooling channel at the highest energies. On the other hand, 
synchrotron and IC in Thomson regime both
soften the radiation spectra, yielding flat SEDs right above the energies in
which non-radiative losses are relevant. Finally, the impact of non-radiative
losses is to harden (unless it had to compete with KN IC cooling) the synchrotron and IC spectra below a given energy, which is
higher when these losses are faster. Given the fixed normalization, it is not
possible to directly compare the levels of emission between plots with
different parameter choices.

In Sect.~\ref{chi2}, we show a method for finding a rough approximation of the best fit parameters, emitter position, and normalization.

\subsection{Maps} \label{sec:maps}

\begin{figure}
  \centering
  \includegraphics[width=0.2\textwidth, angle=270]{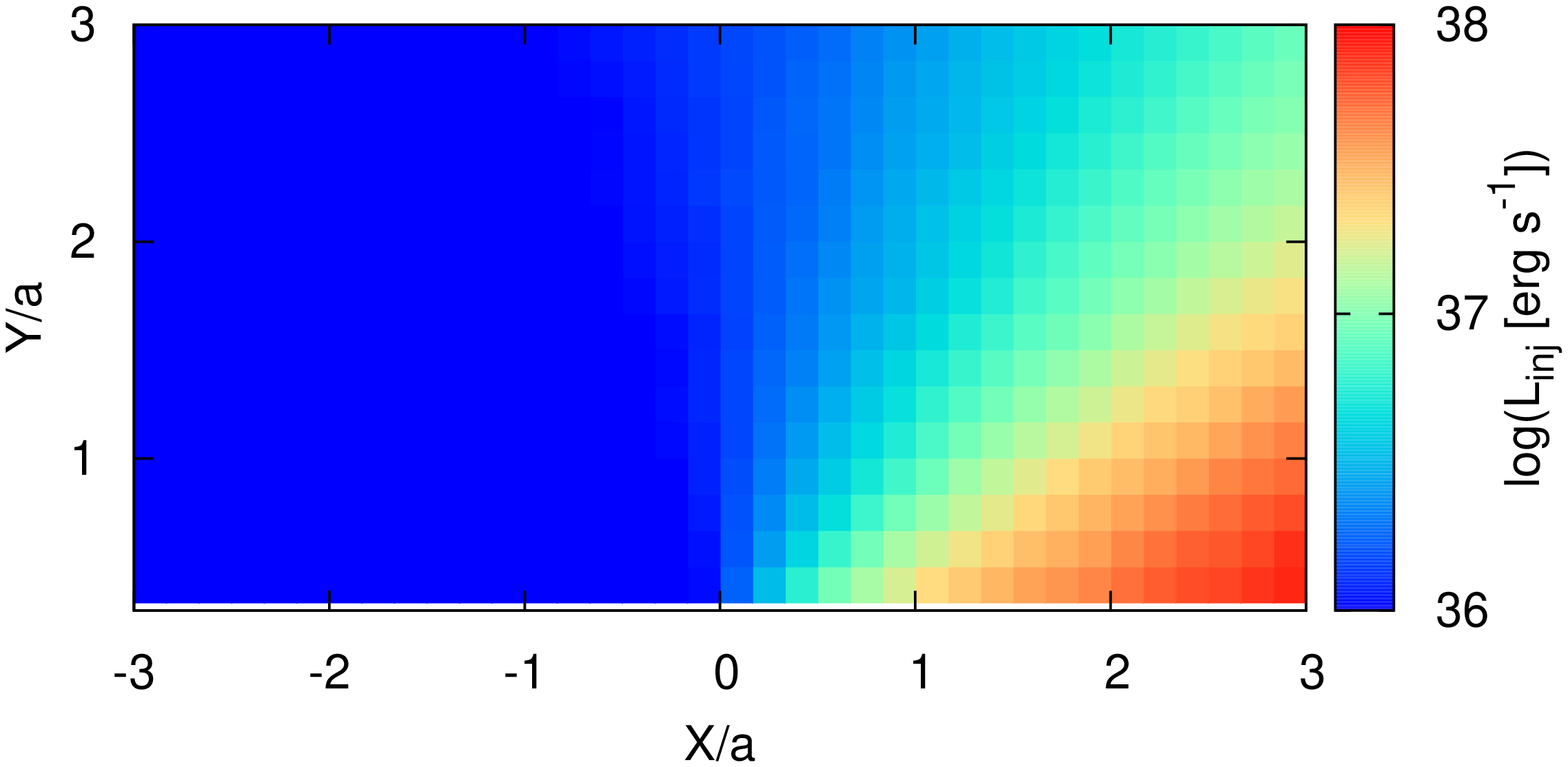}
  \caption[eta=0,delta=0]{Injection luminosity of relativistic particles in the emitter in the case of 
  intermediate non-radiative losses and magnetic field. The normalization was set 
  to reproduce an energy flux in the $0.1$--$10$ GeV range equal to $2.8 \times 10^{-10}$~erg~cm$^{-2}$~s$^{-1}$.}
  \label{fig:gev22inj}

  \centering
  \includegraphics[width=0.2\textwidth, angle=270]{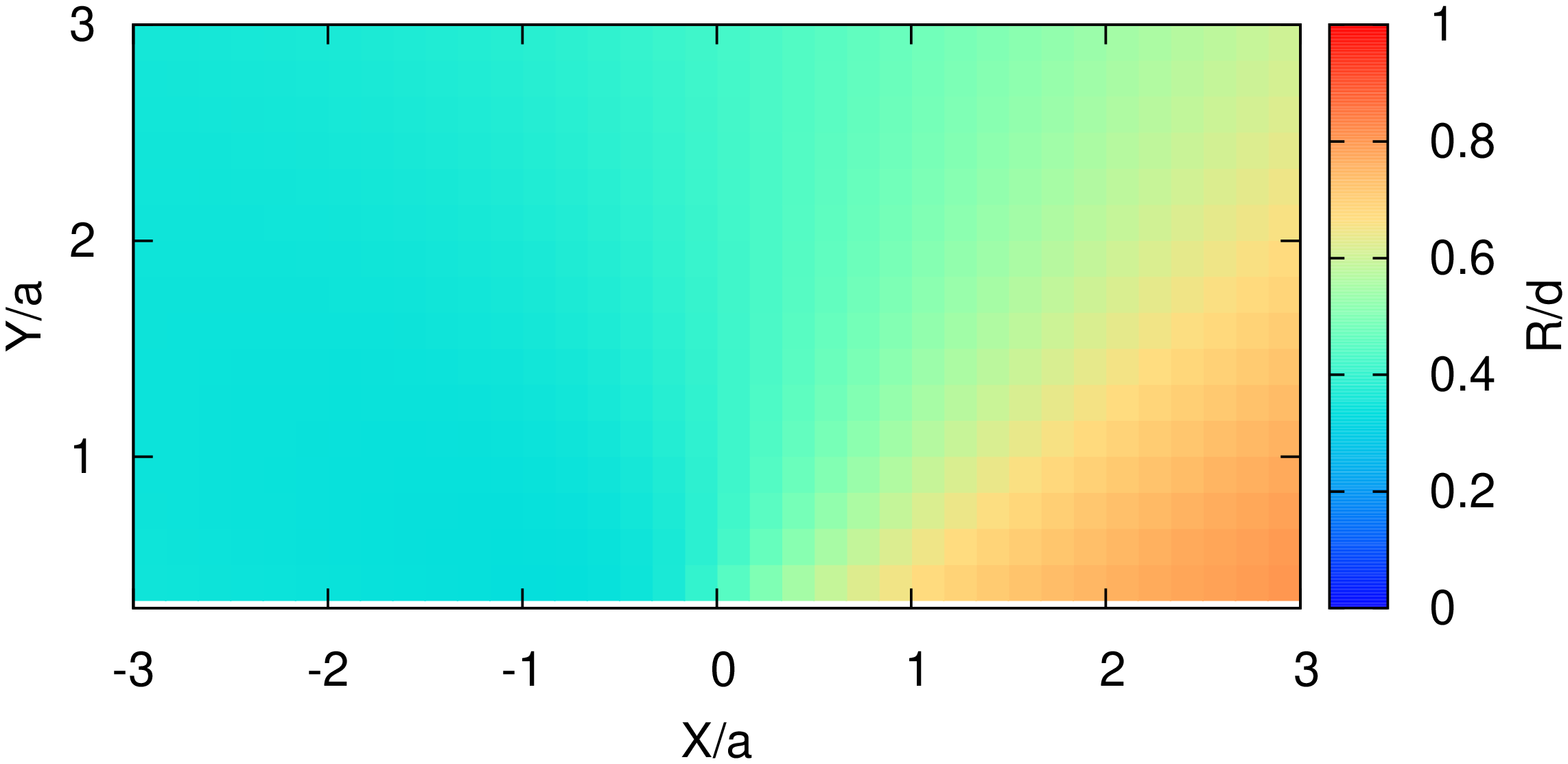}
  \caption[eta=0,delta=0]{As in Fig.~\ref{fig:gev22inj} but showing the emitter's size divided by its distance to the star.}
  \label{fig:gev22confi}

  \centering
  \includegraphics[width=0.2\textwidth, angle=270]{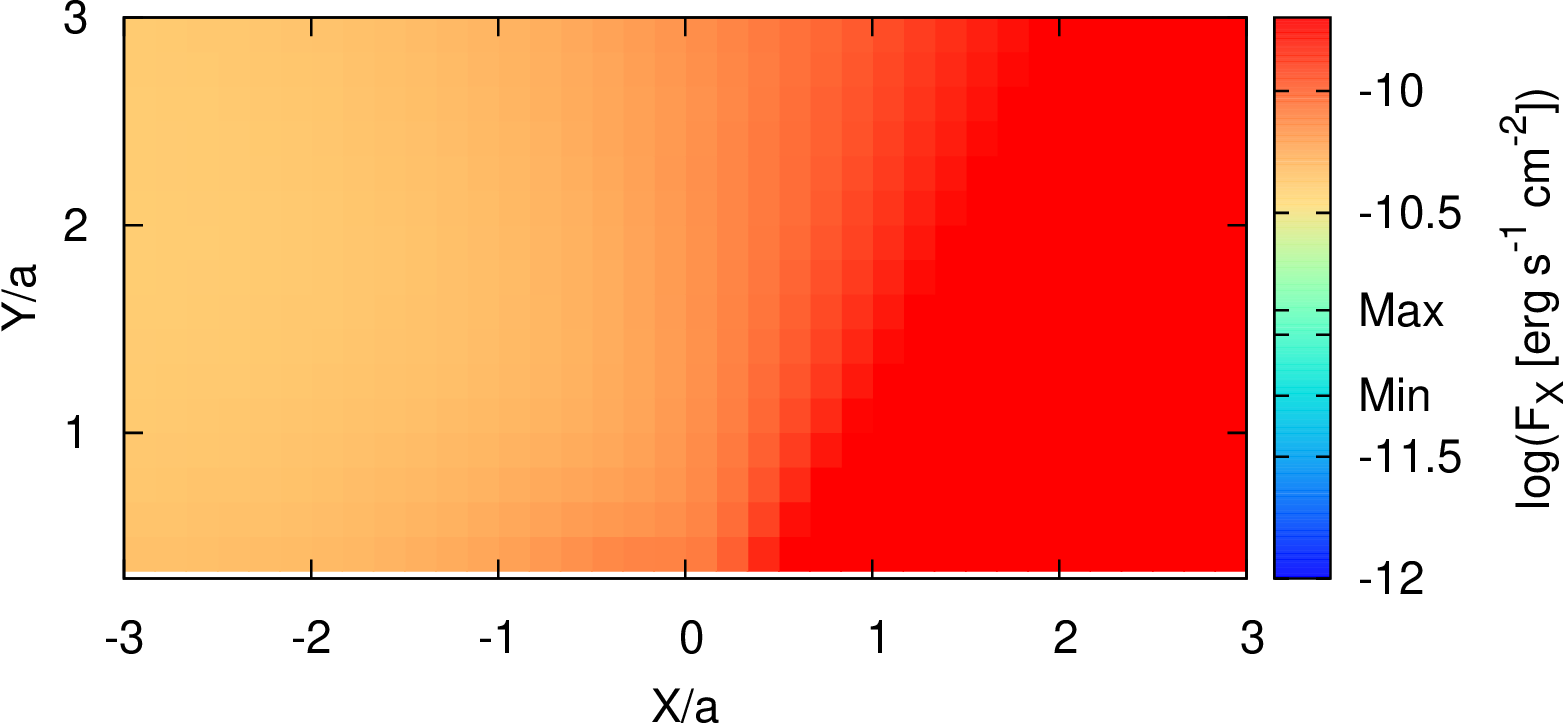}
  \caption[eta=0,delta=0]{As in Fig.~\ref{fig:gev22inj} but showing the integrated energy flux in the 0.3--10 keV energy band.}
  \label{fig:gev22x}

  \centering
  \includegraphics[width=0.2\textwidth, angle=270]{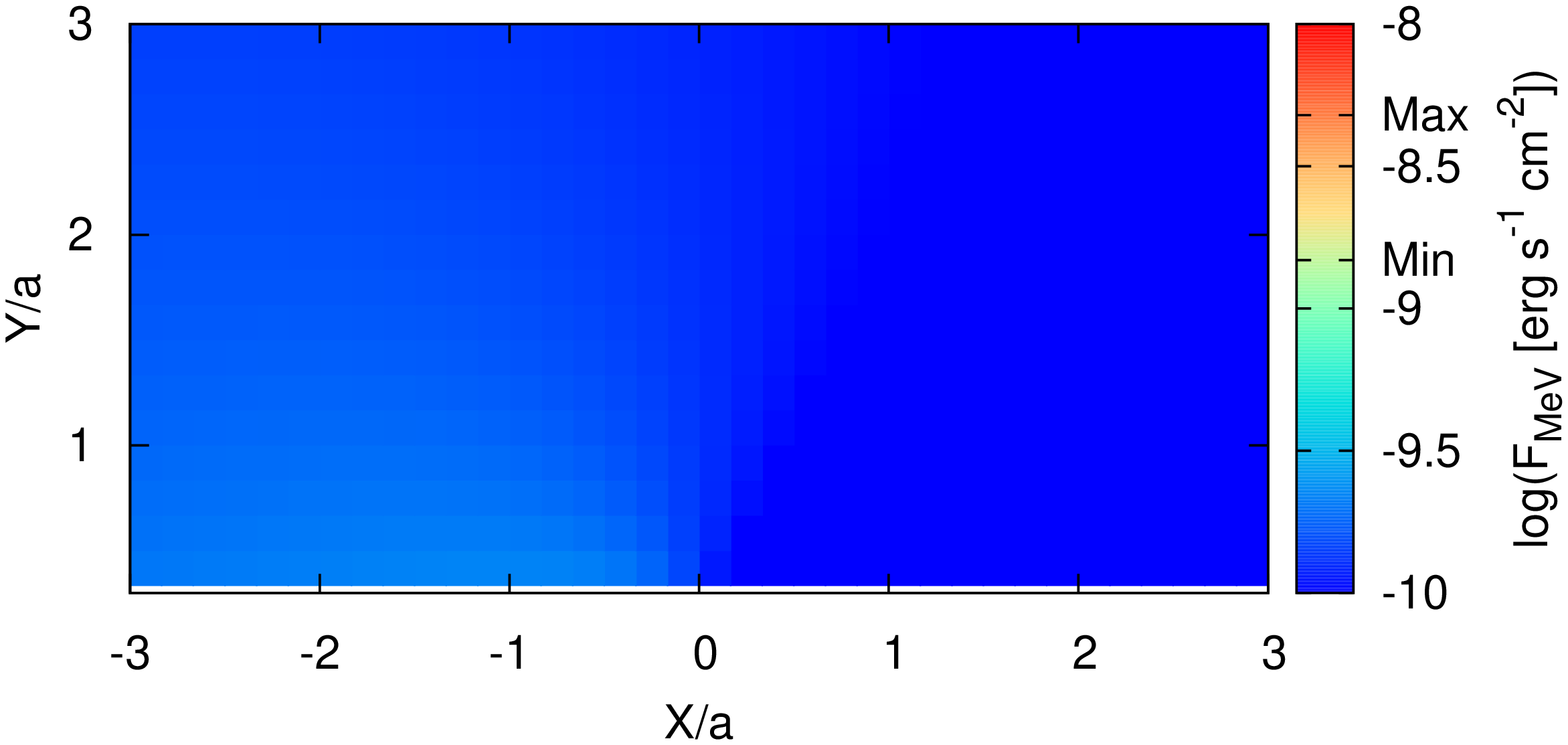}
  \caption[eta=0,delta=0]{As in Fig.~\ref{fig:gev22inj} but showing the integrated energy flux in the 1--30 MeV energy band.}
  \label{fig:gev22mev}
  
  \centering
  \includegraphics[width=0.2\textwidth, angle=270]{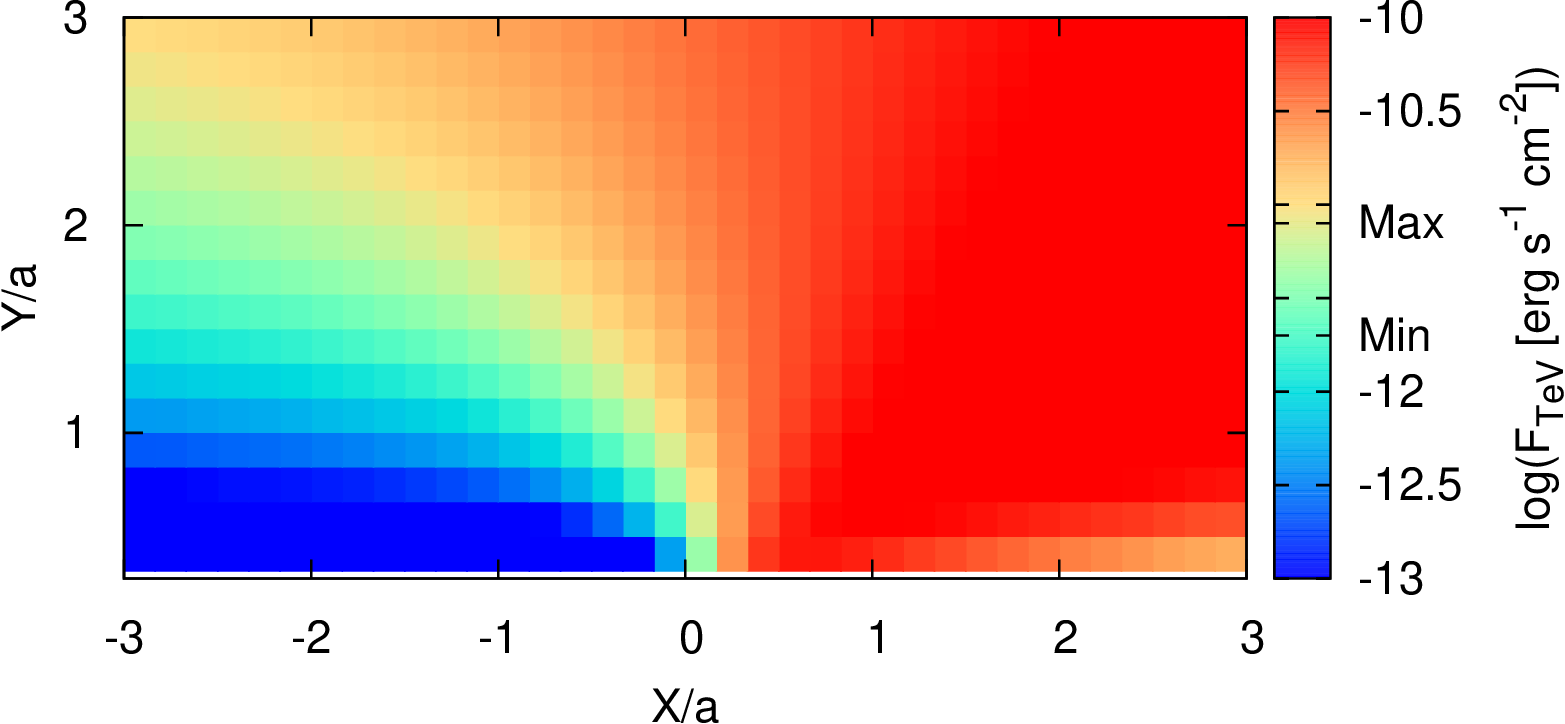}
  \caption[eta=0,delta=0]{As in Fig.~\ref{fig:gev22inj} but showing the integrated energy flux in the 0.1--10 TeV energy band.}
  \label{fig:gev22tev}
  \end{figure}

\begin{figure}
  \centering
  \includegraphics[width=0.2\textwidth, angle=270]{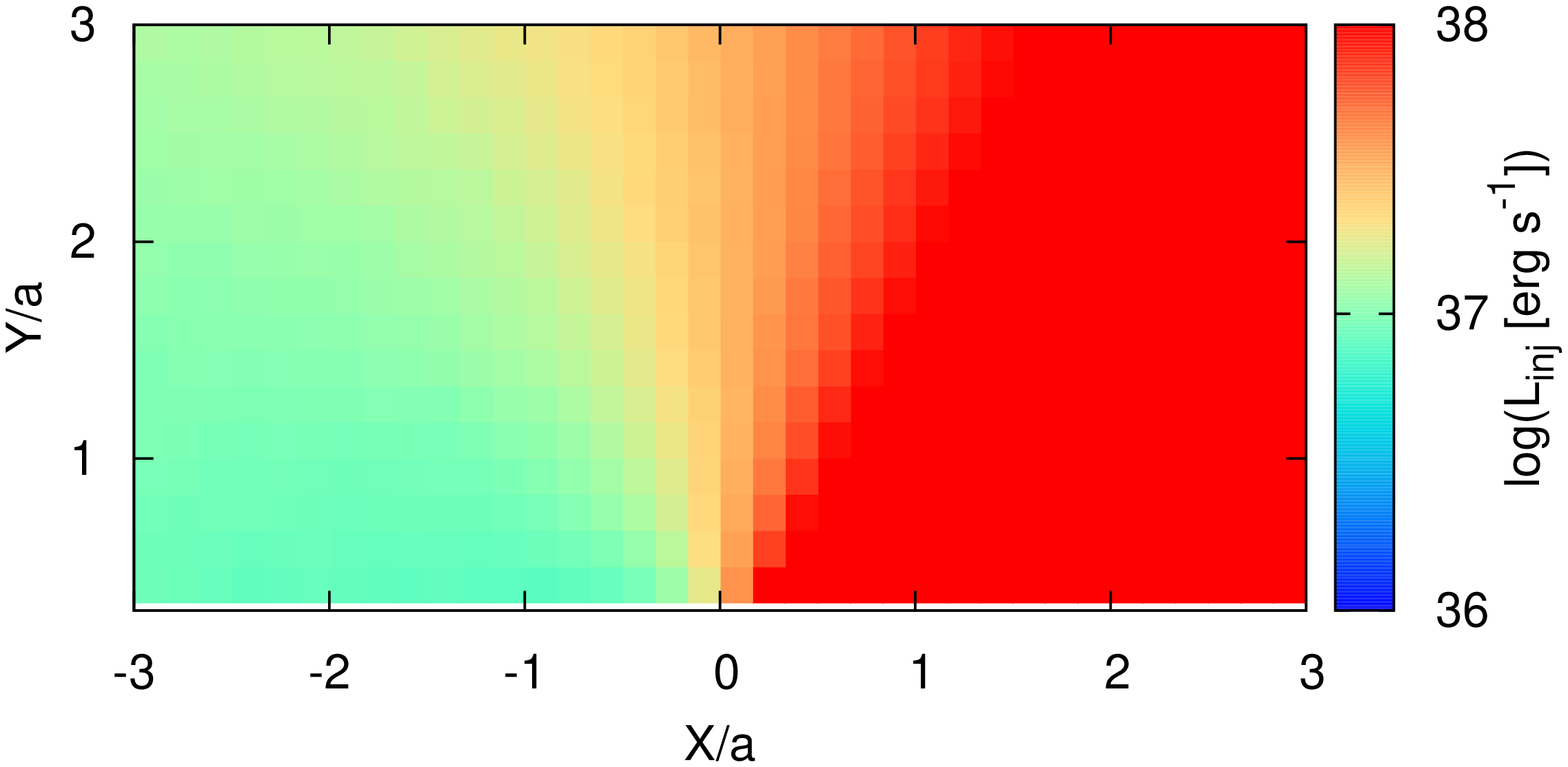}
  \caption[eta=1,delta=0]{Injection luminosity of relativistic particles in the emitter in the case of 
  intermediate non-radiative losses and magnetic field. The normalization was set 
  to reproduce an energy flux in the $1$--$30$ MeV range equal to $2.6 \times 10^{-9}$~erg~cm$^{-2}$~s$^{-1}$}
  \label{fig:mev22inj}

  \centering
  \includegraphics[width=0.2\textwidth, angle=270]{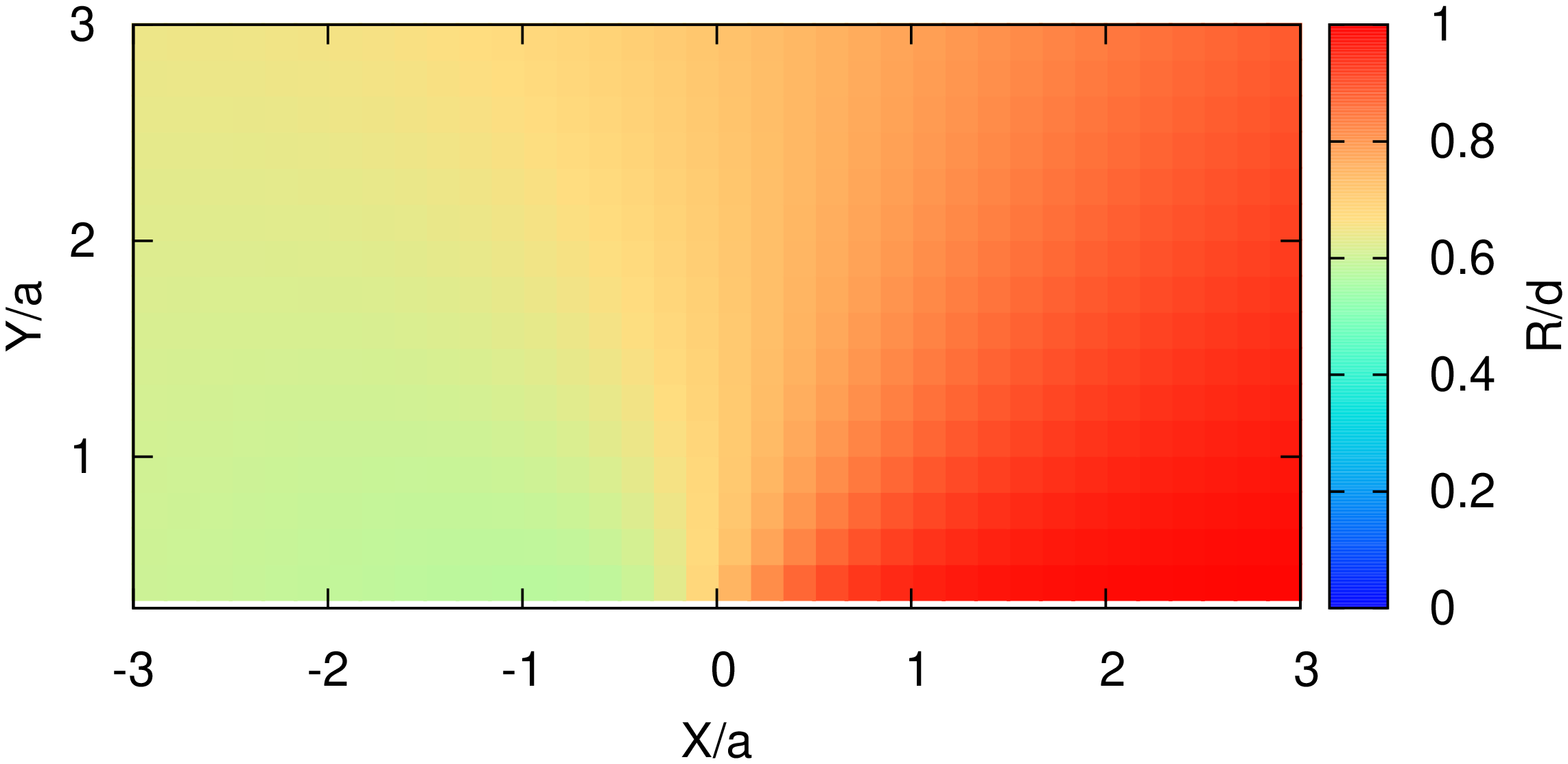}
  \caption[eta=1,delta=0]{As in Fig.~\ref{fig:mev22inj} but showing the emitter's size divided by its distance to the star.}
  \label{fig:mev22confi}

  \centering
  \includegraphics[width=0.2\textwidth, angle=270]{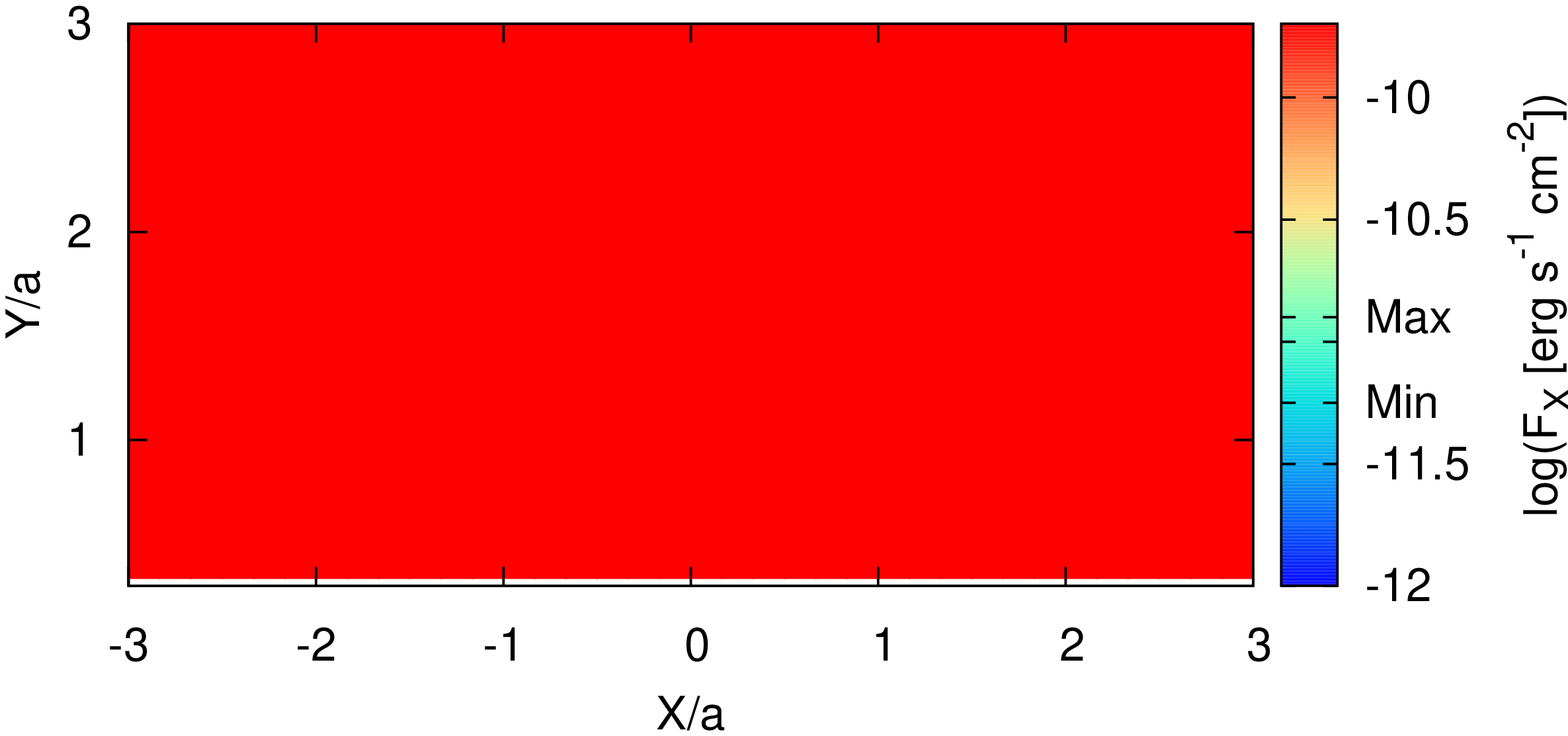}
  \caption[eta=1,delta=0]{As in Fig.~\ref{fig:mev22inj} but showing the integrated energy flux in the 0.3--10 keV energy band.}
  \label{fig:mev22x}

  \centering
  \includegraphics[width=0.2\textwidth, angle=270]{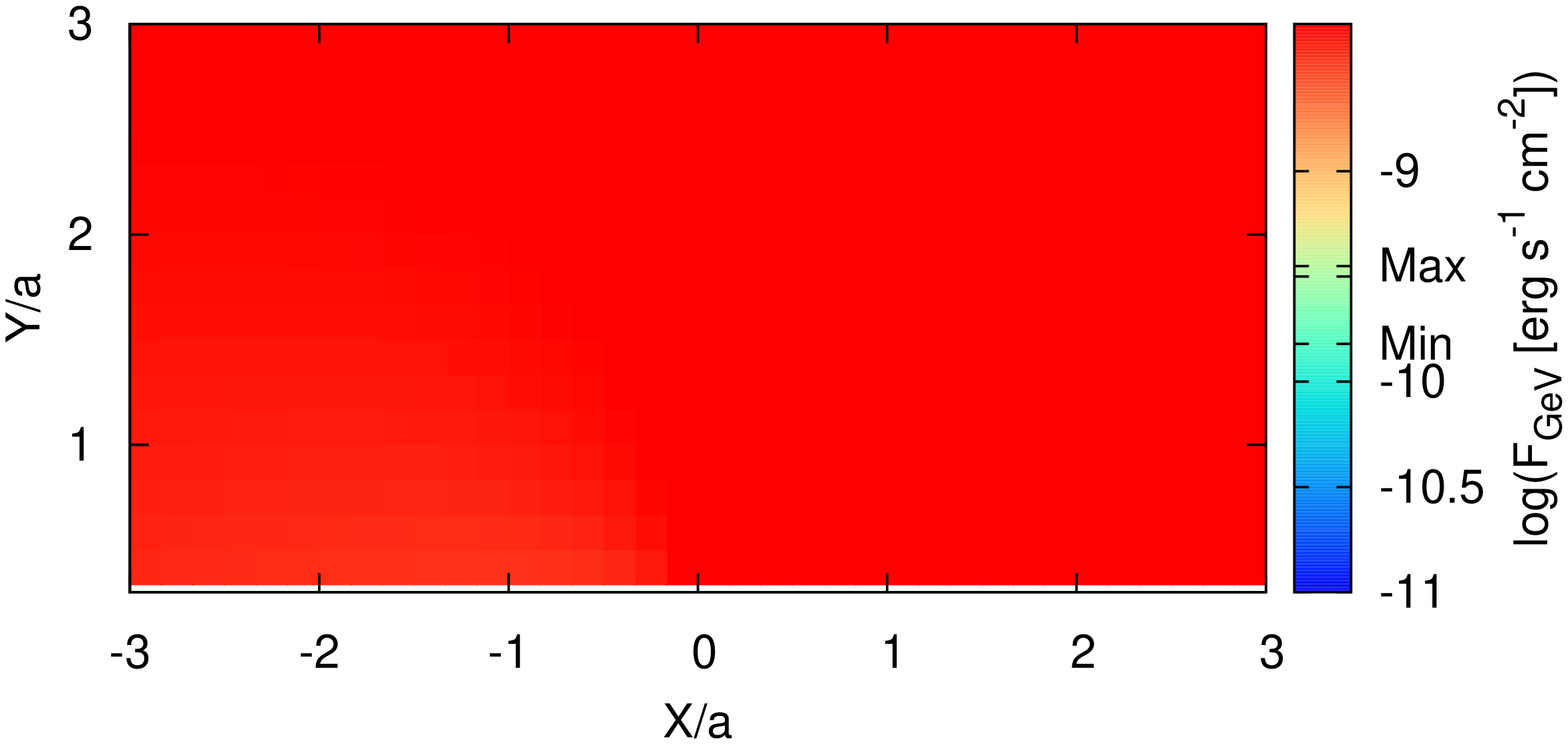}
  \caption[eta=1,delta=0]{As in Fig.~\ref{fig:mev22inj} but showing the integrated energy flux in the 0.1--10 GeV energy band.}
  \label{fig:mev22gev}

  \centering
  \includegraphics[width=0.2\textwidth, angle=270]{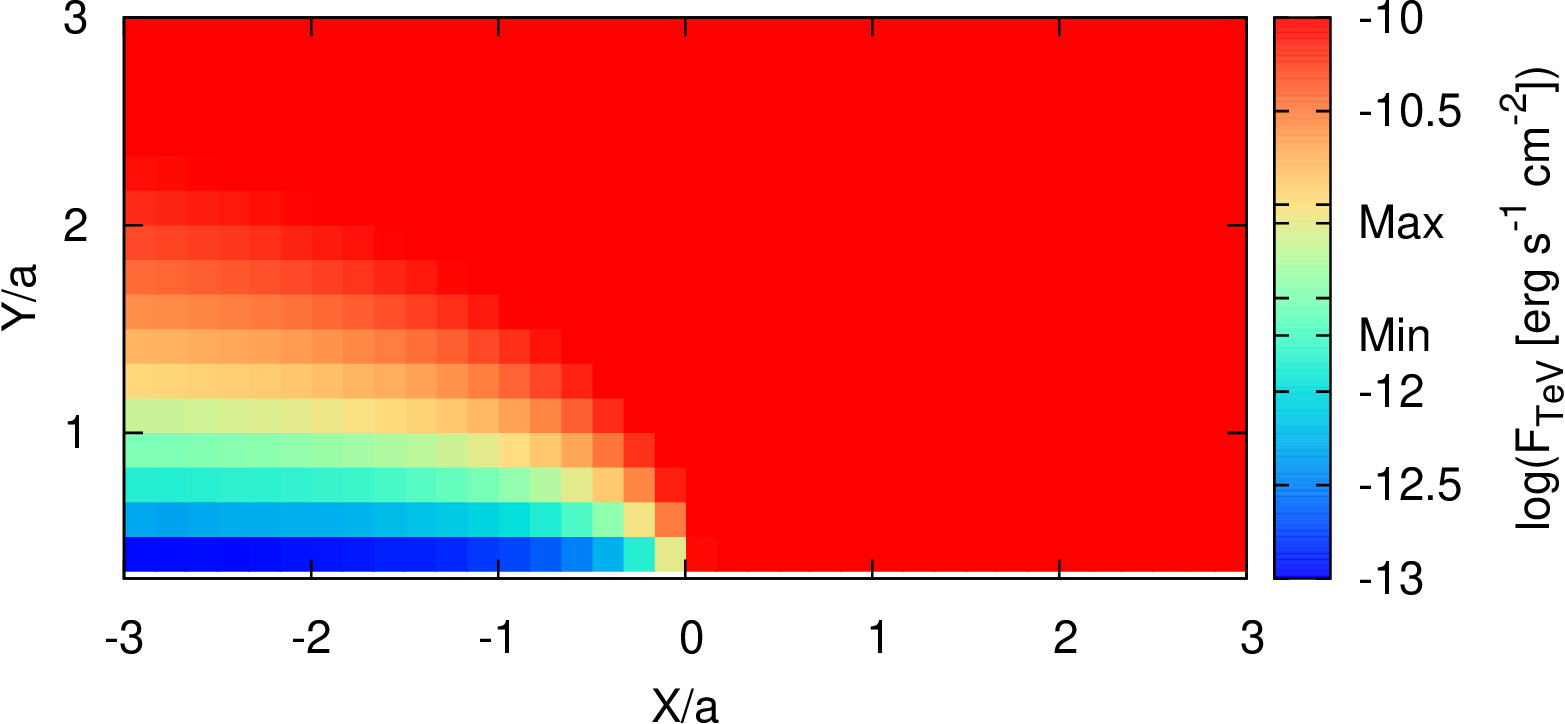}
  \caption[eta=1,delta=0]{As in Fig.~\ref{fig:mev22inj} but showing the integrated energy flux in the 0.1--10 TeV energy band.}
  \label{fig:mev22tev}
  \end{figure}

Figures~\ref{fig:gev22inj}-\ref{fig:mev22tev} show the maps with intermediate non-radiative losses ($v=10^9$~cm~s$^{-1}$) 
and an intermediate magnetic field ($\xi=10^{-2}$) for both $F_{\rm MeV}$ and $F_{\rm GeV}$ normalizations. The color 
scale of all the maps has been chosen such that the colour of areas with values about ten
times above and below the established limits for each quantity are intense red and blue, respectively.

The calculation results displayed as maps of $L_{\rm inj}$, $R/d$, $F_{\rm X}$, $F_{\rm MeV}$ ($F_{\rm GeV}$), and $F_{\rm TeV}$ tend 
to give values for $L_{\rm inj}$, $R/d$, $F_{\rm X}$, $F_{\rm TeV}$, and $F_{\rm GeV}$ when fixing $F_{\rm MeV}$, that are too large for 
half or more of the possible emitter locations. This occurs because of the very high energetic needs to explain the (MeV) 
GeV fluxes when the emitter is in {\it front} ($x>0$) of the star. The maps presented in this section, which are obtained by 
fixing $v$ and 
$\xi$ to intermediate values, already show the general trends and allow us to investigate the disparities between fixing 
$F_{\rm MeV}$ or $F_{\rm GeV}$. These disparities are basically stronger energetic requirements, a larger emitter, and a larger  
departure from the observed fluxes at other bands when fixing $F_{\rm MeV}$. 

When looking at extreme parameter choices in the Appendix, we find that the larger qualitative changes in the represented
 quantities come from magnetic field variations, whereas those produced by modifying the non-radiation timescale are quantitative
 and rather moderate in most of the cases. This happens because a different synchrotron-to-IC cooling ratio severely modifies the energy
 distribution of the radiation, unlike non-radiative cooling, which affects both the synchrotron and the IC components in a more 
uniform manner. As seen from the SEDs above, the GeV fluxes for $x=-a$ are dominated by IC radiation in most of the cases, 
whereas synchrotron (IC) dominates for high (low) $B$-values for $x=a$. On the other hand, the MeV fluxes are mostly of synchrotron
 origin for $x=a$, whereas IC becomes dominant for low $B$-values for $x=-a$. The slopes of the 
spectra for each component also vary more under radiative losses than under non-radiative ones. All this explains the strong
 changes in the color maps when going from low to high $B$ cases.


 \subsection{Identification of the best model}\label{chi2}
 
We are interested in finding the parameters, emitter position, and normalization that best reproduce observations. At this stage, 
we focus only on flux levels of different energy bands, ignoring orbital phase information. We find that it is still possible to estimate
 how close the model and the set of parameters adopted from matching the observations are. Using a minimum deviation method for this task,
 a quantitative assessment can
 be made to illustrate which emitter conditions are closer to the observational values. To provide such an estimate, we first search 
 in the 
literature observations of LS~5039 along its orbit, and then we calculate the average, minimum, and maximum energy flux in the X, 
MeV, GeV,
 and TeV energy band. The average energy fluxes can be considered as observational points with $1\sigma$ errors given by the 
respective maximum and minimum energy flux. Then, we calculate the SED for a given scenario and emitter location and compare the 
theoretical energy fluxes with the observational data. We estimate the deviation of this fit by doing a simple $\chi^2$ test with 
four observational points (actually three, as one is fixed by the normalization). We repeat this procedure for every possible emitter 
location for each magnetic field and escape velocity values considered. Then we identify the parameters and emitter location that 
yielded the best fit (i.e. the lowest $\chi^2$ value). Using this procedure, we find that the minimum deviation ($\chi^2 = 7.5$)
is achieved for an emitter with a low magnetic field ($B=0.5$~G), fast adiabatic losses ($v=c$), a position of $(x,y) = (-1.2a,0.8a)$, and 
normalization according to the GeV energy flux.
 The corresponding SED is shown in Fig.~\ref{fig:bestfit}, along with the observational fluxes in the different energy ranges.

The escape velocity found is similar to the values one would expect for the jet or the shocked pulsar wind in both the
 microquasar and the pulsar binary scenarios. The magnetic field is consistent with values found by previous studies
 \citep[e.g.,][]{khangulyan2008,dubus2008,takahashi2009,zabalza2013}. In particular, a direct comparison can be made with \cite{dubus2008} 
and \cite{takahashi2009}, as they adopted one-zone models to explain the non-thermal emission from LS~5039 and derived magnetic fields 
of $\approx 1$ and $3$~G, respectively. Regarding the location of the emitter, our results are consistent with an emitter off the orbital
plane
 \citep{khangulyan2008,takahashi2009}, and they also seem to point to a more natural explanation for the X-ray, GeV, and TeV 
fluxes in the context of a one-zone model than when the MeV data are included. We note that our calculations do account for different 
emitter locations, which allows for more detailed comparisons with the observed fluxes, although we neglect orbital phase specific 
information in the likelihood analysis described above. In any case, even simply adopting the observed orbital variations of the fluxes in 
different bands as statistical errors, large regions of the source are much worse at reproducing this coarse data presentation than the 
results from the {\it best fit}.
    
  \begin{figure}	
  \centering	
  \includegraphics[width=0.2\textwidth, angle=270]{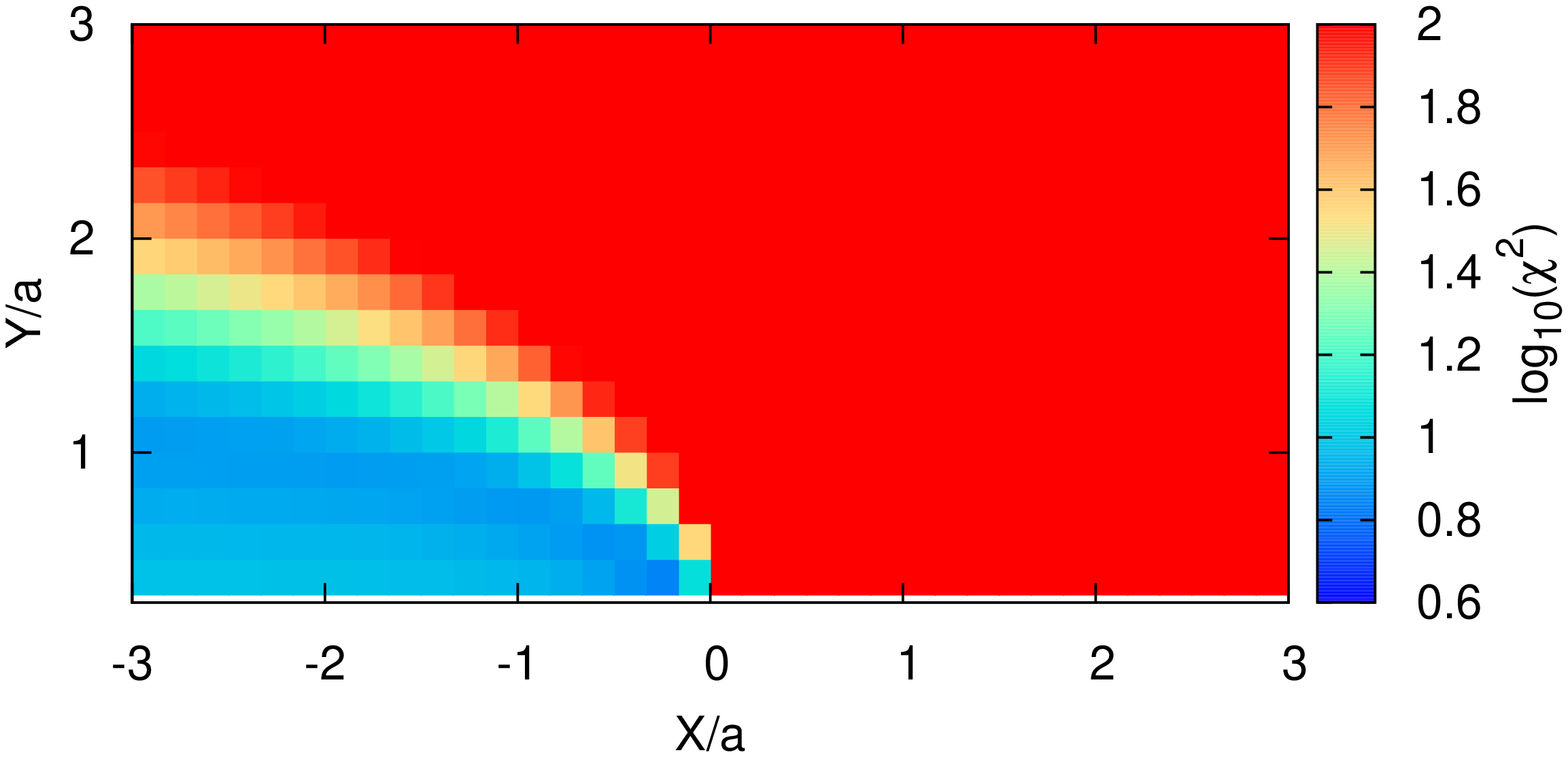}
  \caption{$\chi^2$ map for an emitter with fast non-radiative losses and a weak magnetic field. The normalization was set 
  to reproduce an energy flux in the $0.1$--$10$ GeV range equal to $2.8 \times 10^{-10}$ erg cm$^{-2}$ s$^{-1}$.}
  \label{fig:chi2}
  \end{figure}

  \begin{figure}	
  \centering	
  \includegraphics[width=0.29\textwidth, angle=270]{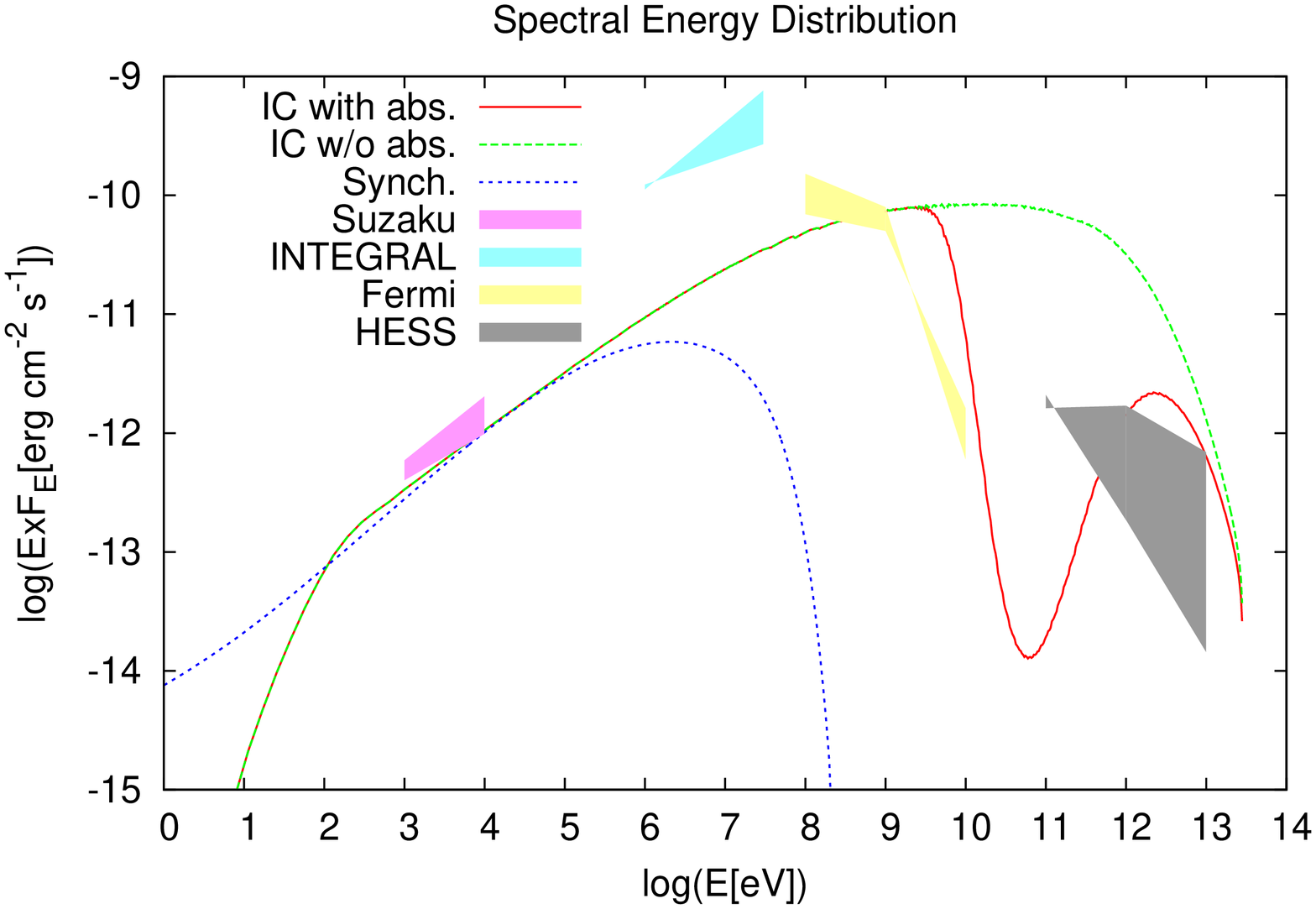}
  \caption{Spectral energy distribution for the best fit scenario ($\chi^2 = 7.5$), which consists of an emitter with fast 
  non-radiative losses and a weak ($B=0.5$G) magnetic field that is located at $x=-1.2a$ and $y=0.8a$. The normalization was set 
  to reproduce an energy flux in the $0.1$--$10$ GeV range equal to $2.8 \times 10^{-10}$~erg~cm$^{-2}$~s$^{-1}$. Observational 
  constraints in X-rays, MeV, GeV, and TeV energies are also presented.} 
  \label{fig:bestfit}
  \end{figure}

\section{Discussion}\label{disc}

 The failure of the one-zone model to globally explain the X-ray, MeV (GeV), and TeV fluxes fixing the GeV (MeV) flux
 to its observed value implies that some of the assumptions adopted for LS~5039 are incorrect: (i) adopting a pure 
leptonic radiation model in a dilute magnetized medium with just synchrotron and IC losses and non-radiative cooling
that is independent of energy; (ii) accounting for gamma-ray absorption in the stellar photon field but neglecting the 
role of electromagnetic IC cascading or the synchrotron emission from the produced pairs; (iii) assuming one population of 
particles follows a power-law and an exponential cutoff in energy under homogeneous conditions in a point-like 
accelerator/emitter; and (iv) neglecting Doppler boosting. Some of these assumptions appear more robust than others when 
looking at the known physical or phenomenological properties of the source. These include the following:

(i) Radiation from hadronic interactions, such as \textit{pp} interactions \citep{romero2003}, cannot be discarded. However, as noted in 
\cite{boschramon2009}, they are not likely 
to be important for relativistic protons/nuclei in the environment of LS~5039 with relatively low matter (for $pp$), X-ray 
photon (for photomeson production and photodisintegration) densities, and very long proton synchrotron timescales. Pulsar
 wind comptonization or pulsar magnetospheric radiation have been discussed as potential sources of GeV-TeV emission 
in LS~5039 \citep[e.g.,][]{sierpowska2008,cerutti2008,takata2014}. Considering these components would imply a 
multi-population, non-homogeneous, non-point-like model (see iii) and the presence of a powerful non-accreting pulsar in the system.

(ii) As discussed by \cite{boschramon2008b}, the magnetic field in the stellar surroundings could be 
low, $\lesssim 1$~G. This would allow the emitter to be located close to the companion, as the effective opacity to TeV 
photons would be much lower for an emitter {\it behind} the star due to electromagnetic IC cascading. Otherwise, the emitter 
could not be located deep within the system and close to the companion, as the production of the required TeV fluxes, and therefore, the
 injection luminosity, would be very high and imply secondary synchrotron X-rays above the observed fluxes.
 However, {\it Fermi} data seem to rule out effective electromagnetic IC cascading in LS~5039, as the predicted emission 
 around 1--100~GeV,
 according to leptonic models, is too high with respect to the observed values \citep[e.g.,][]{cerutti2010,yamaguchi2010,takata2014}.
 This implies again an emitter far from the star and the companion \citep[see]
[for a similar conclusion when studying X-rays]{boschramon2007,szostek2011}. This is a model independent fact, as the 
10--100~GeV range is just around the pair-creation threshold in an UV photon field, and most of the cascade radiation 
should be released there. A low-energy cutoff in the emitting electron distribution above several hundred GeV may 
alleviate this constraint on cascading, but then the GeV and TeV components would have to come from different emitting 
populations (see iii). 

(iii) The cooling timescales of electrons are short in LS~5039, if they are accelerated within the binary, 
$\sim 10-10^4$~s \citep[see fig. 5 in][]{takahashi2009}, so the one-zone assumption may seem quite natural.
 However, different populations and a very structured accelerator/emitter are physically well motivated \citep{vila2012}. As 
simulations show \citep[see, e.g.][and references therein]{perucho2012,boschramon2012}, both the microquasar
 and the pulsar scenario present dissipation regions in the periphery of the system and beyond: powerful flows
 interact in the system and produce a large variety of hydrodynamical phenomena, potentially generating large amounts
 of non-thermal energy that is carried away outside of the binary \citep[e.g.,][]{zabalza2013}. In addition, as noted, emitting 
sites other than
 those of pure hydrodynamical origin, as the cold pulsar wind or the magnetosphere of the pulsar, could be also relevant 
\citep[see][for a review on different proposals behind the GeV emission in LS~5039]{dubus2013}. 
Therefore, the one-zone assumption is possibly quite inaccurate. 

(iv) Doppler boosting in both the microquasar and the pulsar scenario cannot be neglected in general, as relativistic 
flows are involved in both cases \citep{mirabel1994,mirabel1999,bogovalov2008}. Simulations show that the interaction 
between the microquasar jet or the pulsar wind with the stellar wind can lead to slower regions containing non-thermal 
particles \citep{perucho2012,boschramon2012}, but the shocked flow is expected to be mildly relativistic. Reacceleration
 in the postshock region is expected \citep{bogovalov2008}. Therefore, even when standing shocks can form, Doppler 
boosting may be important at some distance from the shock. The energy flux in the observer frame transforms as 
$F\sim \delta^4F'$ for one emitting region, where $\delta=1/\Gamma(1-\beta\cos\theta)$, $\Gamma$ is the Lorentz 
factor, $\beta=v/c$ ($v$ is the velocity of the flow), and $\theta$ the angle between the emitter motion and the line of sight. 
When most of the injected 
energy is radiated, one can relate the observed luminosity to the injected one as $L\sim \delta^4 L_{\rm inj}/\Gamma^2$, 
so $L\sim (1+\beta)^4\Gamma^2L_{\rm inj}$ for an emitter pointing to the observer ($\theta=0$). This means that even a modest Lorentz 
factor $\lesssim 2$ can already enhance the radiation by more than an order of magnitude. At least under some 
specific source geometries, this shows that adopting $\delta=1$ is far from realistic.

\section{Conclusions}

The tool presented in this paper consists of an exhaustive application of the one-zone model that includes non-radiative cooling to 
non-thermal emitting high-mass binaries. This procedure allows the determination of significant departures 
from a simple model for any source. This may suggest more complex physical schemes, including relativistic motions, unaccounted radiation 
components, or a non-uniform 
emitter. As a test and because of its interesting potential implications, we have exhaustively applied our model to LS~5039, possibly 
the best known and most intriguing gamma-ray (high-mass) binary. For simplicity, this application has been carried out without a detailed 
account of the spectral and orbital behavior of the source but has focused on the energy fluxes at different bands. Our results, as noted 
in Sect.~\ref{disc}, are already suggestive of different extensions of the one-zone model and are also compatible with previous works 
that had already proposed possible improvements beyond that model. Two additional advantages of the 
present approach are that it is both robust and slightly model dependent, as it relies on basic source information and physics. 

In particular, if the MeV (GeV) data are to be explained with our model (one-zone leptonic model without secondary pair radiation 
nor Doppler boosting), our study of LS~5039 shows that this tends to over-predict in most of the configurations the X-ray and TeV fluxes. 
The model also underpredicts (overpredict) the MeV (GeV) fluxes and requires an emitter of size incompatible with the point-like 
assumption, with very demanding energetic requirements. 

Low magnetic fields and high non-radiative losses still yield relatively good results when the emitter is {\it behind} 
the star. Otherwise, when the emitter is in {\it front} of the star, the energetic requirements are too high when trying 
to explain the MeV and GeV fluxes. This is the case for any combination of the magnetic field and the velocity of the flow values. 

Large magnetic fields (regardless of the value of the flow velocity) can be also discarded for almost all configurations, 
as they yield injection luminosities, emitters, and X-ray fluxes that are too large. We note that the disparities between predictions 
and observations are already very apparent without accounting for spectral and orbital behavior, which would just narrow even more
the applicability of the one-zone model.

Our results strongly favor several emitting populations, as the fluxes in different energy bands are incompatible with 
just one population. The large predicted emitter also suggests that this should be extended and inhomogeneous. The large
energetic requirements when the emitter is in {\it front} of the star hints at Doppler boosting as a way to overcome the
low radiation efficiencies in the corresponding orbital phases. Finally, the large relativistic pressures derived for the
non-thermal particles may also hint to relativistic plasma motions. All this indicates that assumptions (iii) and (iv) 
should be ruled out and goes in favor of an emitter with characteristic locations at some distance from the star and the 
companion, as discussed when considering (ii). The assumption of a leptonic model still seems appropriate, but additional 
components, as the mentioned cold pulsar wind or the pulsar magnetosphere, cannot be discarded.

The maps presented in this paper can be applied, after a simple re-scaling, to the study of other binary systems. With the increase 
in the number of known binaries, this simple yet powerful analytical tool may become a guide toward better understanding the 
mechanisms that operate in gamma-ray binaries.

  \section{Appendix (electronically available)}
  
  \subsection{Maps normalizing through $F_{\rm GeV}$:}

\begin{figure} 
  \centering
  \includegraphics[width=0.2\textwidth, angle=270]{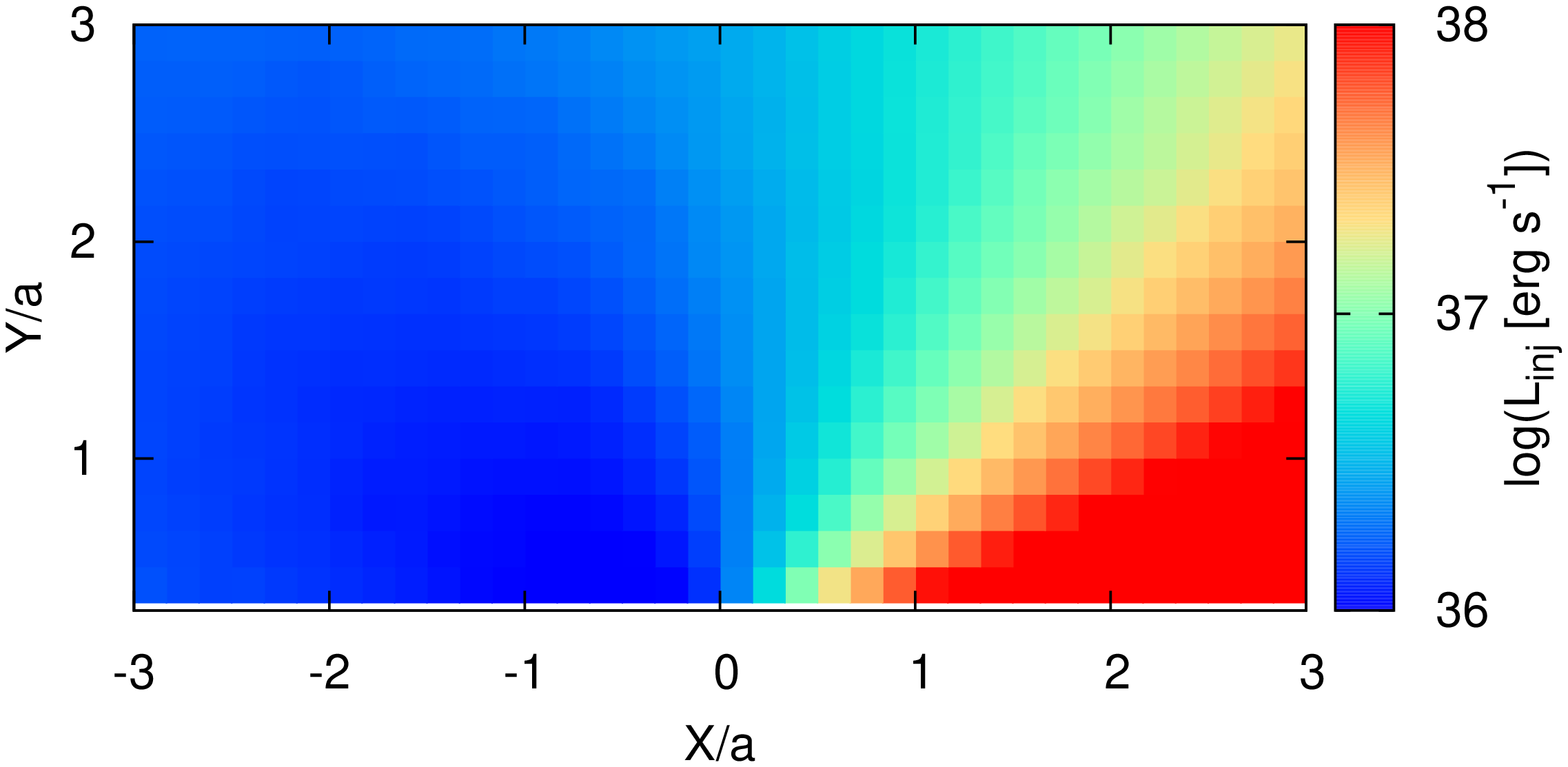}
  \caption[eta=1,delta=0]{Injection luminosity of relativistic particles in the emitter in the case of 
  fast non-radiative losses and a weak magnetic field. The normalization was set 
  to reproduce an energy flux in the $0.1$--$10$ GeV range equal to $2.8 \times 10^{-10}$ erg cm$^{-2}$ s$^{-1}$.}
  \label{fig:gev10inj}

  \centering
  \includegraphics[width=0.2\textwidth, angle=270]{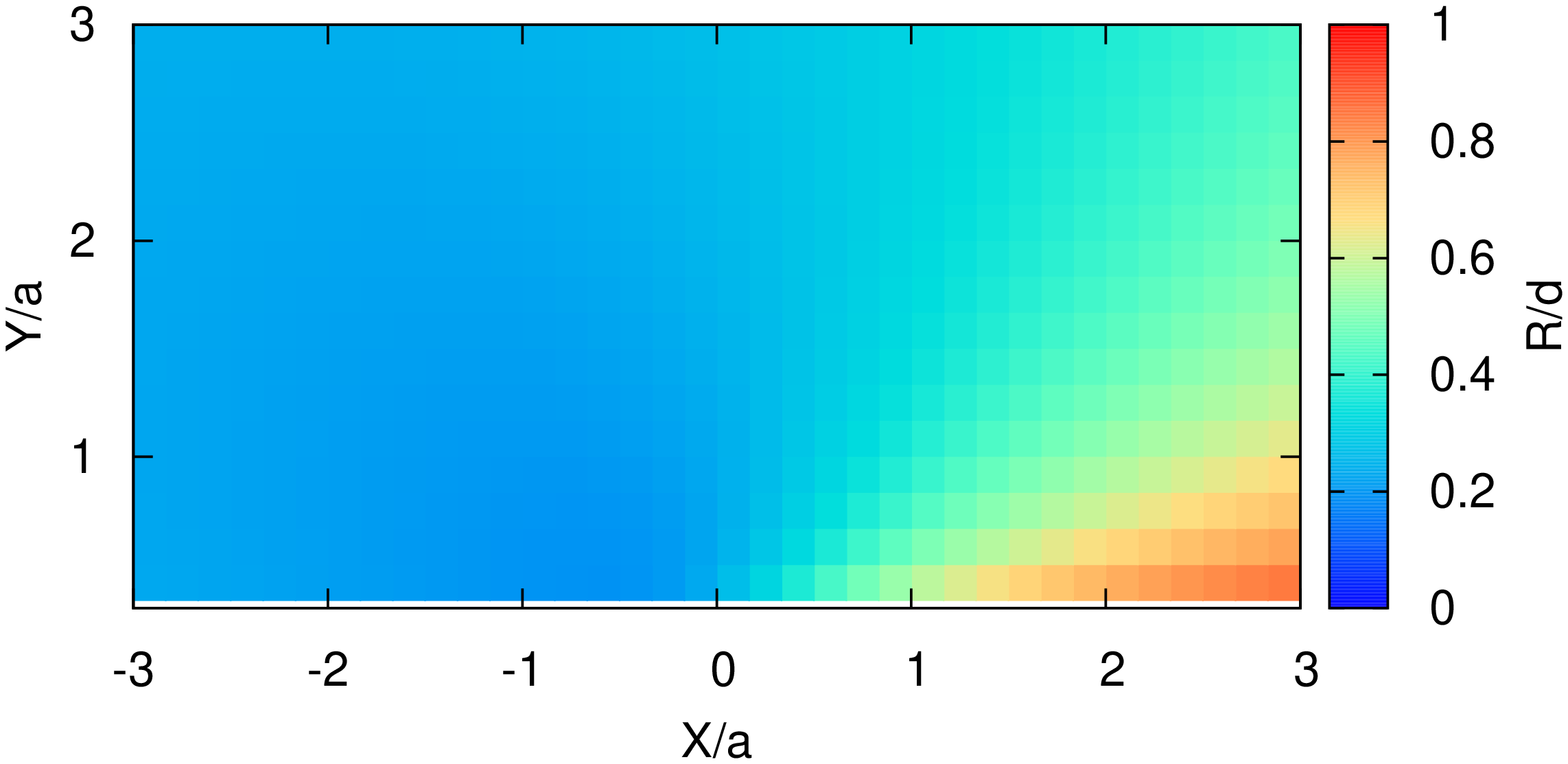}
  \caption[eta=1,delta=0]{As in Fig.~\ref{fig:gev10inj} but showing the emitter's size divided by its distance to the star.}
  \label{fig:gev10confi}

  \centering
  \includegraphics[width=0.2\textwidth, angle=270]{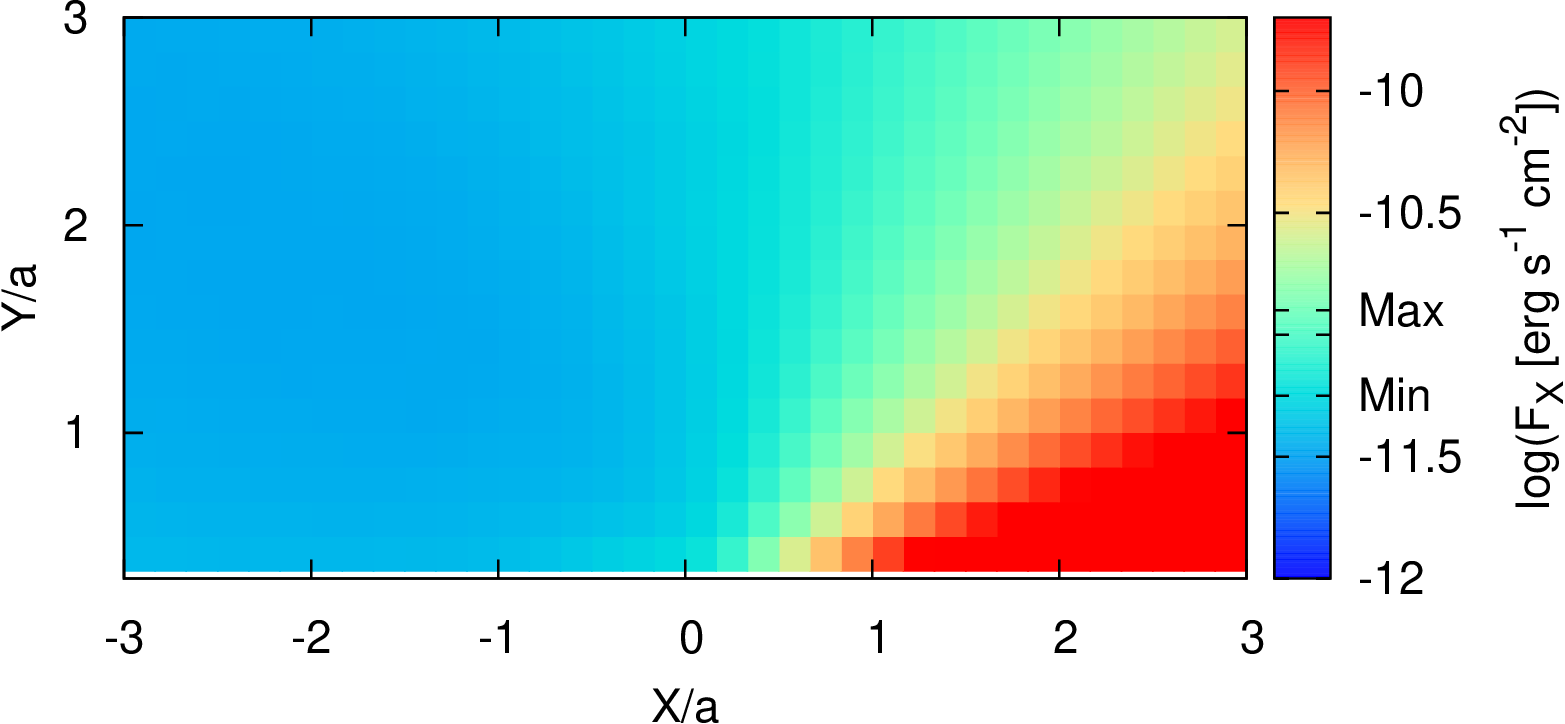}
  \caption[eta=1,delta=0]{As in Fig.~\ref{fig:gev10inj} but showing the integrated energy flux in the 0.3--10 keV energy band.}
  \label{fig:gev10x}

  \centering
  \includegraphics[width=0.2\textwidth, angle=270]{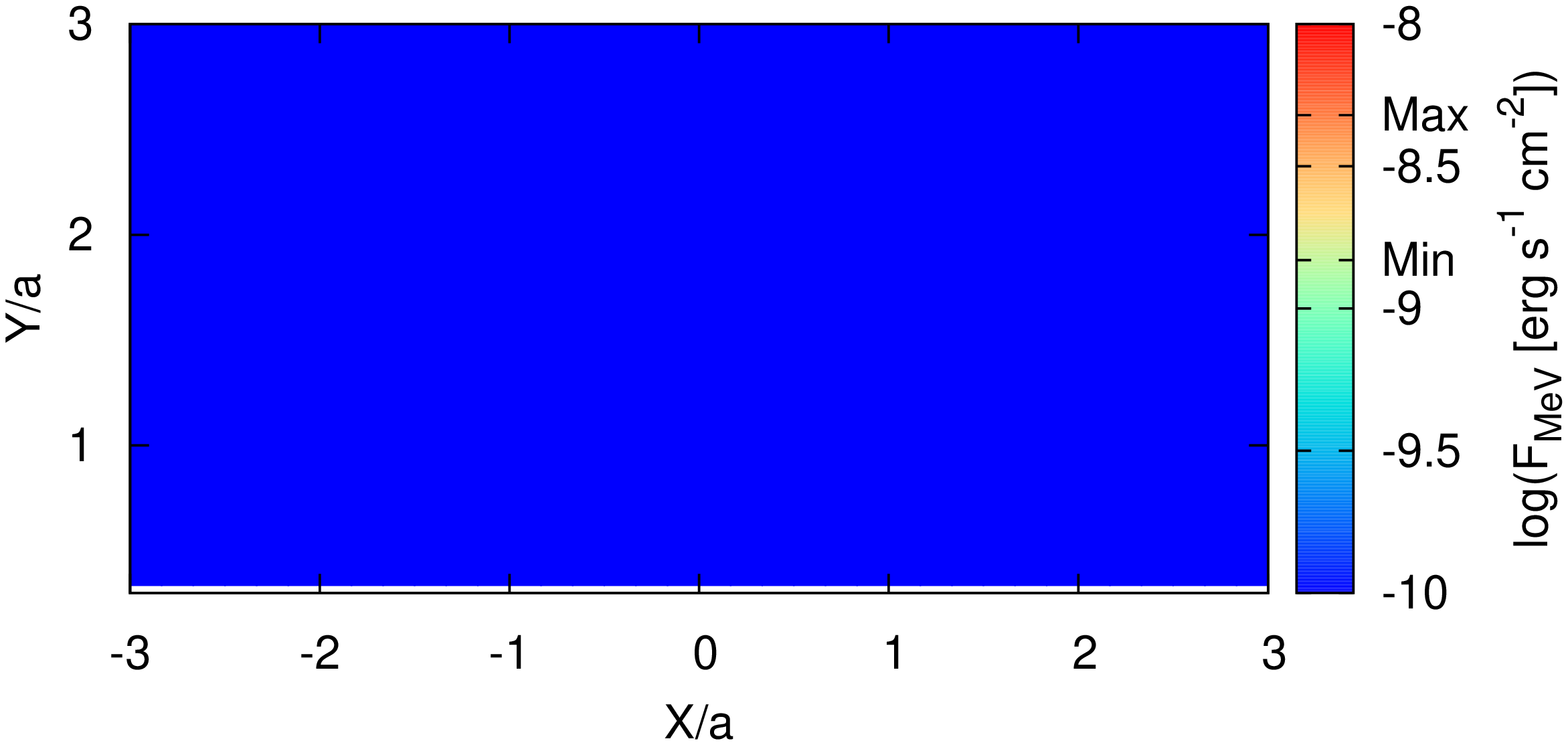}
  \caption[eta=1,delta=0]{As in Fig.~\ref{fig:gev10inj} but showing the integrated energy flux in the 1--30 MeV energy band.}
  \label{fig:gev10mev}

  \centering
  \includegraphics[width=0.2\textwidth, angle=270]{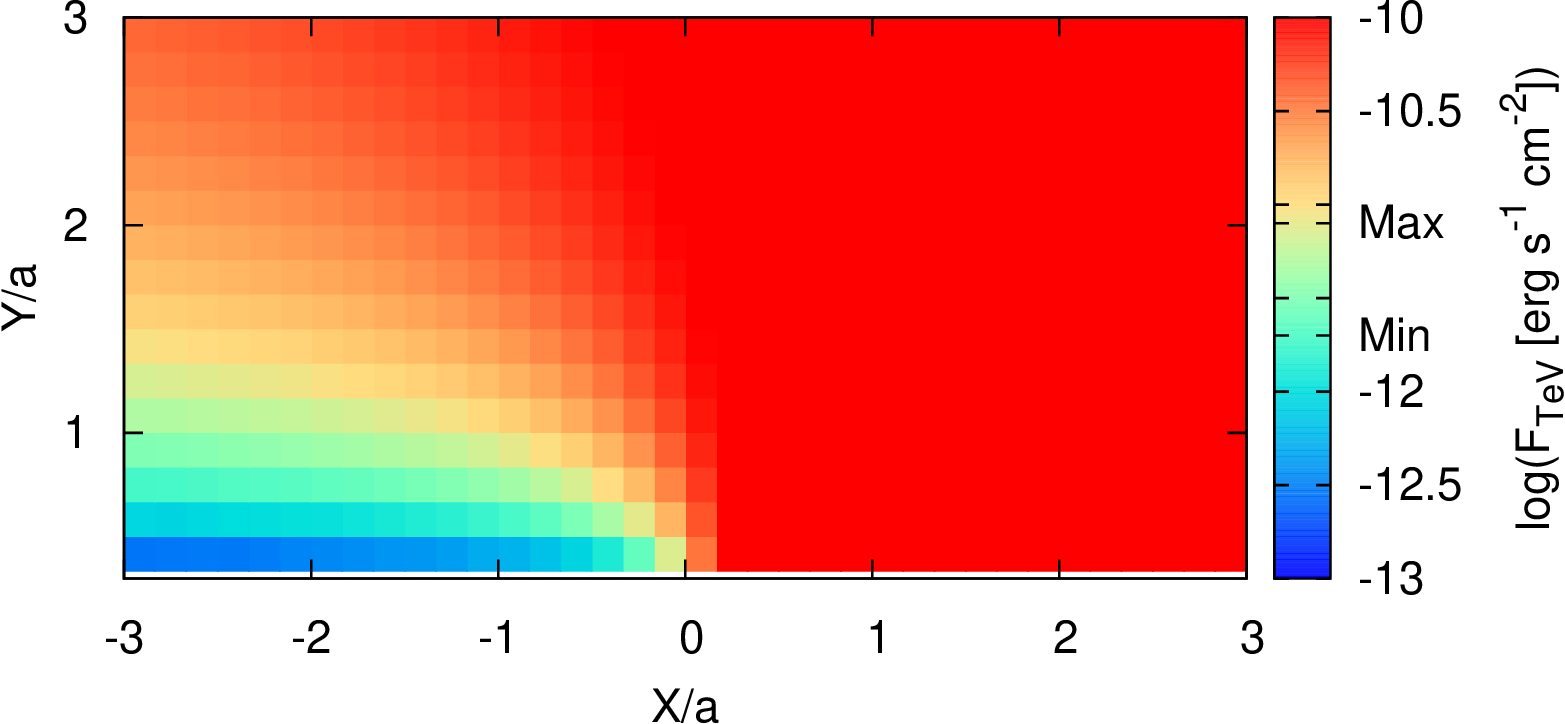}
  \caption[eta=1,delta=0]{As in Fig.~\ref{fig:gev10inj} but showing the integrated energy flux in the 0.1--10 TeV energy band.}
  \label{fig:gev10tev}
  \end{figure}

\begin{figure}
  \centering
  \includegraphics[width=0.2\textwidth, angle=270]{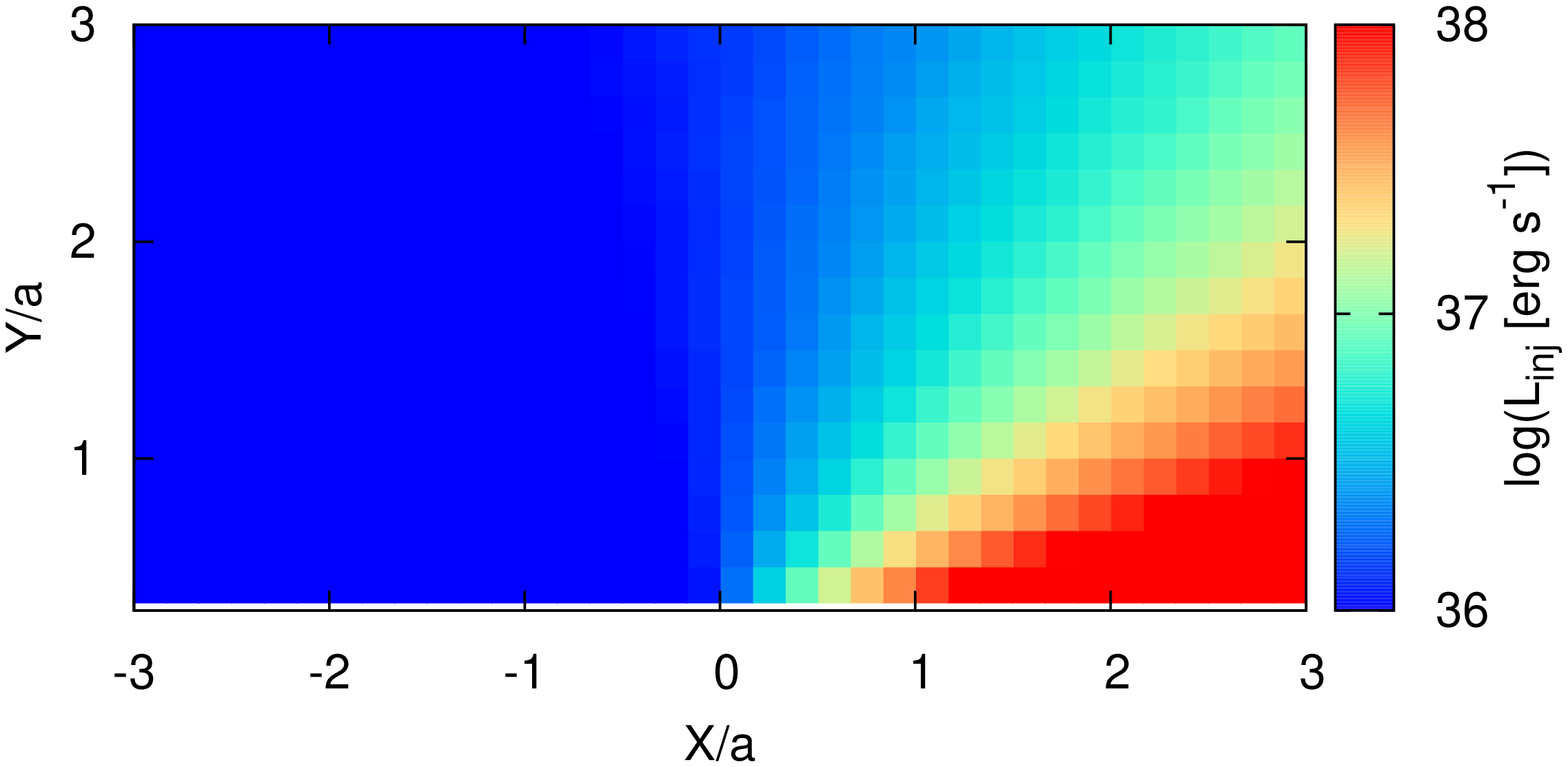}
  \caption[eta=0,delta=0]{Injection luminosity of relativistic particles in the emitter in the case of slow non-radiative 
  losses and a weak magnetic field. The normalization was set 
  to reproduce an energy flux in the $0.1$--$10$ GeV range equal to $2.8 \times 10^{-10}$ erg cm$^{-2}$ s$^{-1}$.}
  \label{fig:gev00inj}

  \centering
  \includegraphics[width=0.2\textwidth, angle=270]{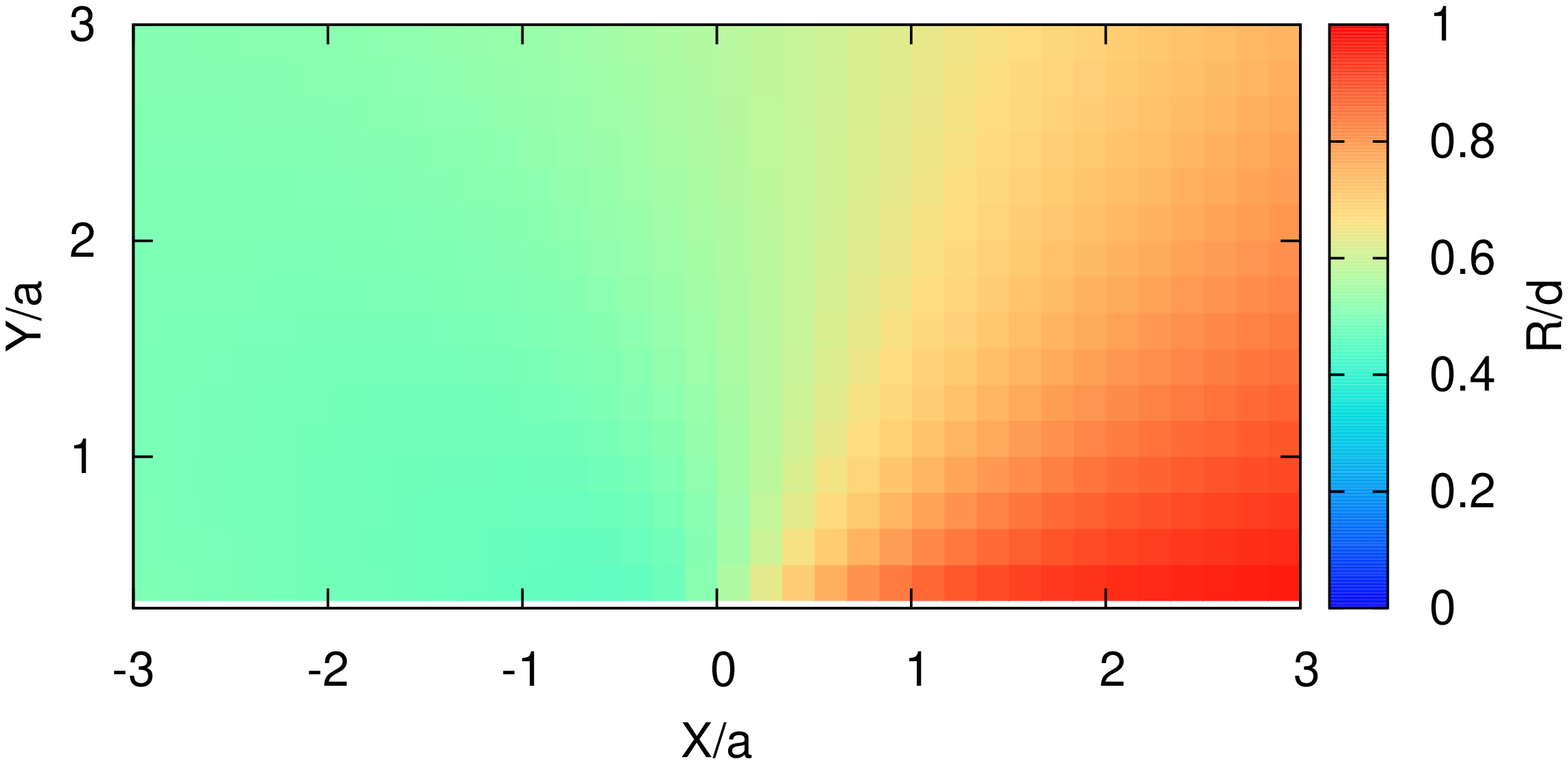}
  \caption[eta=0,delta=0]{As in Fig.~\ref{fig:gev00inj} but showing the emitter's size divided by its distance to the star.}
  \label{fig:gev00confi}

  \centering
  \includegraphics[width=0.2\textwidth, angle=270]{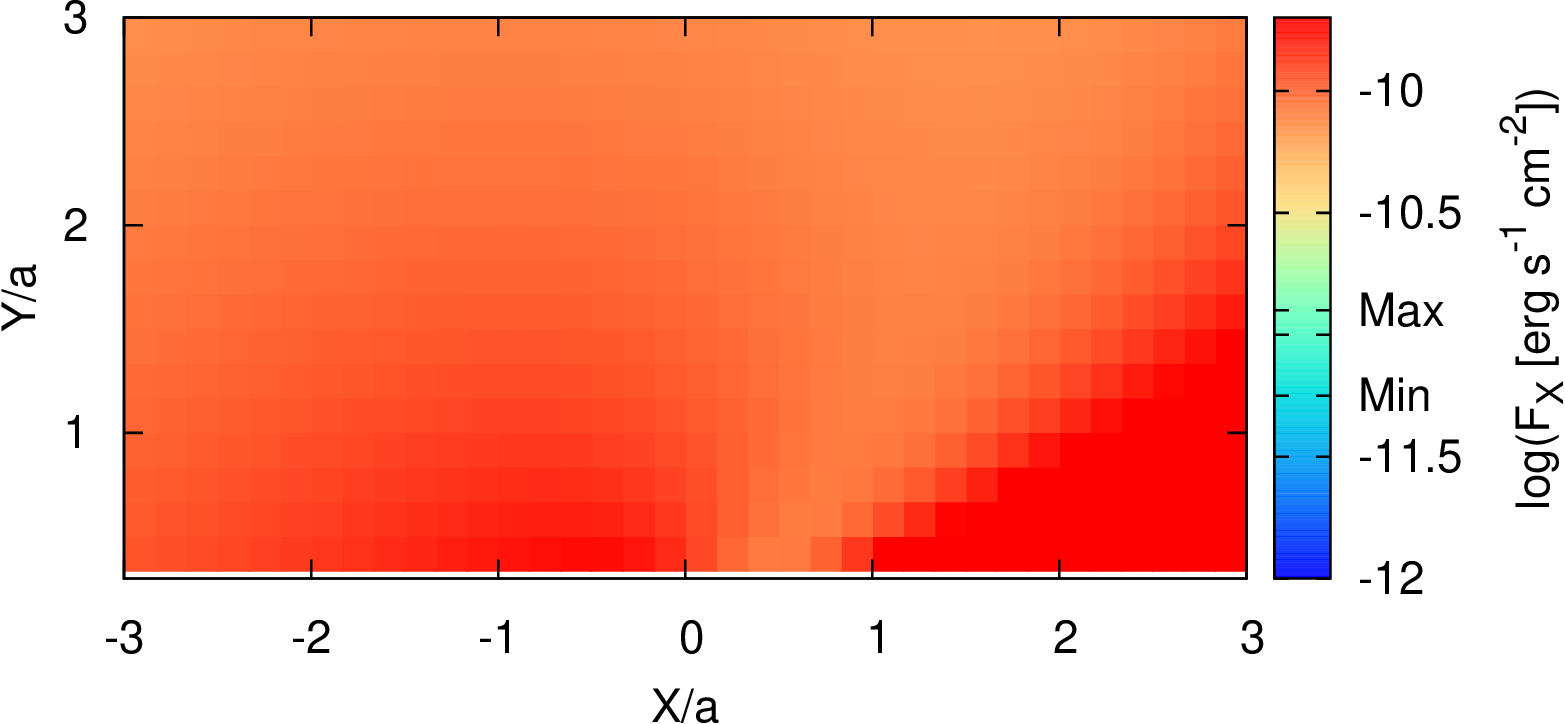}
  \caption[eta=0,delta=0]{As in Fig.~\ref{fig:gev00inj} but showing the integrated energy flux in the 0.3--10 keV energy band.}
  \label{fig:gev00x}

  \centering
  \includegraphics[width=0.2\textwidth, angle=270]{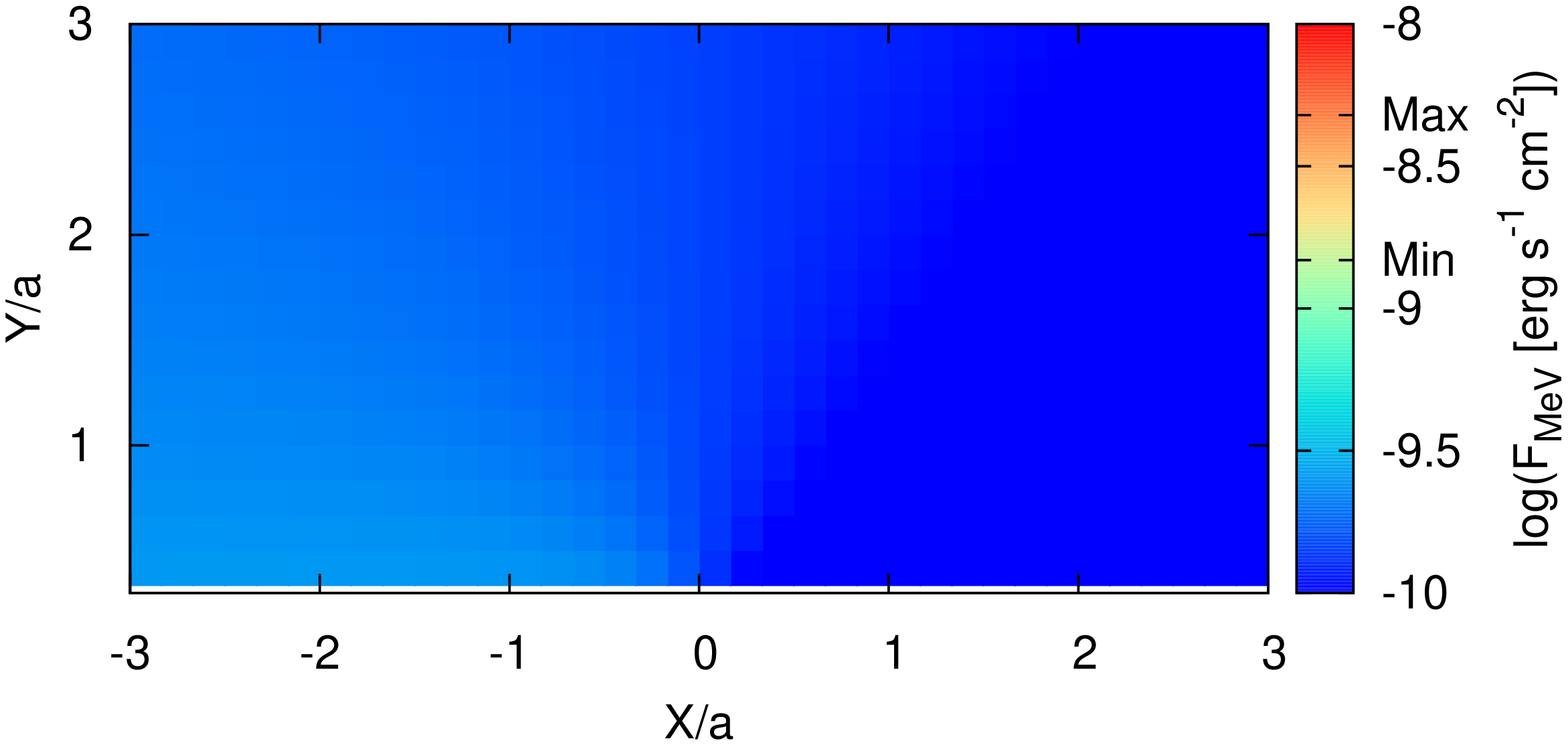}
  \caption[eta=0,delta=0]{As in Fig.~\ref{fig:gev00inj} but showing the integrated energy flux in the 1--30 MeV energy band.}
  \label{fig:gev00mev}
  
  \centering
  \includegraphics[width=0.2\textwidth, angle=270]{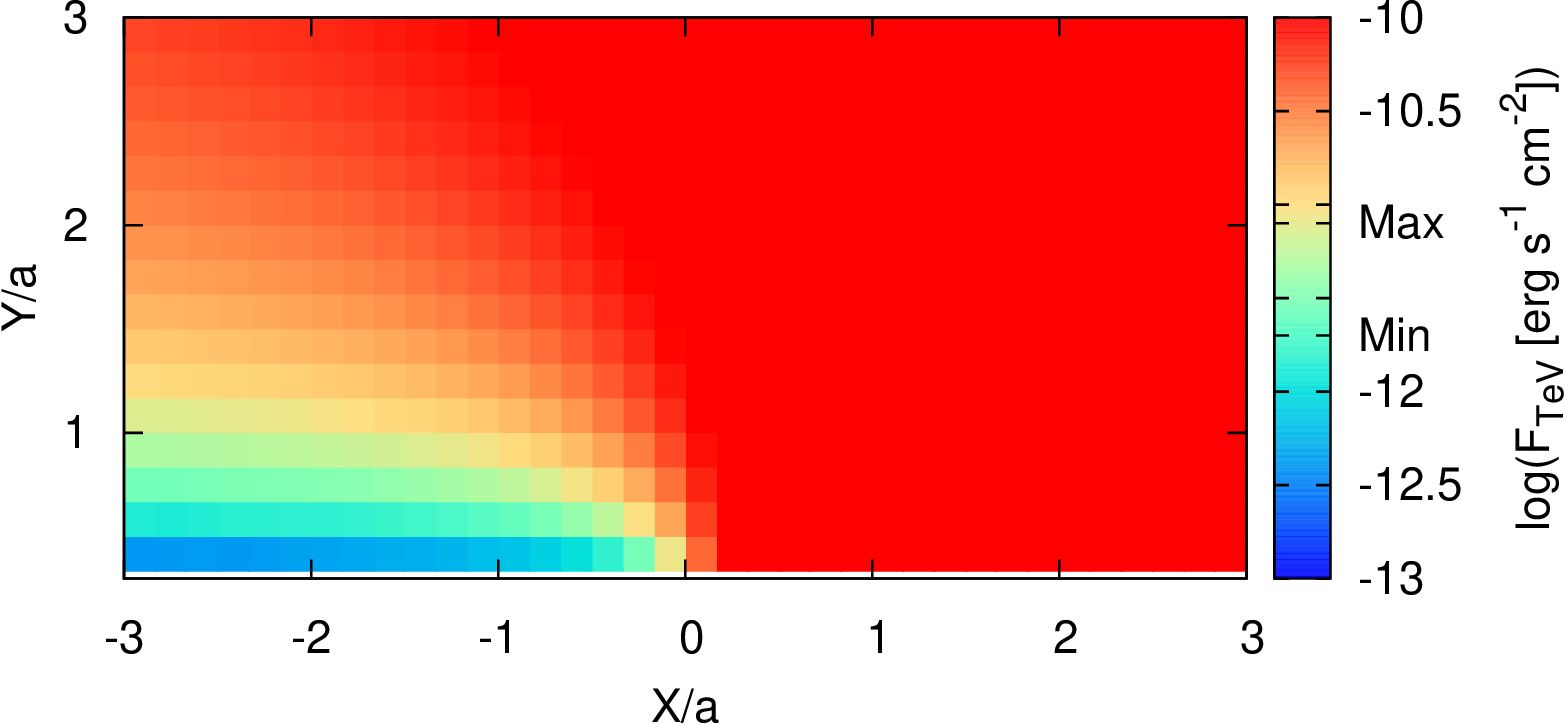}
  \caption[eta=0,delta=0]{As in Fig.~\ref{fig:gev00inj} but showing the integrated energy flux in the 0.1--10 TeV energy band.}
  \label{fig:gev00tev}
  \end{figure}

\begin{figure}
  \centering
  \includegraphics[width=0.2\textwidth, angle=270]{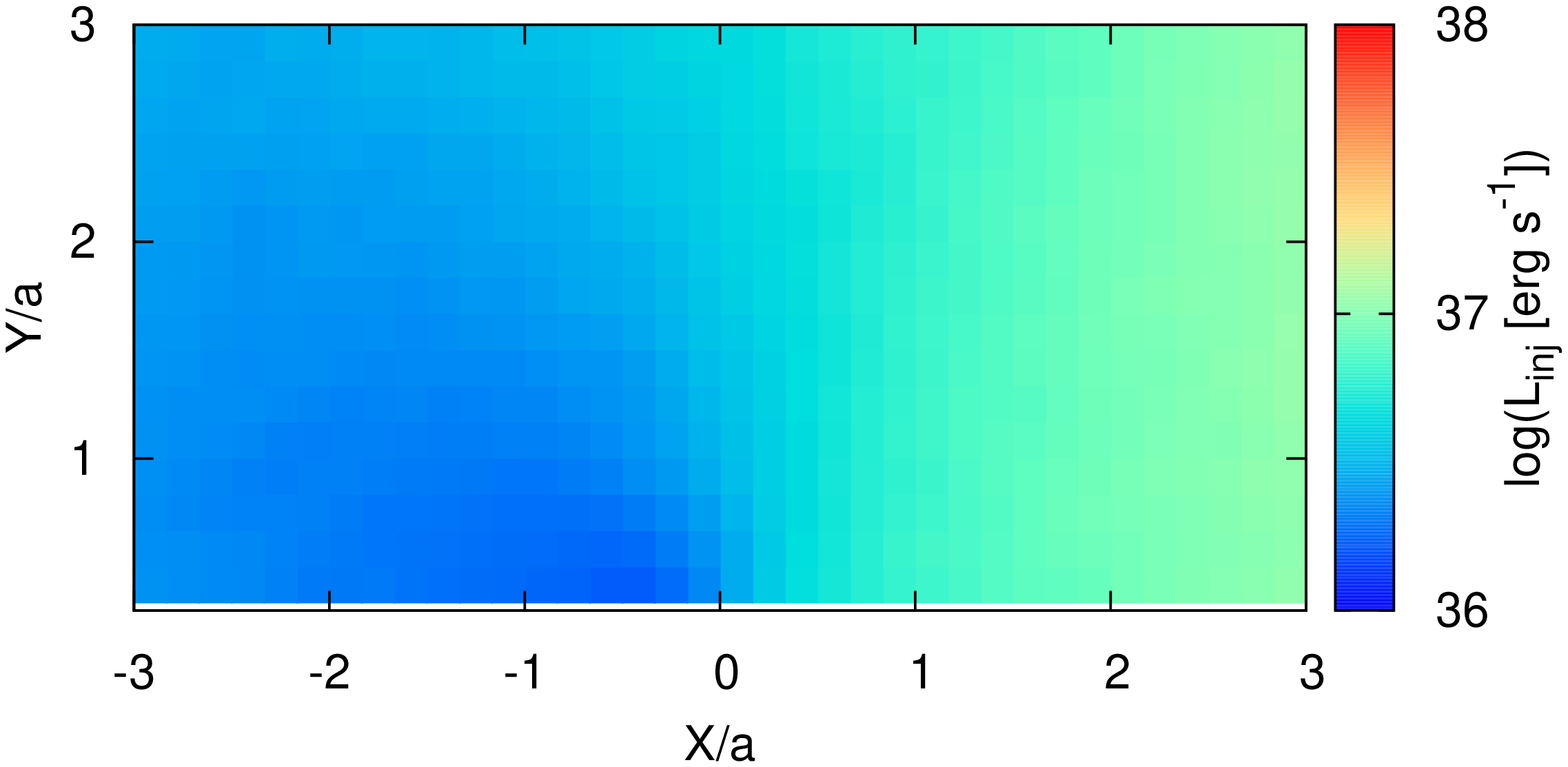}
  \caption[eta=1,delta=1]{Injection luminosity of relativistic particles in the emitter
  in the case of fast non-radiative losses and a strong magnetic field. The normalization was set 
  to reproduce an energy flux in the $0.1$--$10$ GeV range equal to $2.8 \times 10^{-10}$ erg cm$^{-2}$ s$^{-1}$.}
  \label{fig:gev11inj}

  \centering
  \includegraphics[width=0.2\textwidth, angle=270]{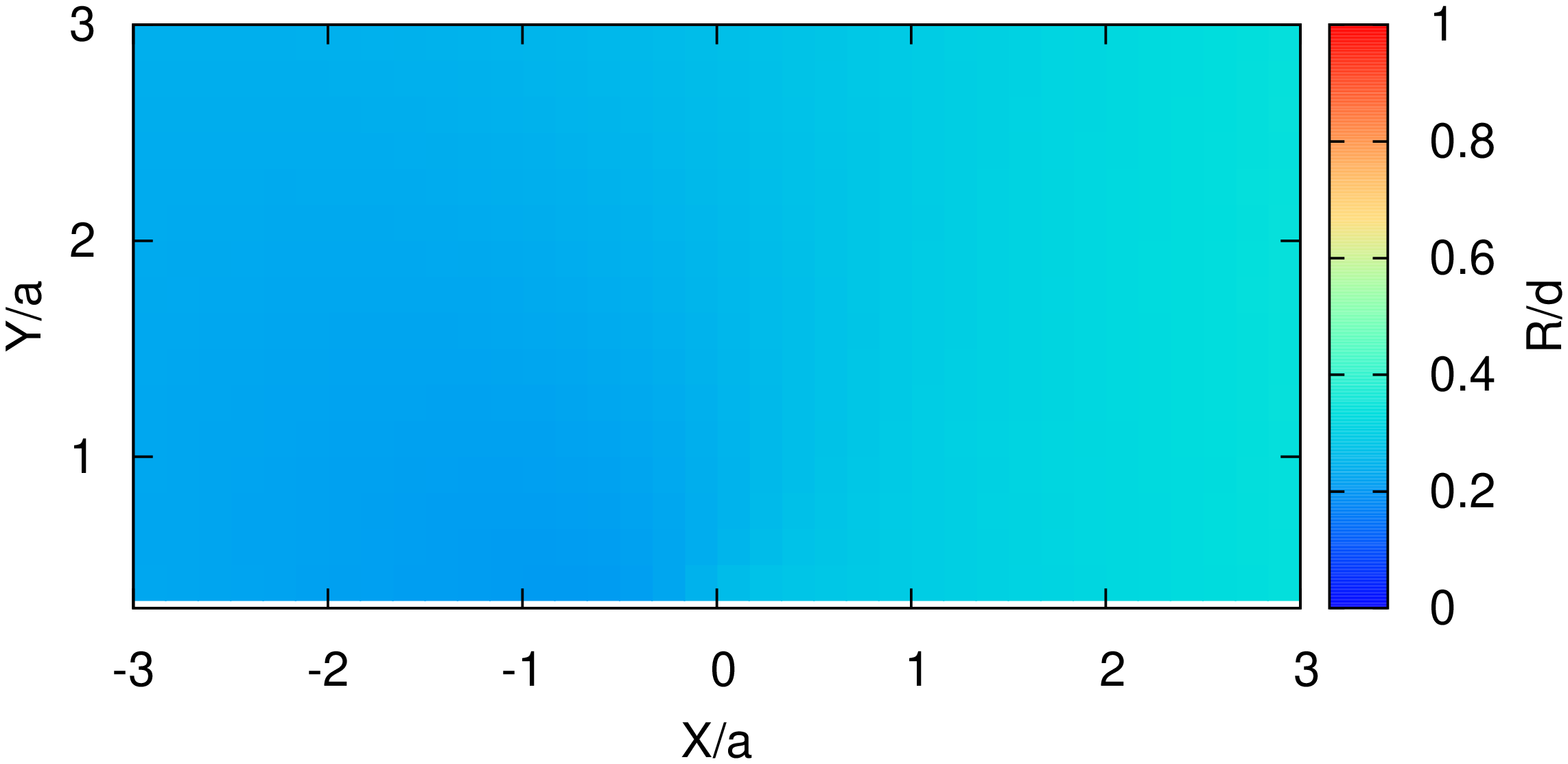}
  \caption[eta=1,delta=1]{As in Fig.~\ref{fig:gev11inj} but showing the emitter's size divided by its distance to the star.}
  \label{fig:gev11confi}

  \centering
  \includegraphics[width=0.2\textwidth, angle=270]{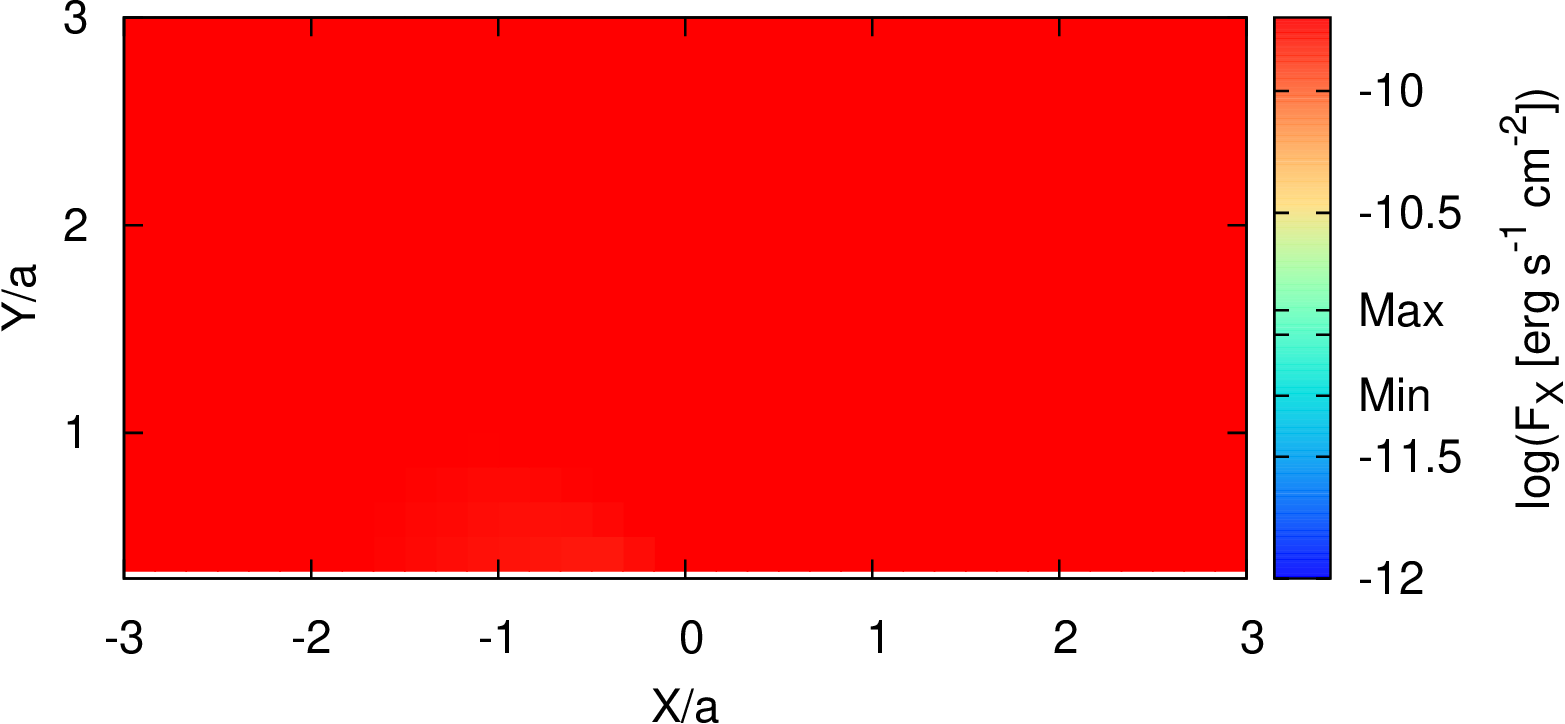}
  \caption[eta=1,delta=1]{As in Fig.~\ref{fig:gev11inj} but showing the integrated energy flux in the 0.3--10 keV energy band.}
  \label{fig:gev11x}

  \centering
  \includegraphics[width=0.2\textwidth, angle=270]{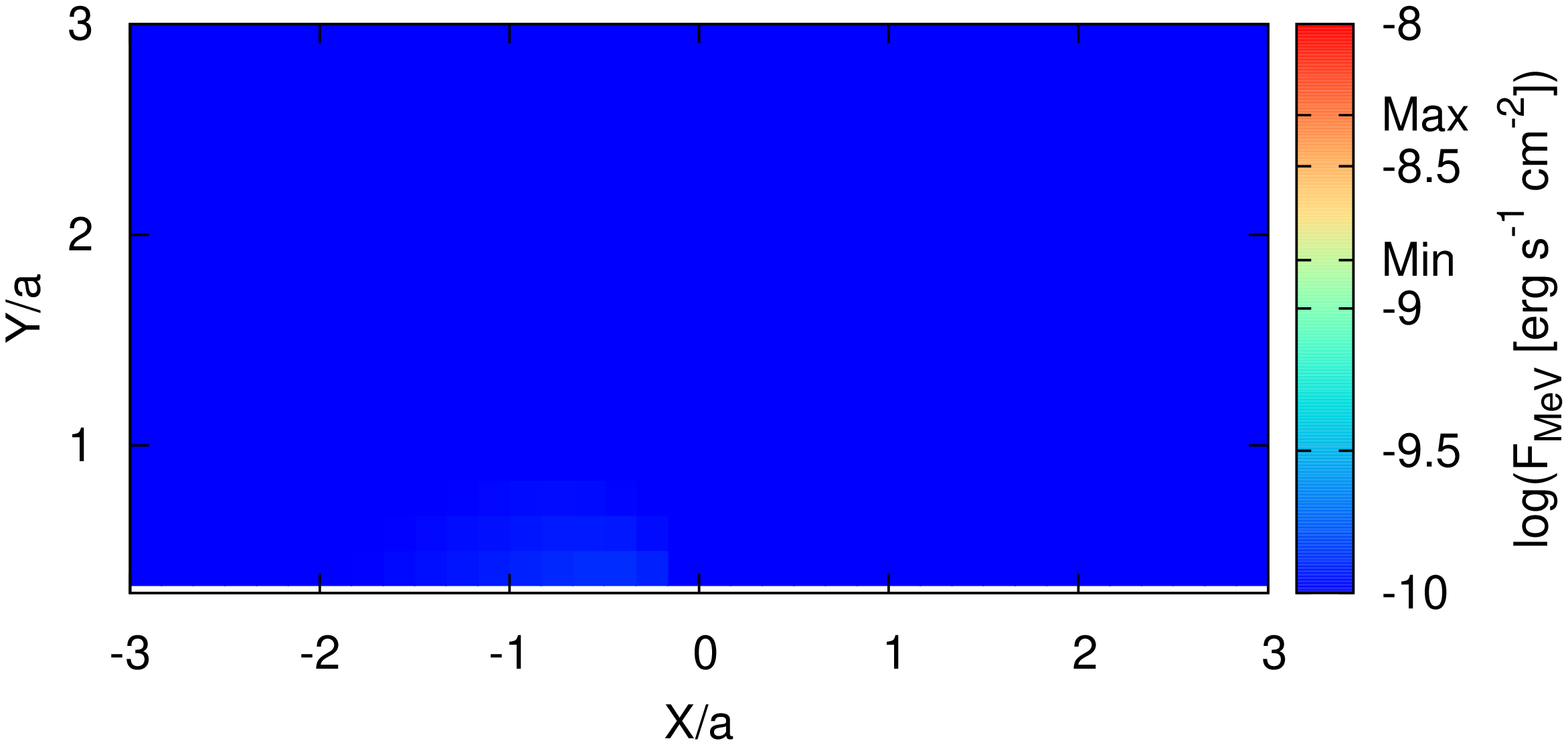}
  \caption[eta=1,delta=1]{As in Fig.~\ref{fig:gev11inj} but showing the integrated energy flux in the 1--30 MeV energy band.}
  \label{fig:gev11mev}
  
  \centering
  \includegraphics[width=0.2\textwidth, angle=270]{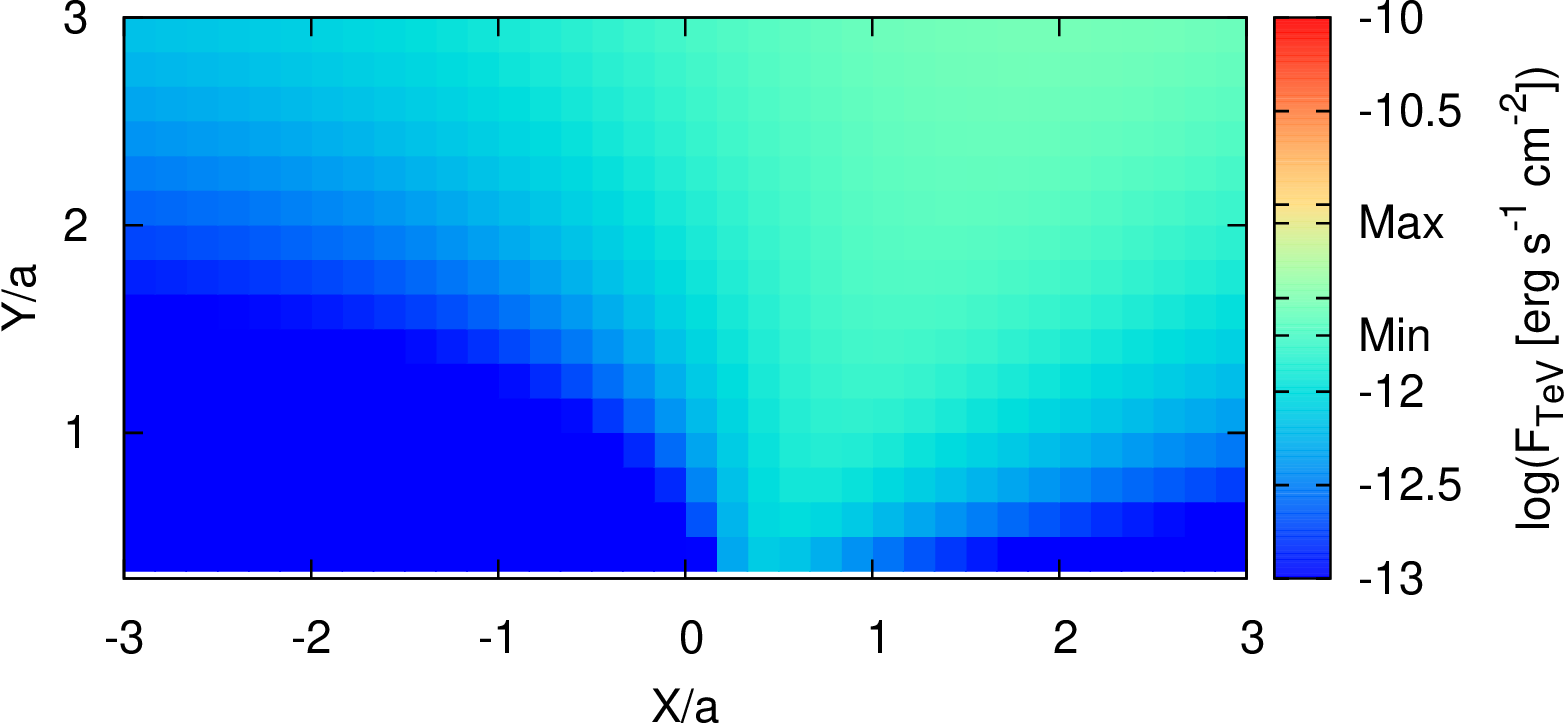}
  \caption[eta=1,delta=1]{As in Fig.~\ref{fig:gev11inj} but showing the integrated energy flux in the 0.1--10 TeV energy band.}
  \label{fig:gev11tev}
  \end{figure}

\begin{figure}
  \centering
  \includegraphics[width=0.2\textwidth, angle=270]{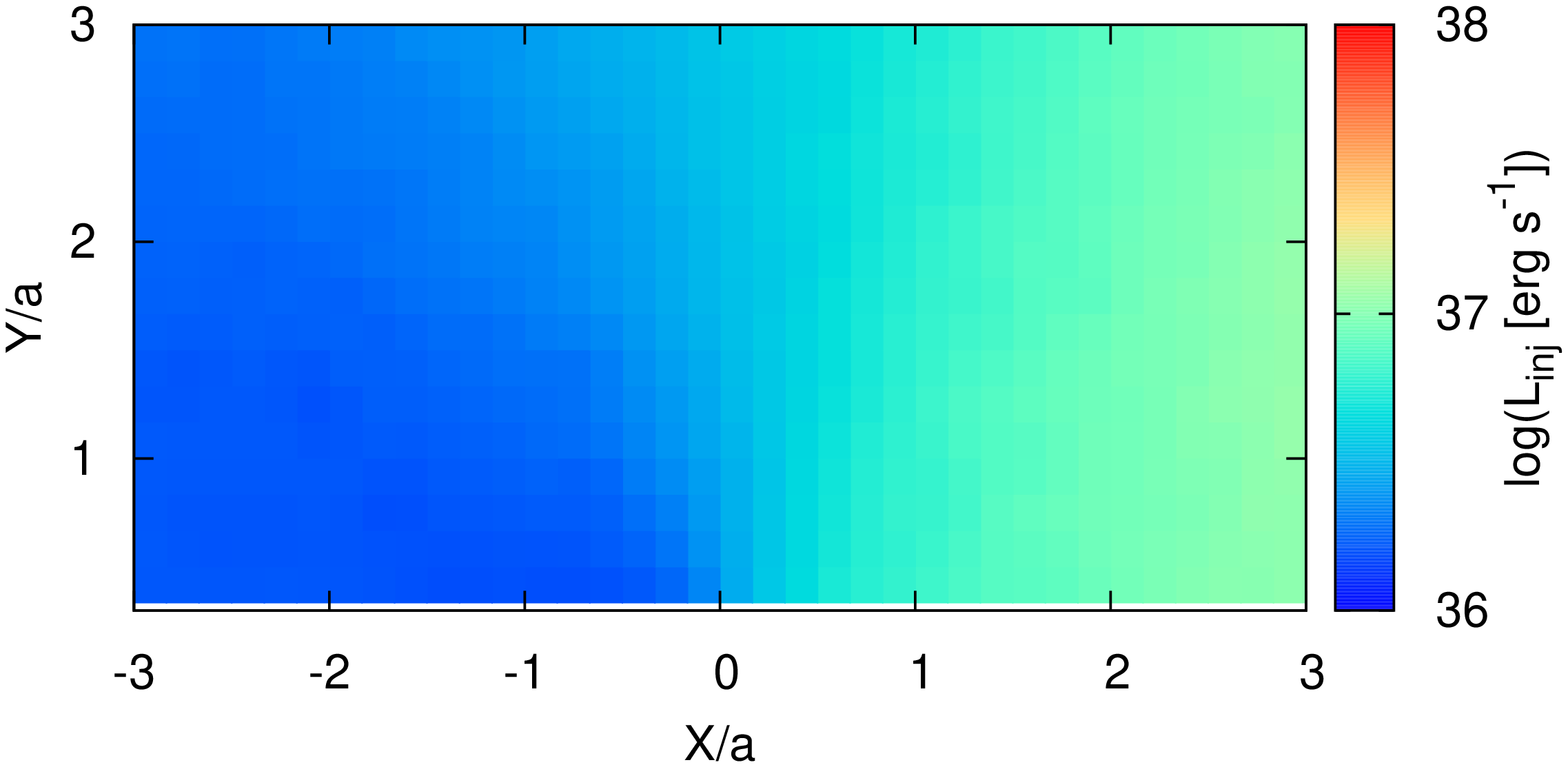}
  \caption[eta=1,delta=0]{Injection luminosity of relativistic particles in the emitter in the case of slow non-radiative 
  losses and a strong magnetic field. The normalization was set  to reproduce an energy flux in the $0.1$--$10$ GeV
  range equal to $2.8 \times 10^{-10}$ erg cm$^{-2}$ s$^{-1}$.}
  \label{fig:gev01inj}

  \centering
  \includegraphics[width=0.2\textwidth, angle=270]{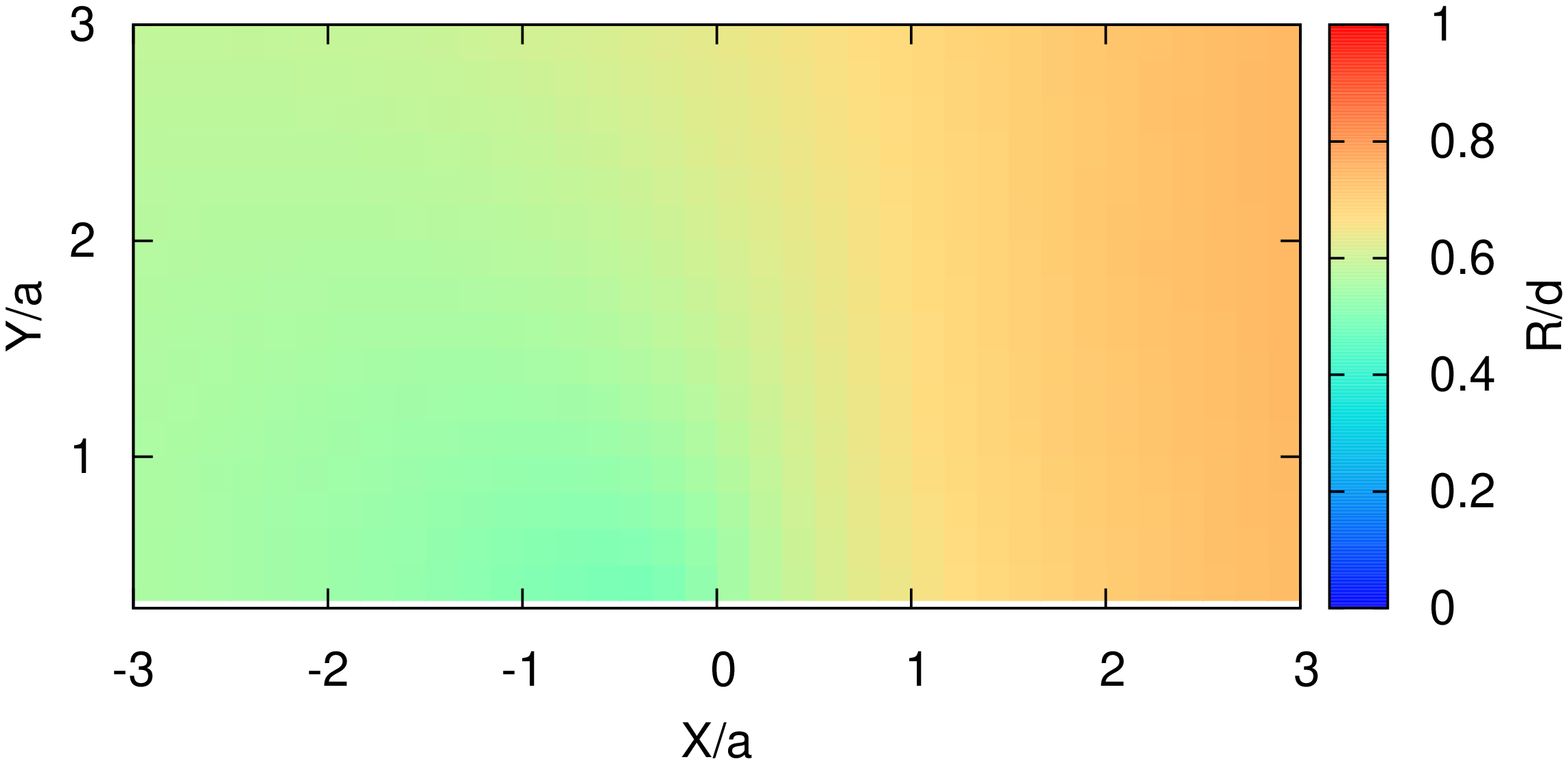}
  \caption[eta=1,delta=0]{As in Fig.~\ref{fig:gev01inj} but showing the emitter's size divided by its distance to the star.}
  \label{fig:gev01confi}

  \centering
  \includegraphics[width=0.2\textwidth, angle=270]{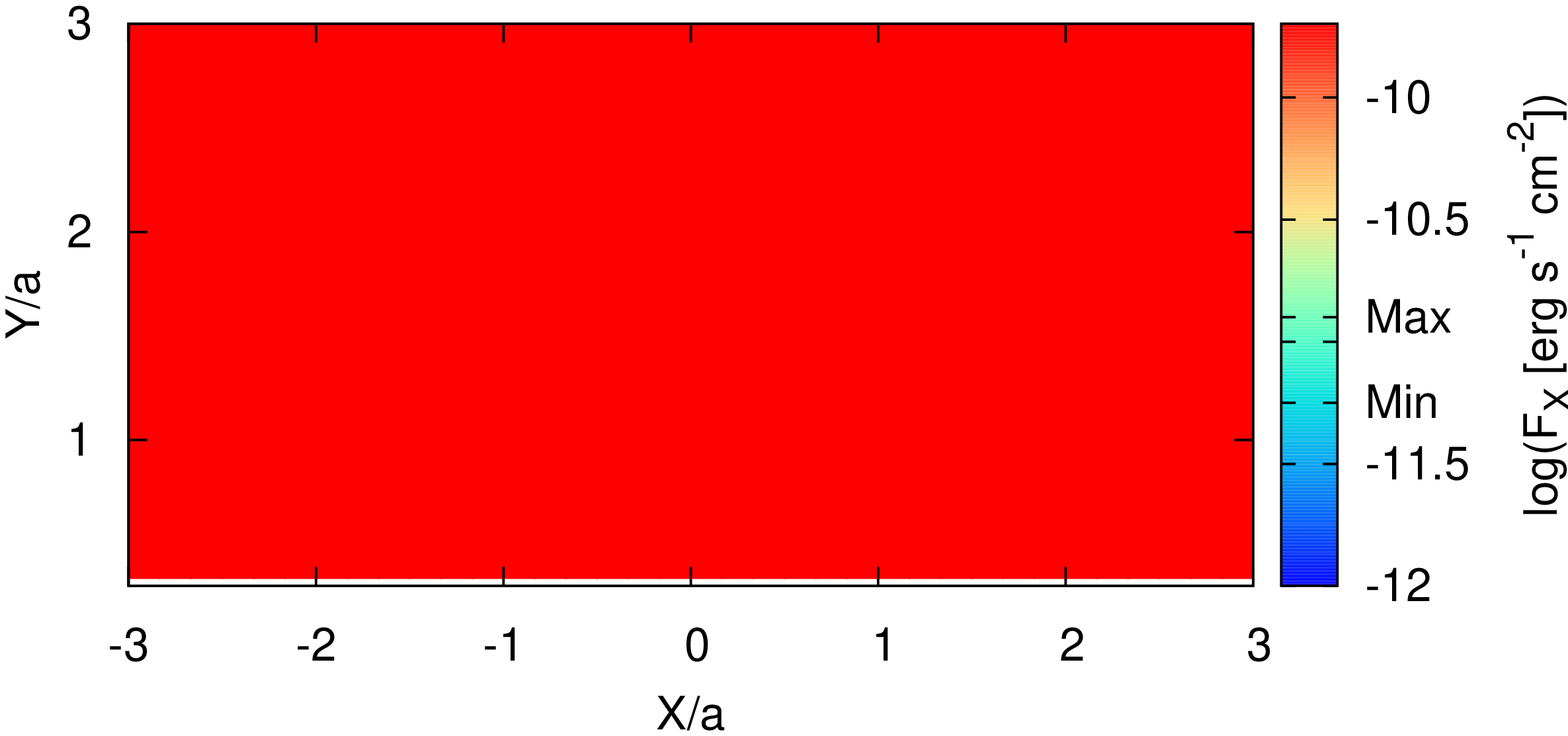}
  \caption[eta=1,delta=0]{As in Fig.~\ref{fig:gev01inj} but showing the integrated energy flux in the 0.3--10 keV energy band.}
  \label{fig:gev01x}

  \centering
  \includegraphics[width=0.2\textwidth, angle=270]{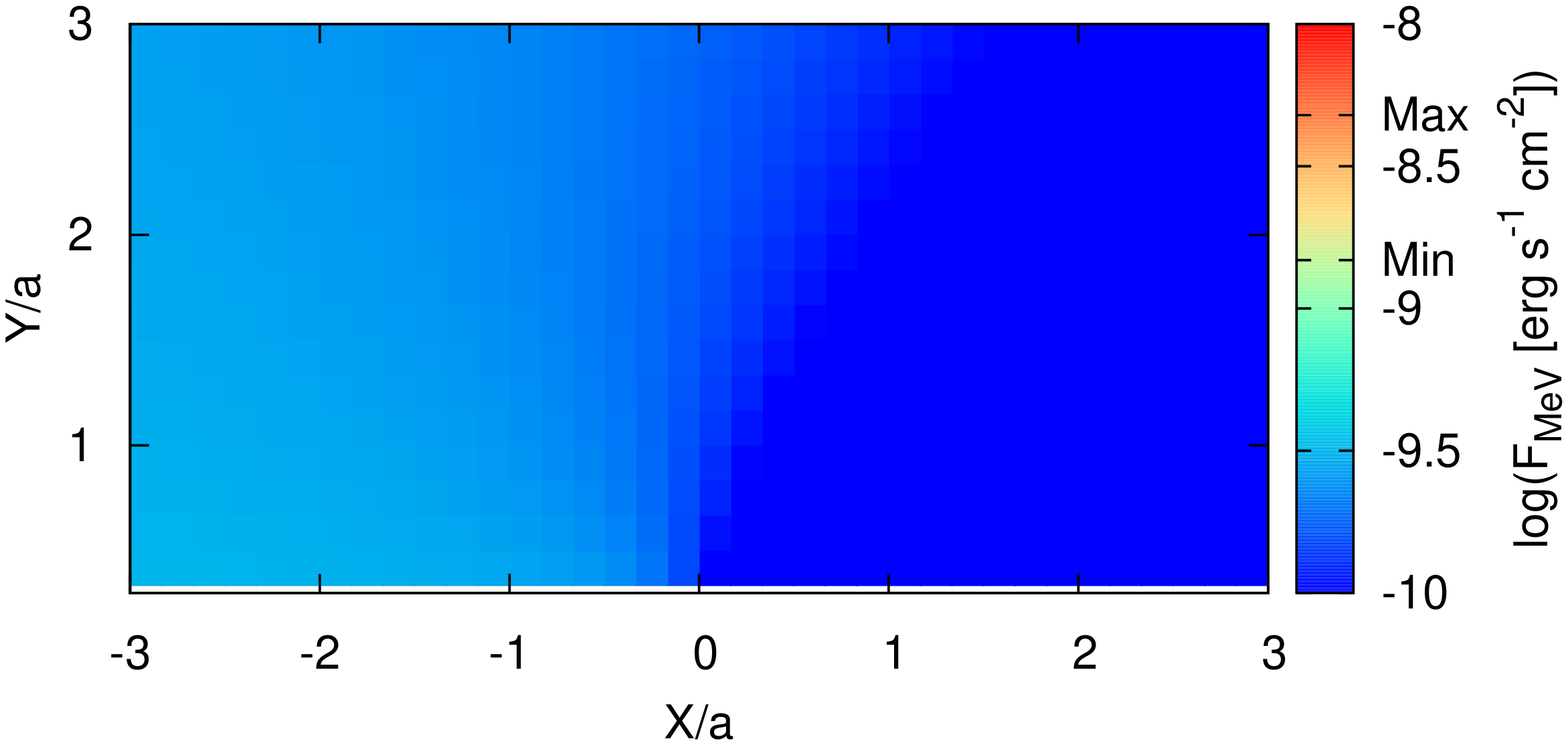}
  \caption[eta=1,delta=0]{As in Fig.~\ref{fig:gev01inj} but showing the integrated energy flux in the 1--30 MeV energy band.}
  \label{fig:gev01mev}
  
  \centering
  \includegraphics[width=0.2\textwidth, angle=270]{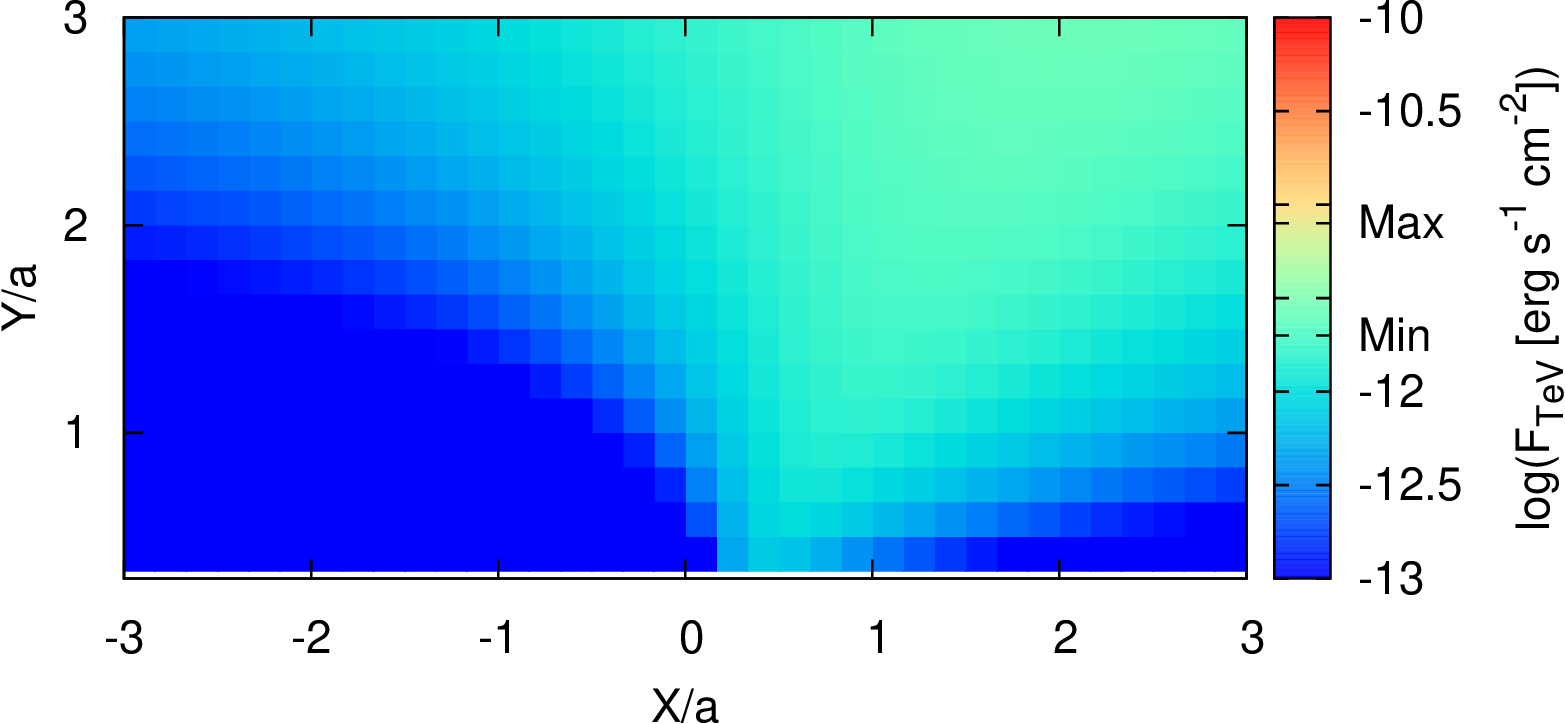}
  \caption[eta=1,delta=0]{As in Fig.~\ref{fig:gev01inj} but showing the integrated energy flux in the 0.1--10 TeV energy band.}
  \label{fig:gev01tev}
  \end{figure}

  In Sec.~\ref{sec:maps}, we presented the maps for intermediate adiabatic losses ($v=10^{9}$ cm s$^{-1}$) 
  and intermediate magnetic fields ($\xi=10^{-2}$, which yields $B$ fields of $\sim$ 10 G close to the massive star to
  $\sim$ 1 G far from it). In this section, we present the maps for the normalization set to reproduce the observed GeV flux 
  for four extreme
  scenarios, varying between fast/slow adiabatic losses ($v=c$ and $v=10^8$ cm s$^{-1}$, respectively) and high/low magnetic fields 
  ($\xi=1$ -- $B$ between 10--10$^2$ G -- and $\xi=10^{-4}$ -- $B$ between 0.1--1 G --, respectively).
  The results are shown in Figs.~\ref{fig:gev10inj} -- \ref{fig:gev01tev}.

  \subsection{Maps normalizing through $F_{\rm MeV}$}

\begin{figure}
  \centering
  \includegraphics[width=0.2\textwidth, angle=270]{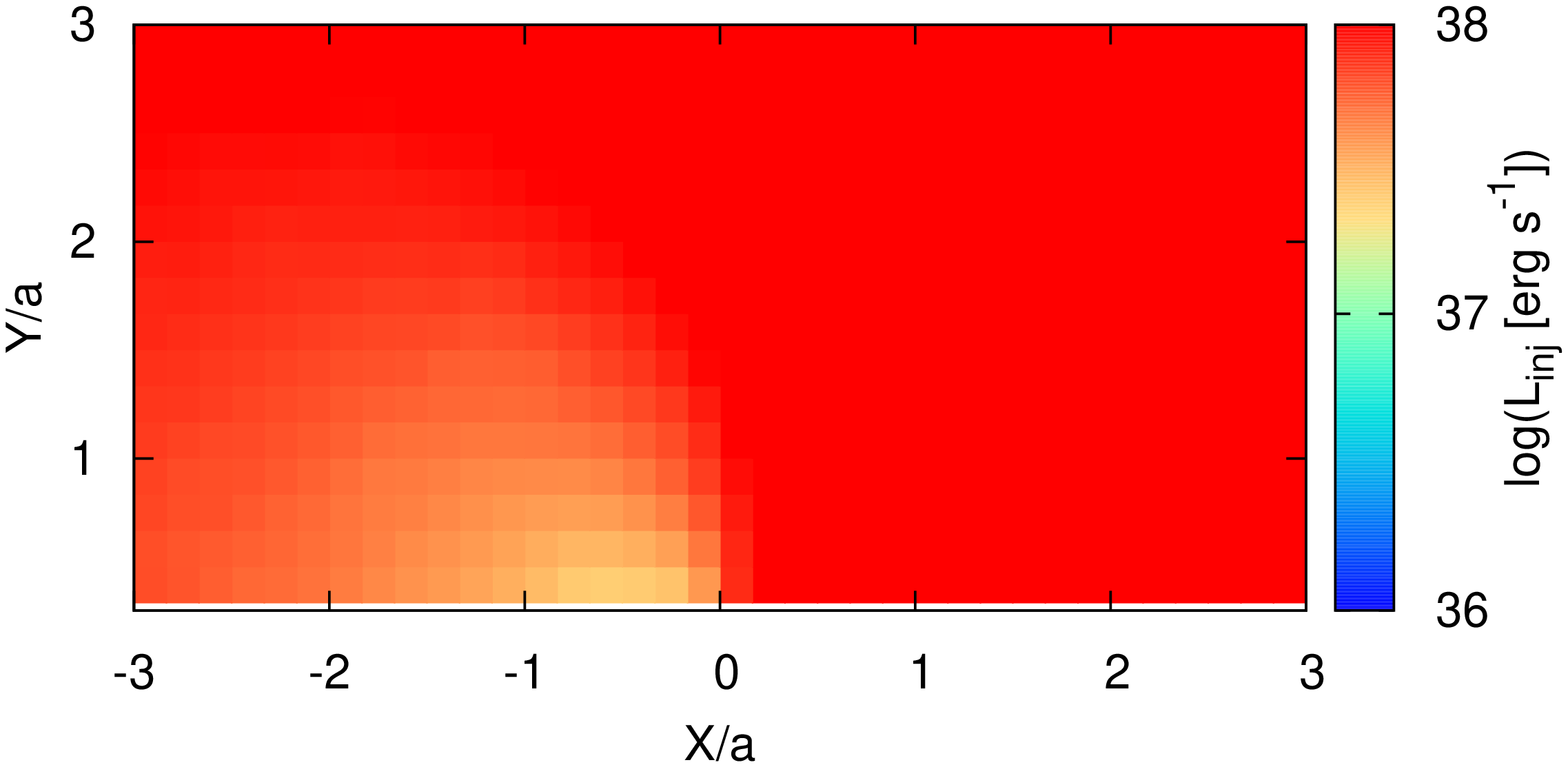}
  \caption[eta=1,delta=0]{Injection luminosity of relativistic particles in the emitter in the case of 
  fast non-radiative losses and a weak magnetic field. The normalization was set 
  to reproduce an energy flux in the $1$--$30$ MeV range equal to $2.6 \times 10^{-9}$ erg cm$^{-2}$ s$^{-1}$.}
  \label{fig:mev10inj}

  \centering
  \includegraphics[width=0.2\textwidth, angle=270]{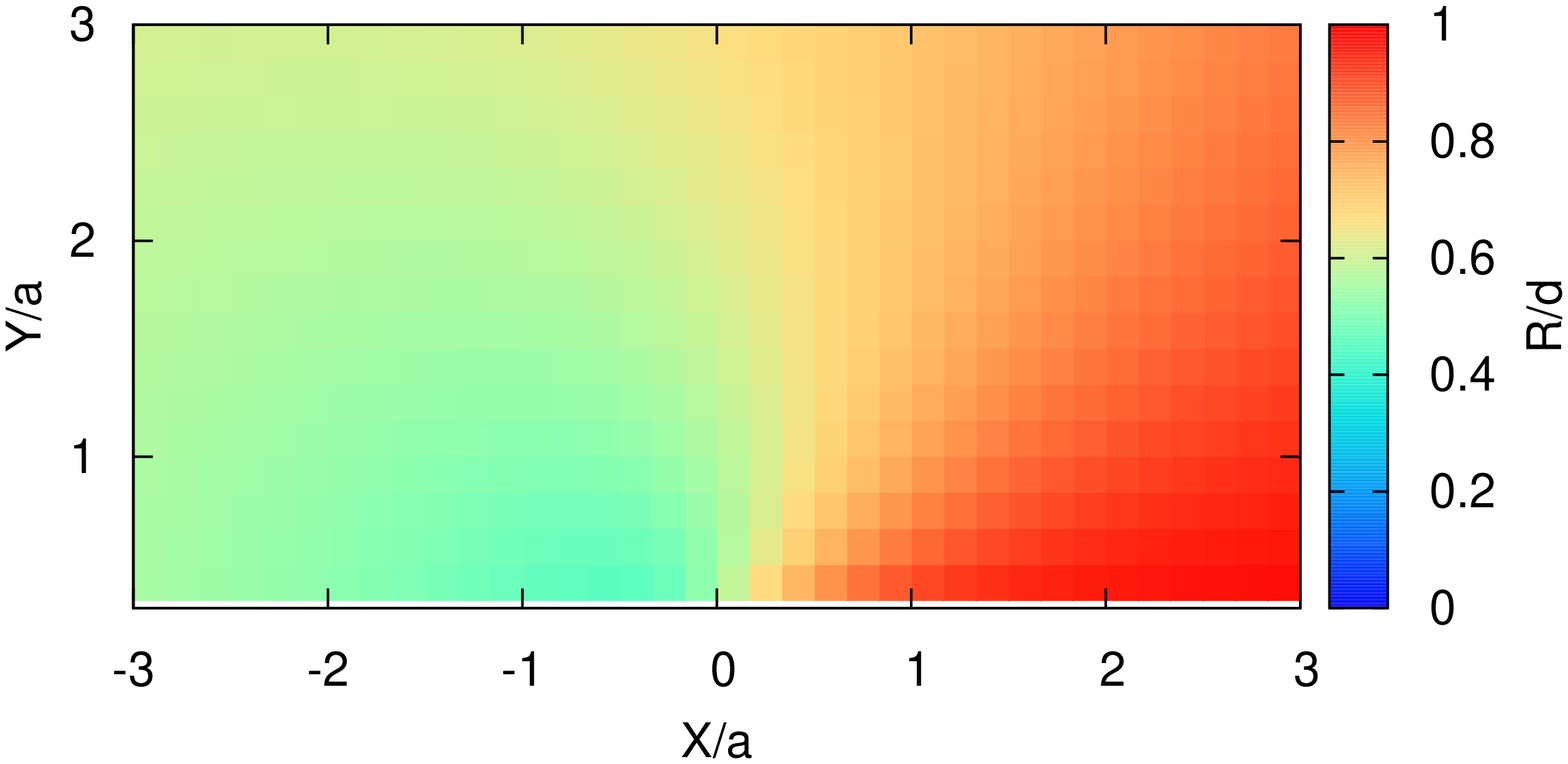}
  \caption[eta=1,delta=0]{As in Fig.~\ref{fig:mev10inj} but showing the emitter's size divided by its distance to the star.}
  \label{fig:mev10confi}

  \centering
  \includegraphics[width=0.2\textwidth, angle=270]{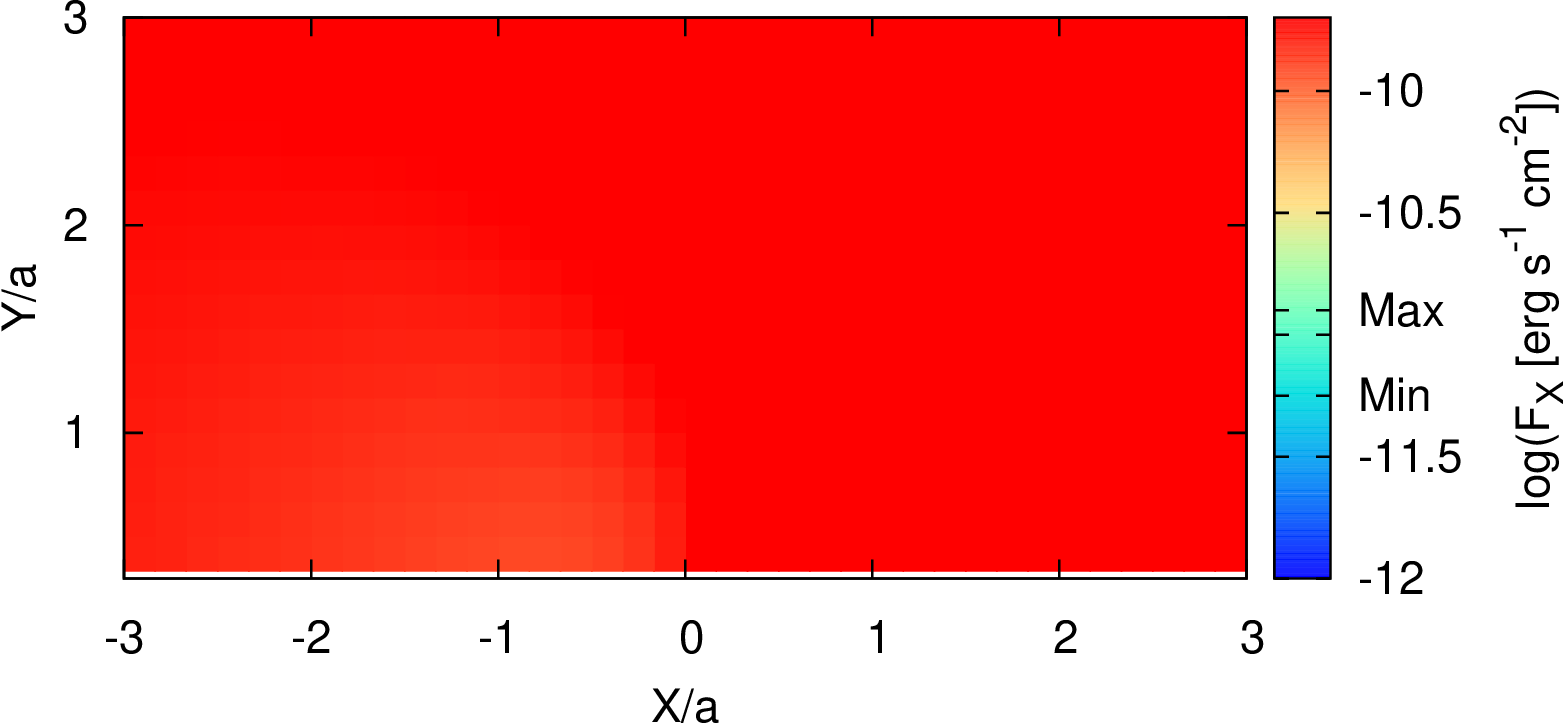}
  \caption[eta=1,delta=0]{As in Fig.~\ref{fig:mev10inj} but showing the integrated energy flux in the 0.3--10 keV energy band.}
  \label{fig:mev10x}

  \centering
  \includegraphics[width=0.2\textwidth, angle=270]{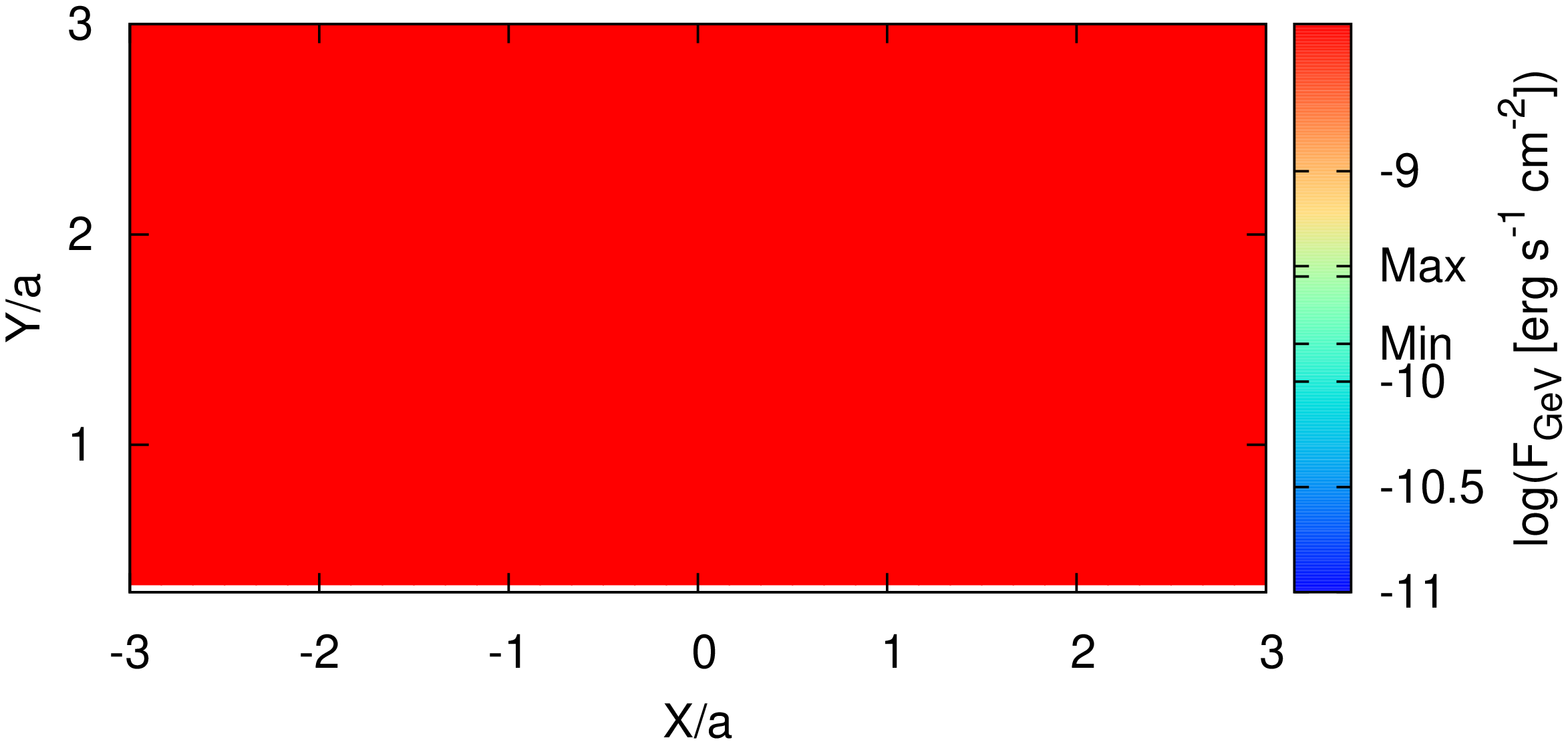}
  \caption[eta=1,delta=0]{As in Fig.~\ref{fig:mev10inj} but showing the integrated energy flux in the 0.1--10 GeV energy band.}
  \label{fig:mev10gev}

  \centering
  \includegraphics[width=0.2\textwidth, angle=270]{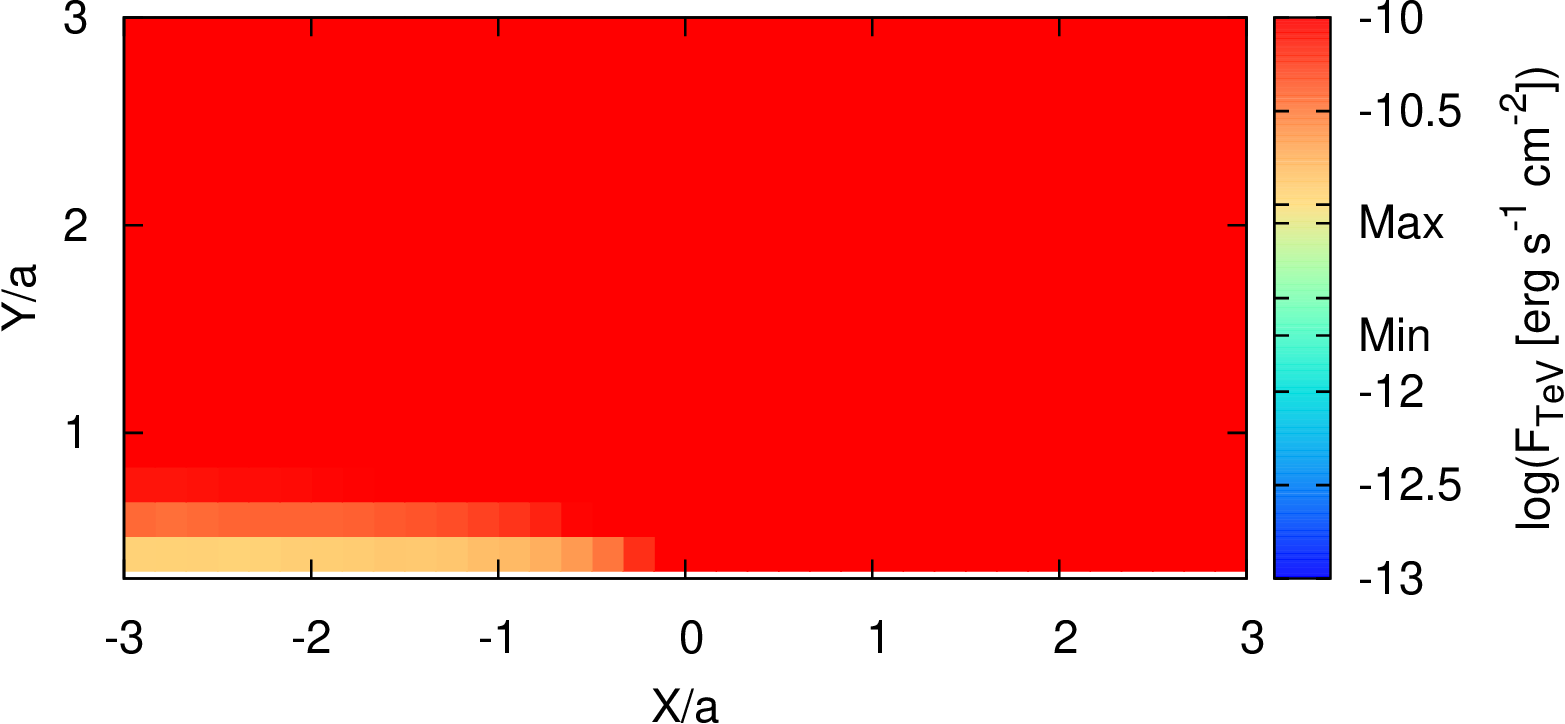}
  \caption[eta=1,delta=0]{As in Fig.~\ref{fig:mev10inj} but showing the integrated energy flux in the 0.1--10 TeV energy band.}
  \label{fig:mev10tev}
  \end{figure}

\begin{figure}
  \centering
  \includegraphics[width=0.2\textwidth, angle=270]{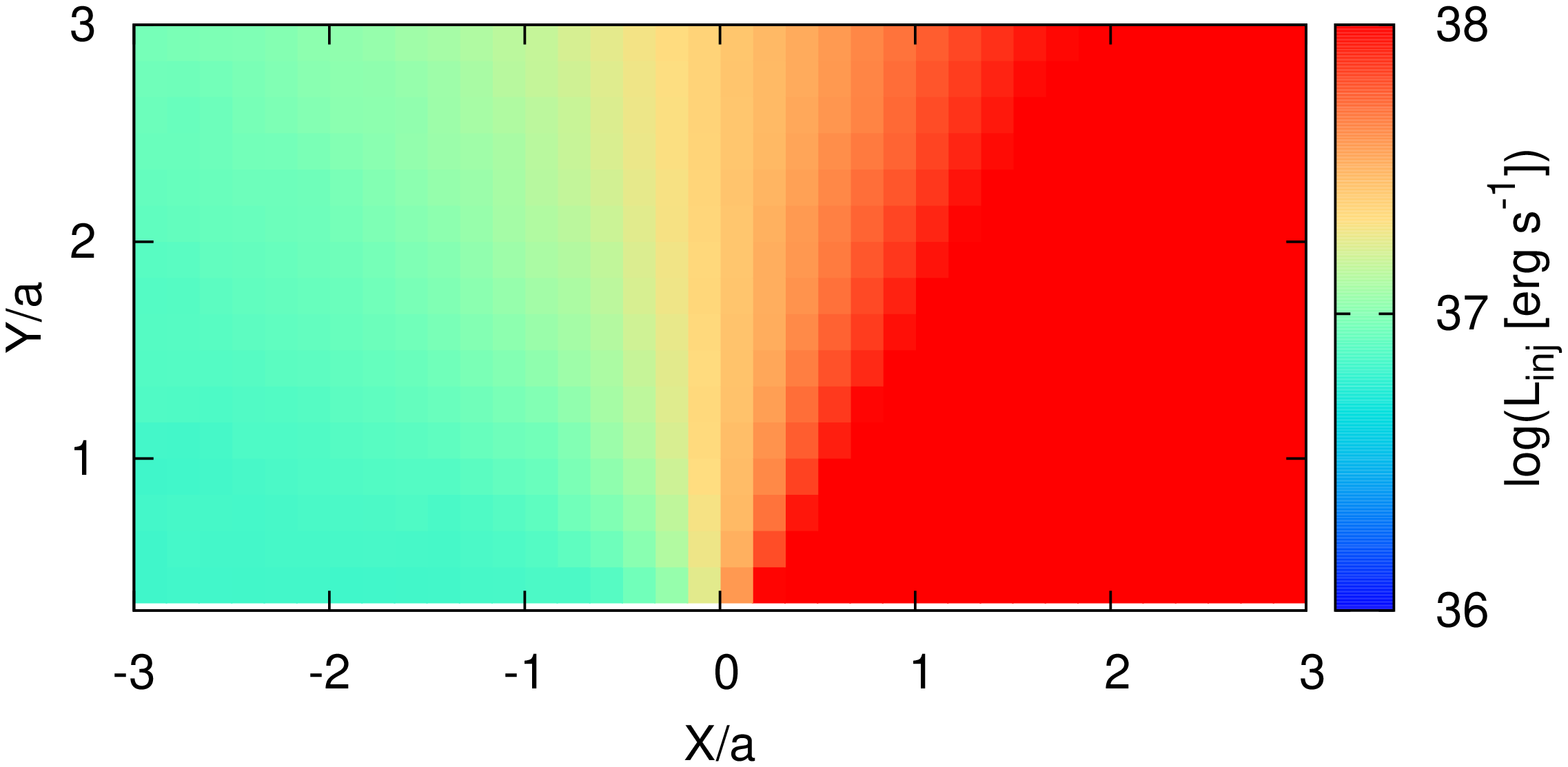}
  \caption[eta=1,delta=0]{Injection luminosity of relativistic particles in the emitter in the case of slow 
  non-radiative losses and a weak magnetic field. The normalization was set 
  to reproduce an energy flux in the $1$--$30$ MeV range equal to $2.6 \times 10^{-9}$ erg cm$^{-2}$ s$^{-1}$.}
  \label{fig:mev00inj}

  \centering
  \includegraphics[width=0.2\textwidth, angle=270]{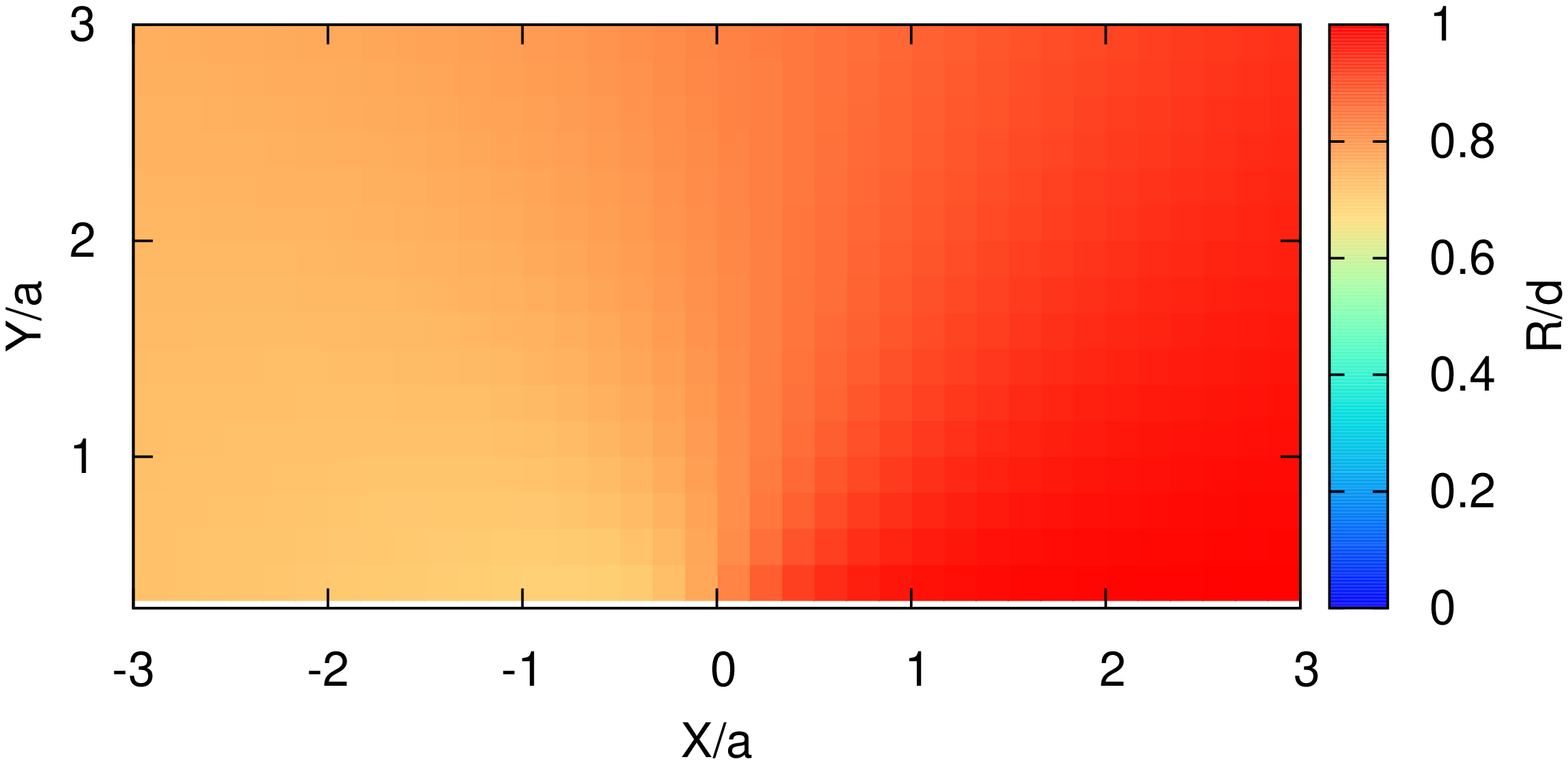}
  \caption[eta=0,delta=0]{As in Fig.~\ref{fig:mev00inj} but showing the emitter's size divided by its distance to the star.}
  \label{fig:mev00confi}

  \centering
  \includegraphics[width=0.2\textwidth, angle=270]{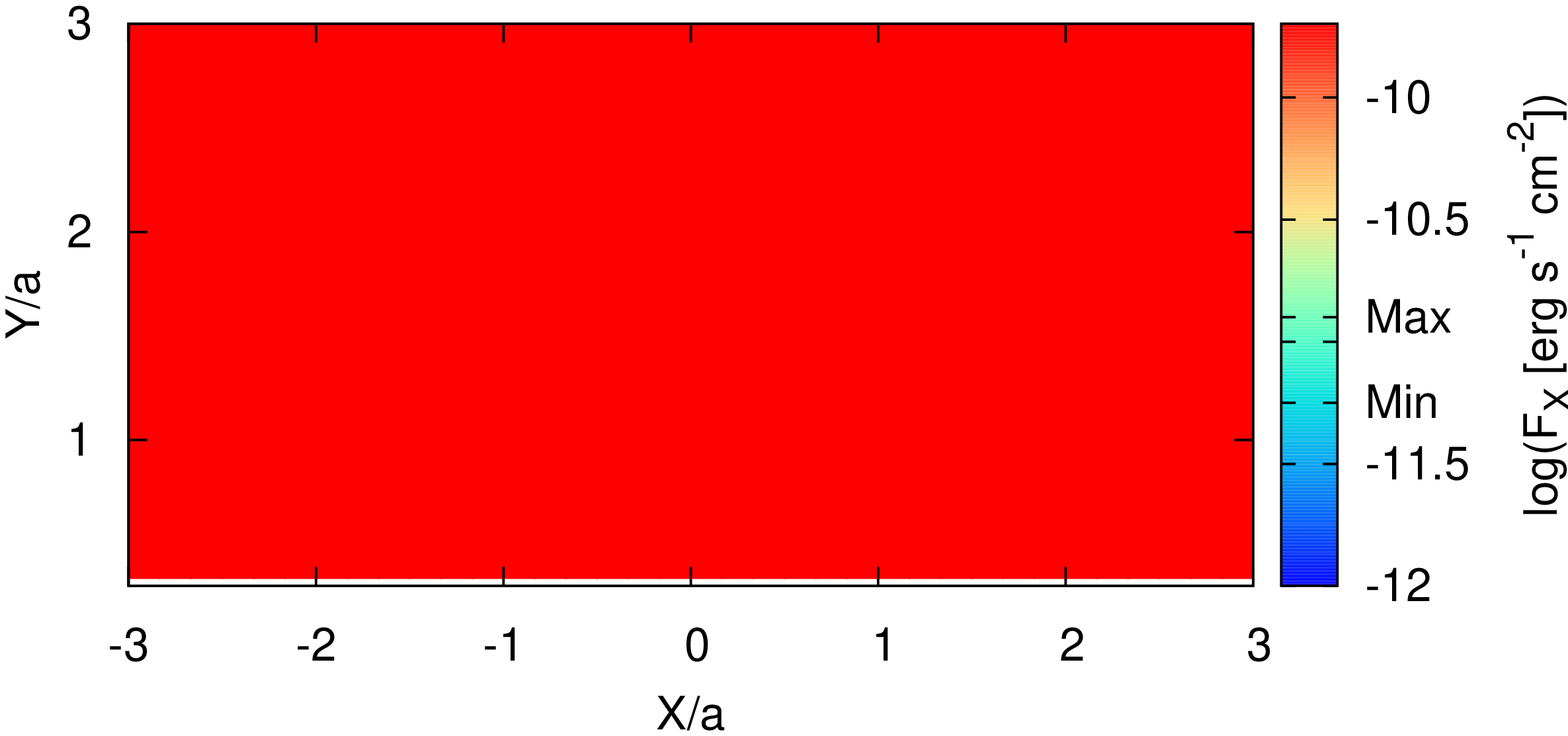}
  \caption[eta=0,delta=0]{As in Fig.~\ref{fig:mev00inj} but showing the integrated energy flux in the 0.3--10 keV energy band.}
  \label{fig:mev00x}

  \centering
  \includegraphics[width=0.2\textwidth, angle=270]{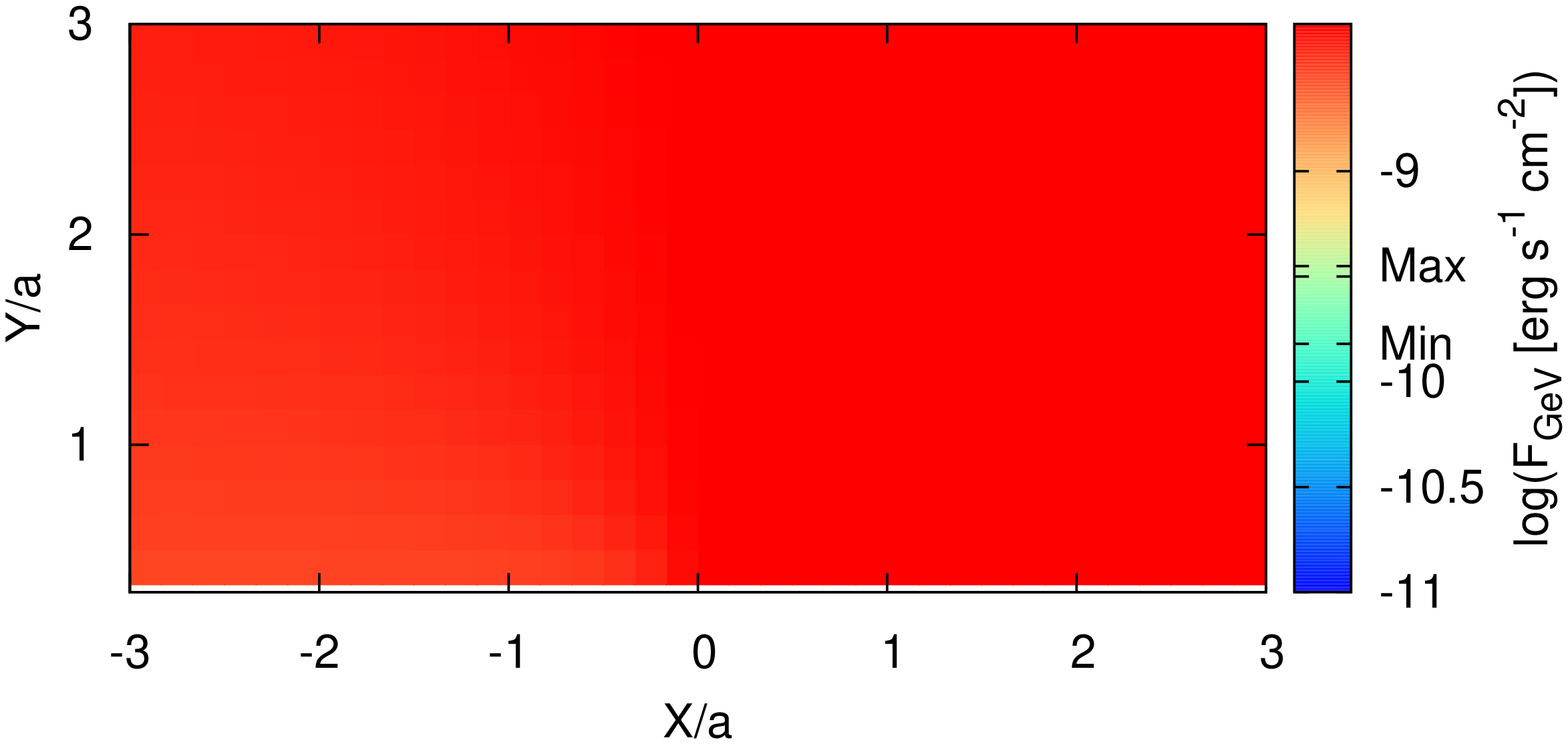}
  \caption[eta=0,delta=0]{As in Fig.~\ref{fig:mev00inj} but showing the integrated energy flux in the 0.1--10 GeV energy band.}
  \label{fig:mev00gev}
  
  \centering
  \includegraphics[width=0.2\textwidth, angle=270]{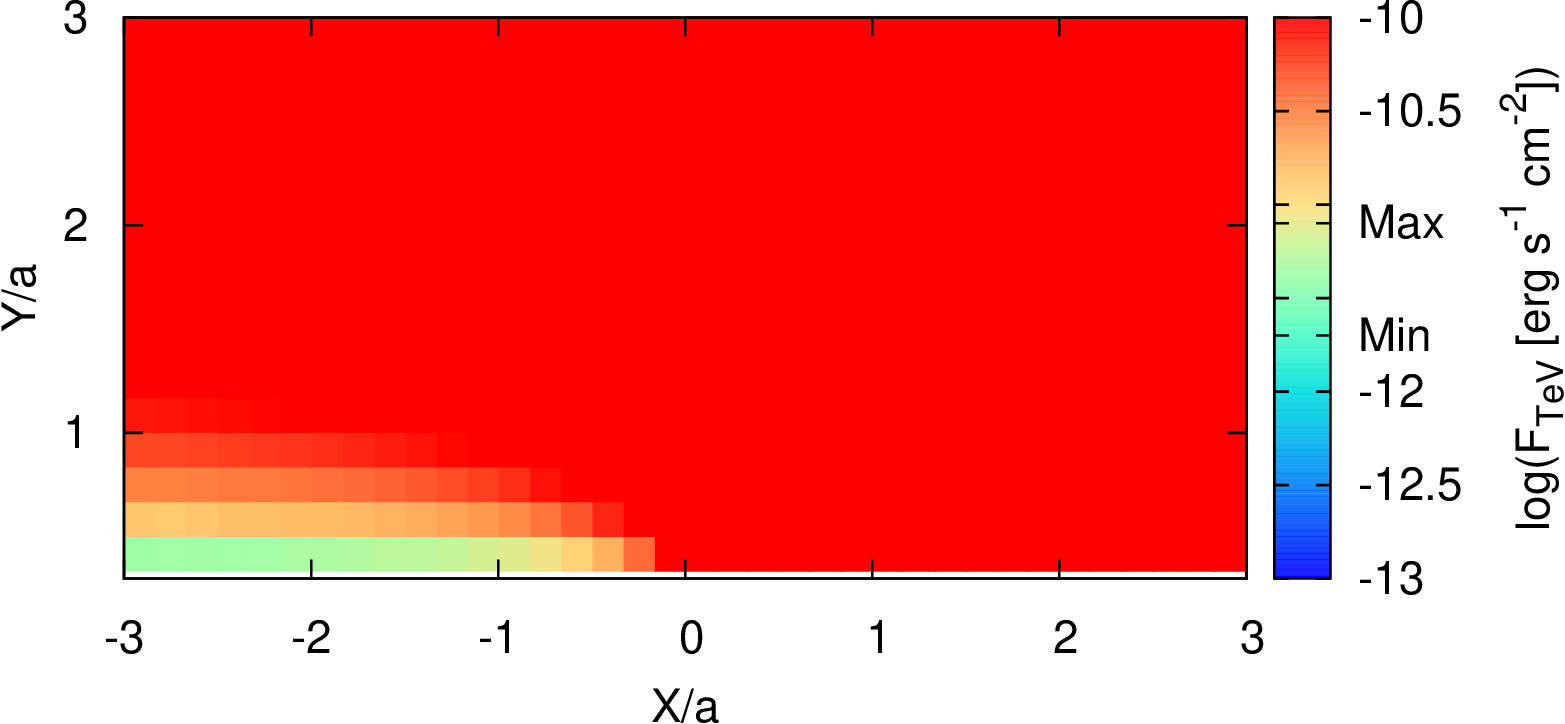}
  \caption[eta=0,delta=0]{As in Fig.~\ref{fig:mev00inj} but showing the integrated energy flux in the 0.1--10 TeV energy band.}
  \label{fig:mev00tev}
  \end{figure}

\begin{figure}
  \centering
  \includegraphics[width=0.2\textwidth, angle=270]{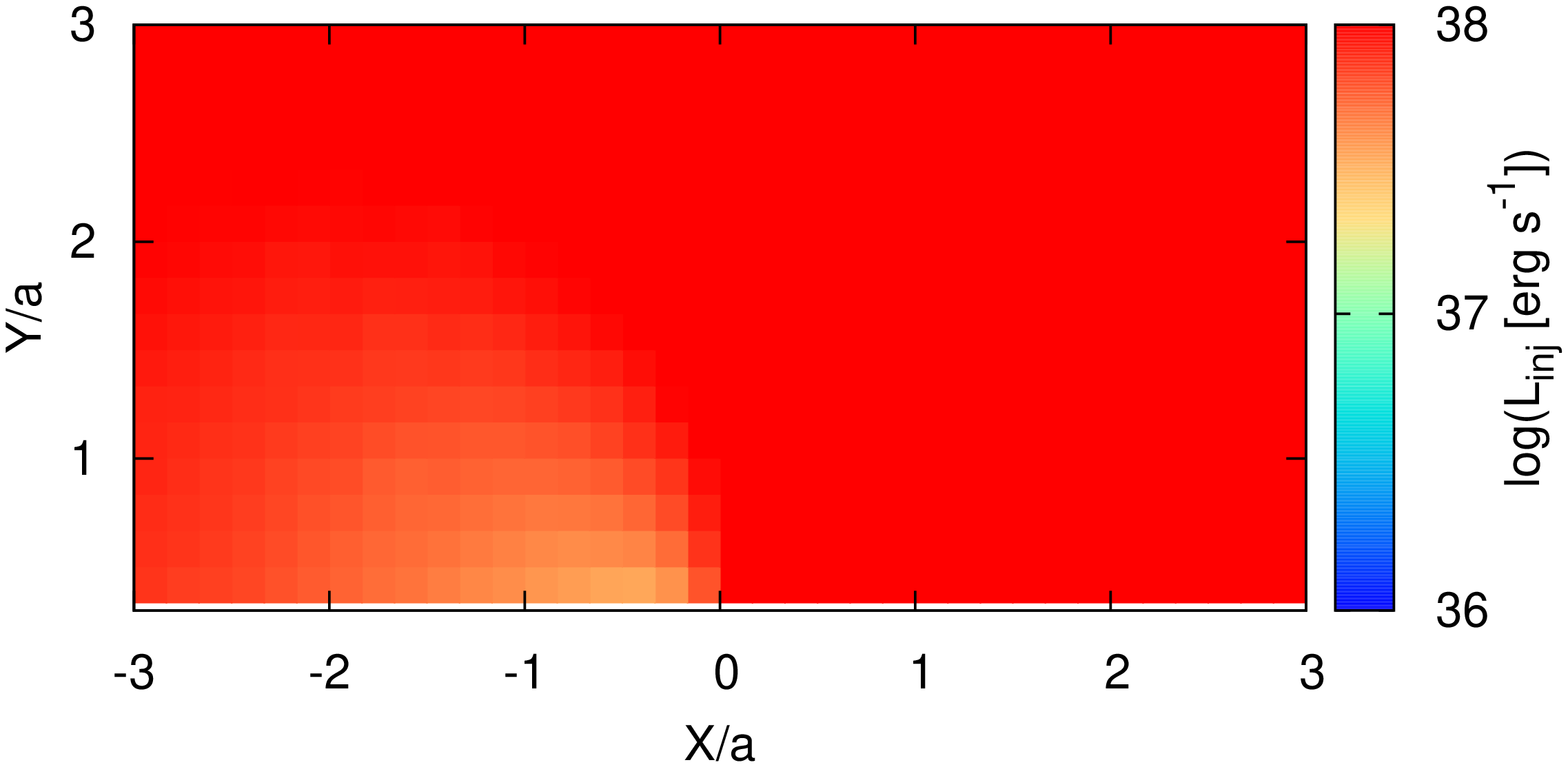}
  \caption[eta=1,delta=0]{Injection luminosity of relativistic particles in the emitter in the case of fast non-radiative 
  losses and a strong magnetic field. The normalization was set 
  to reproduce an energy flux in the $1$--$30$ MeV range equal to $2.6 \times 10^{-9}$ erg cm$^{-2}$ s$^{-1}$.}
  \label{fig:mev11inj}

  \centering
  \includegraphics[width=0.2\textwidth, angle=270]{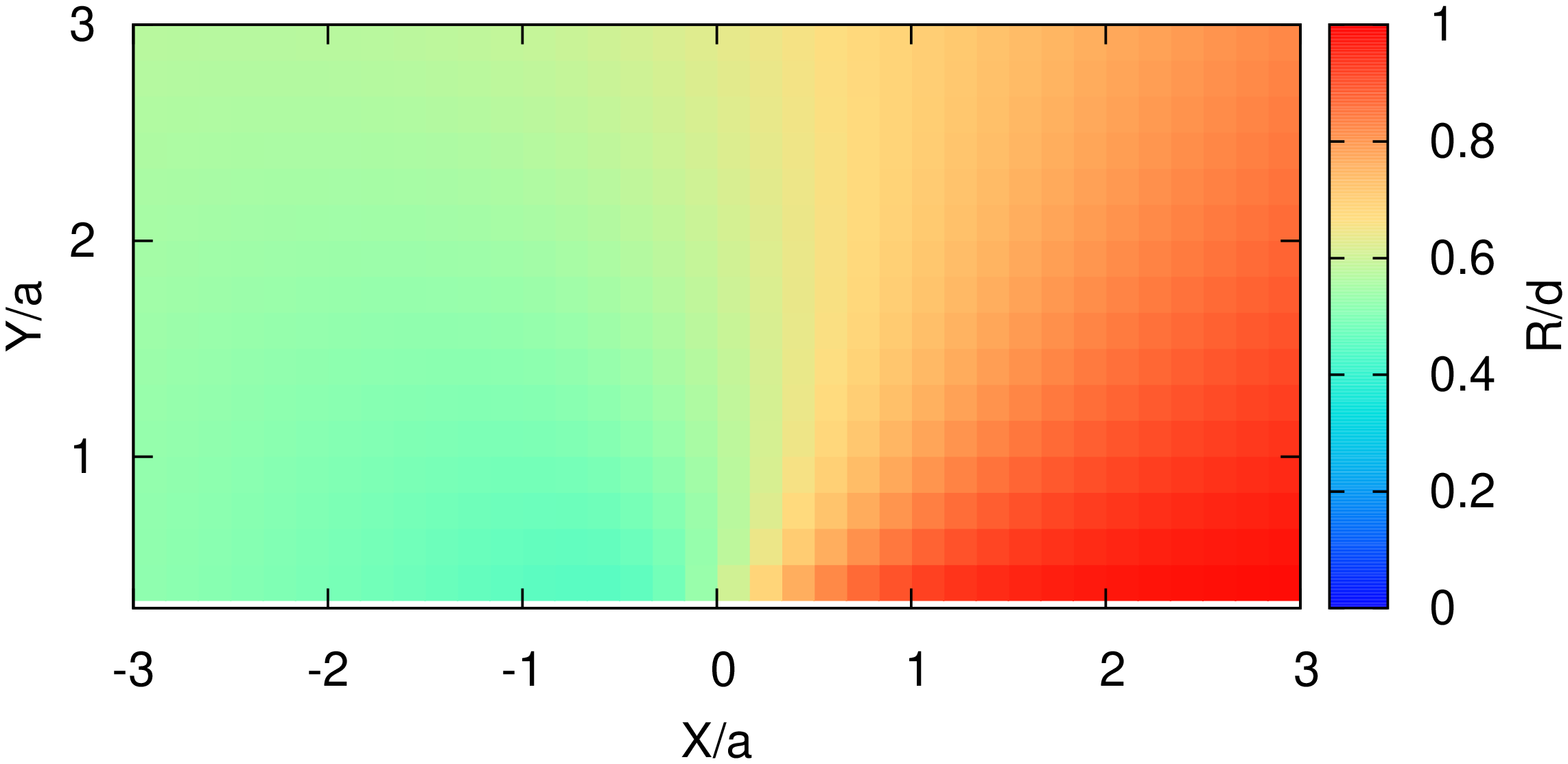}
  \caption[eta=1,delta=1]{As in Fig.~\ref{fig:mev11inj} but showing the emitter's size divided by its distance to the star.}
  \label{fig:mev11confi}

  \centering
  \includegraphics[width=0.2\textwidth, angle=270]{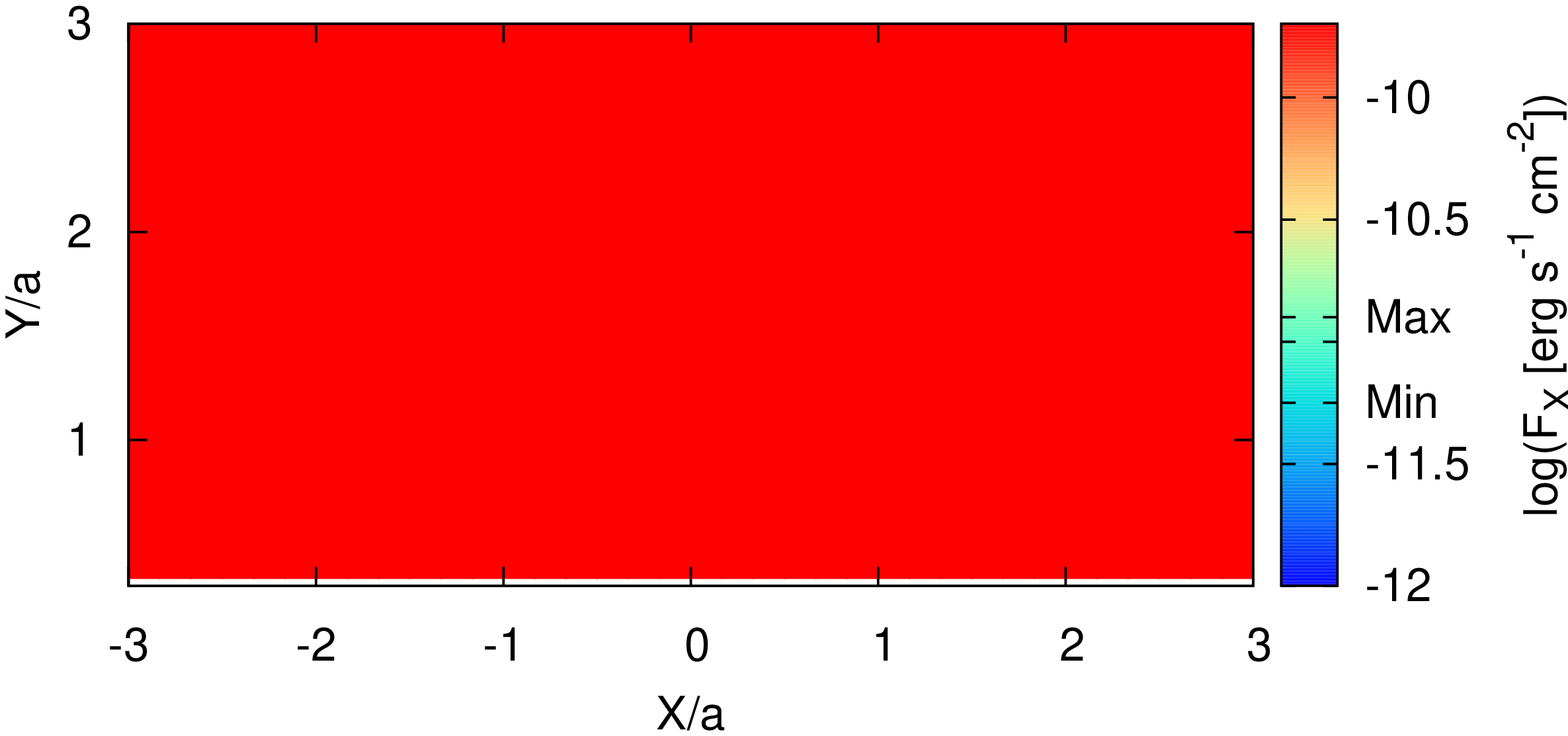}
  \caption[eta=1,delta=1]{As in Fig.~\ref{fig:mev11inj} but showing the integrated energy flux in the 0.3-10 keV energy band.}
  \label{fig:mev11x}

  \centering
  \includegraphics[width=0.2\textwidth, angle=270]{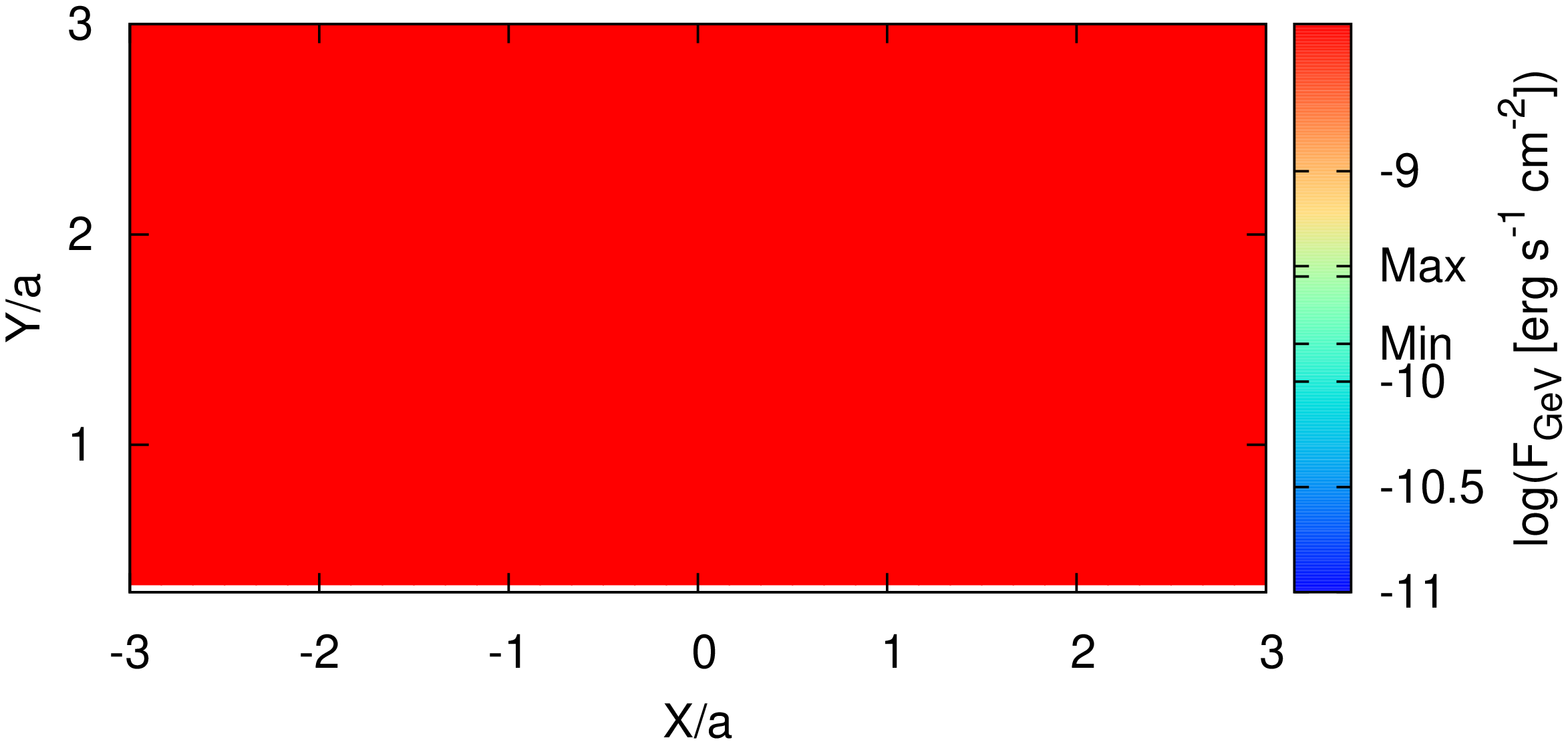}
  \caption[eta=1,delta=1]{As in Fig.~\ref{fig:mev11inj} but showing the integrated energy flux in the 0.1--10 GeV energy band.}
  \label{fig:mev11gev}
  
  \centering
  \includegraphics[width=0.2\textwidth, angle=270]{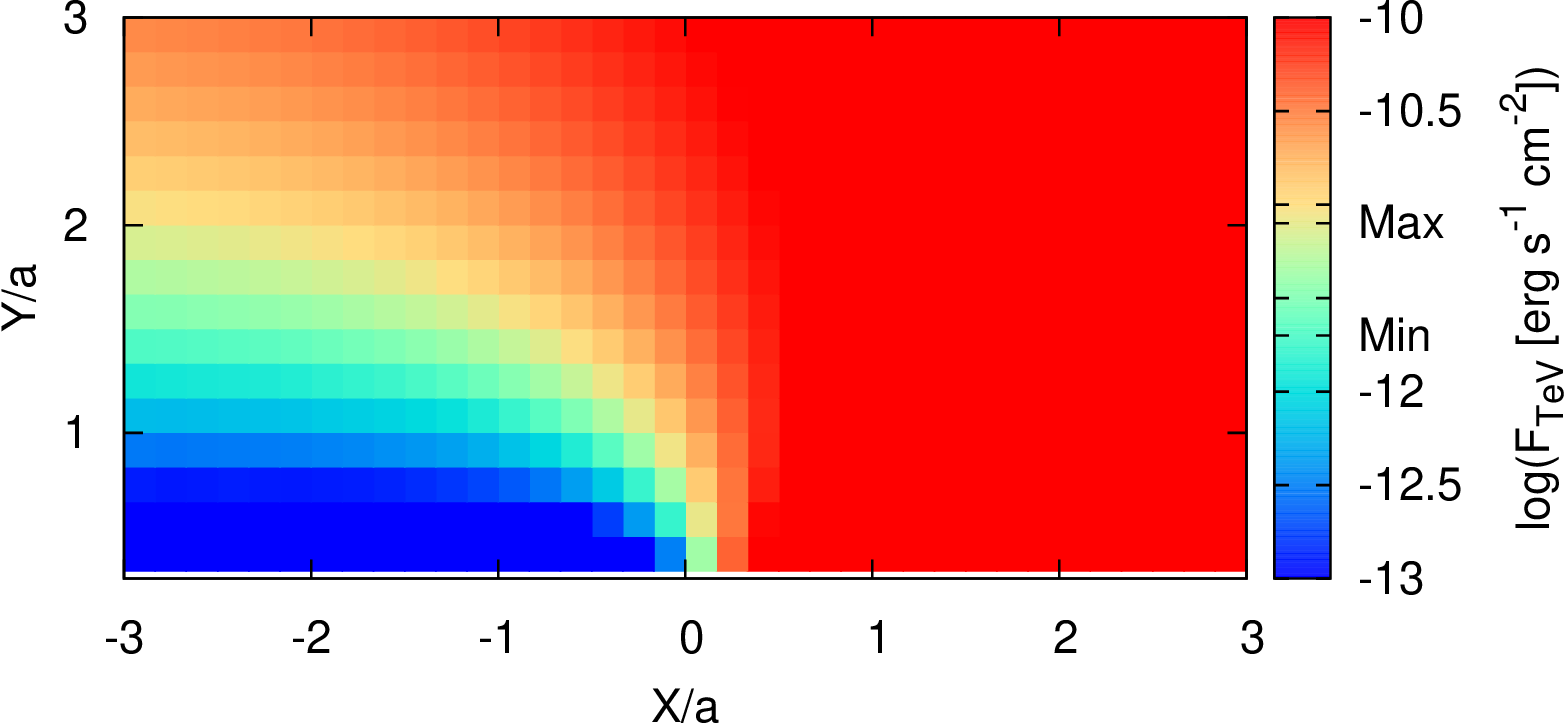}
  \caption[eta=1,delta=1]{As in Fig.~\ref{fig:mev11inj} but showing the integrated energy flux in the 0.1--10 TeV energy band.}
  \label{fig:mev11tev}
  \end{figure}

\begin{figure}
  \centering
  \includegraphics[width=0.2\textwidth, angle=270]{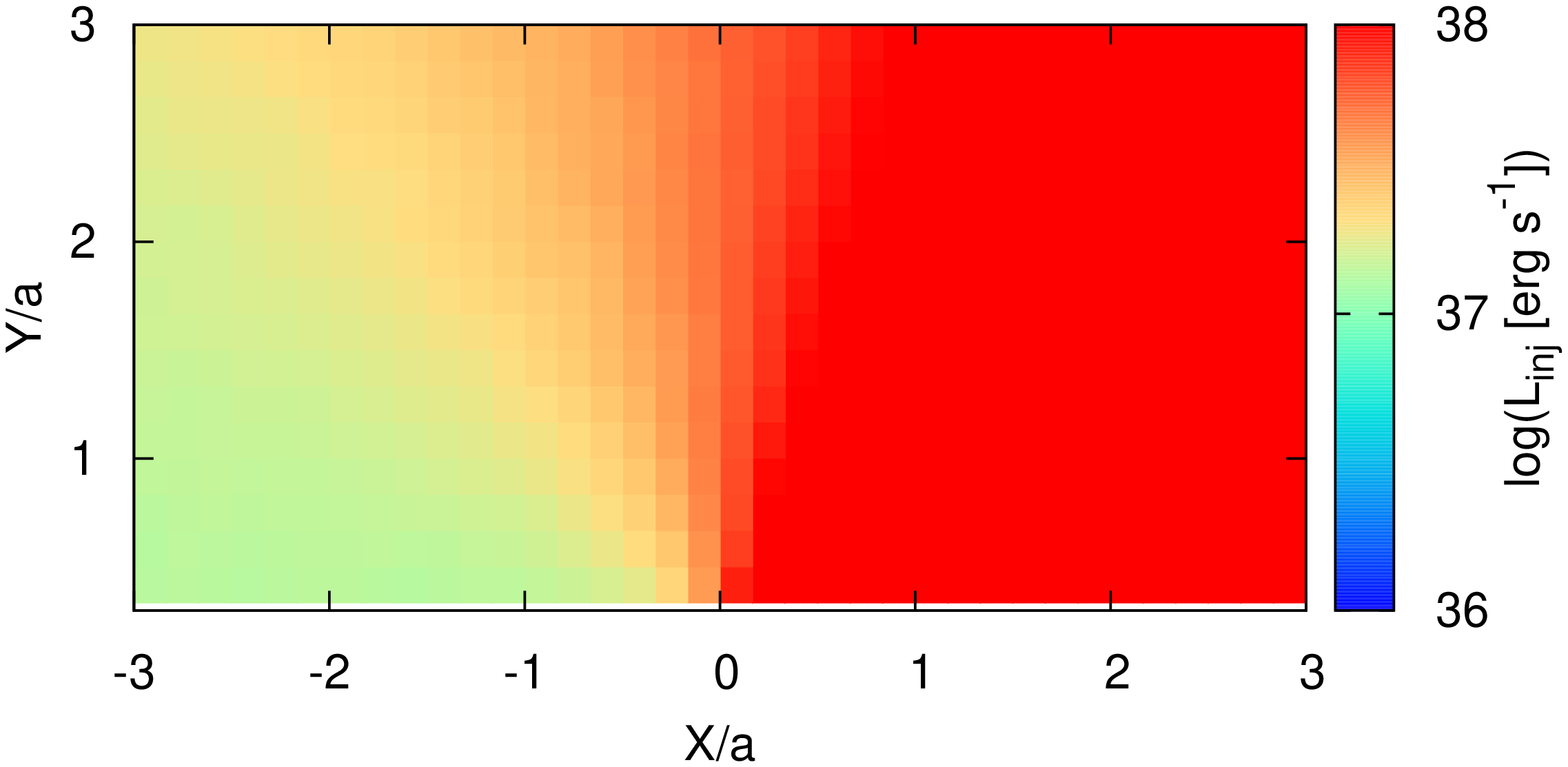}
  \caption[eta=1,delta=0]{Injection luminosity of relativistic particles in the emitter in the case of slow non-radiative 
  losses and a strong magnetic field. The normalization was set to reproduce an energy 
  flux in the $1$--$30$ MeV range equal to $2.6 \times 10^{-9}$ erg cm$^{-2}$ s$^{-1}$.}
  \label{fig:mev01inj}

  \centering
  \includegraphics[width=0.2\textwidth, angle=270]{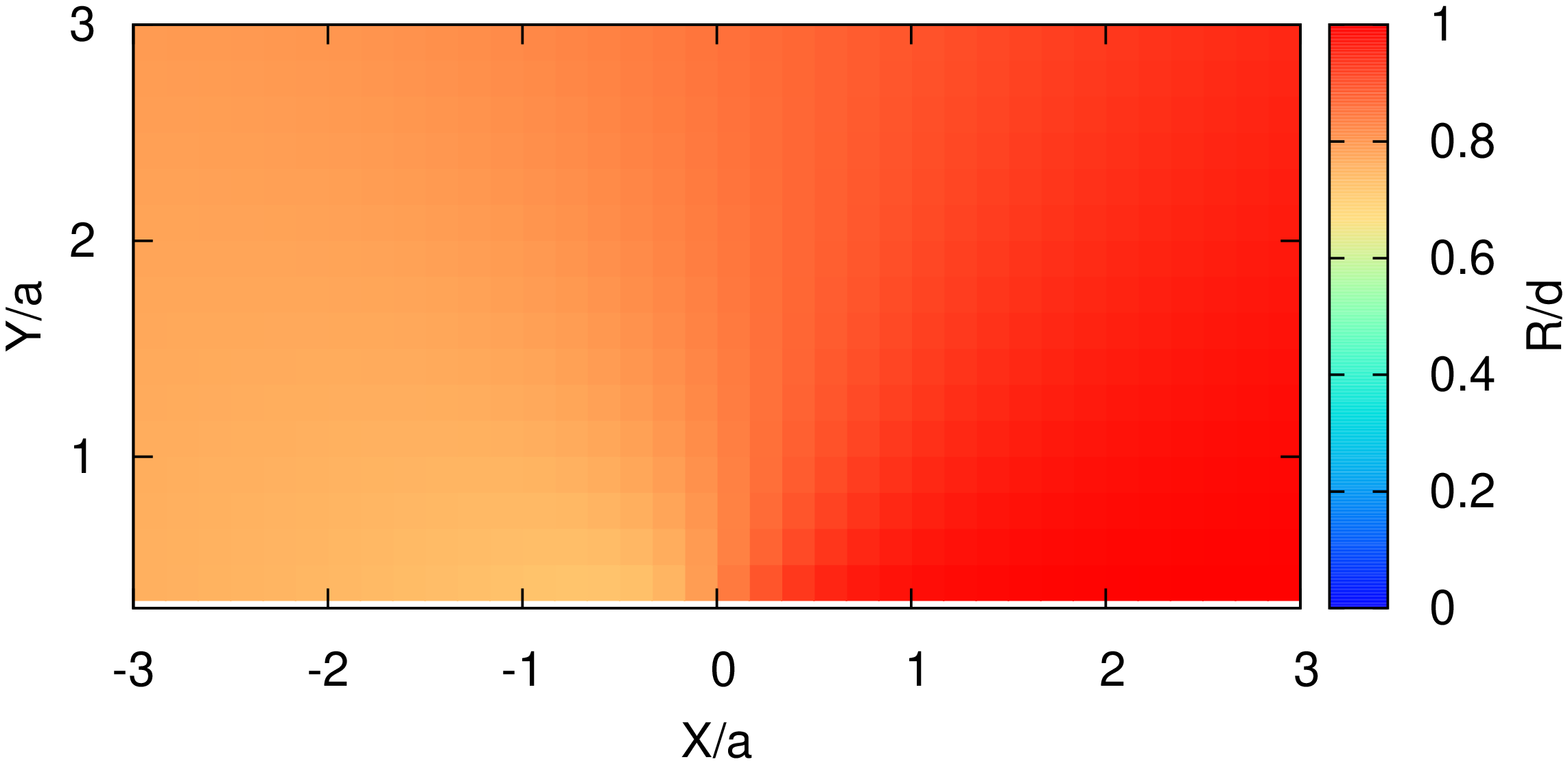}
  \caption[eta=1,delta=0]{As in Fig.~\ref{fig:mev01inj} but showing the emitter's size divided by its distance to the star.}
  \label{fig:mev01confi}
  
  \centering
  \includegraphics[width=0.2\textwidth, angle=270]{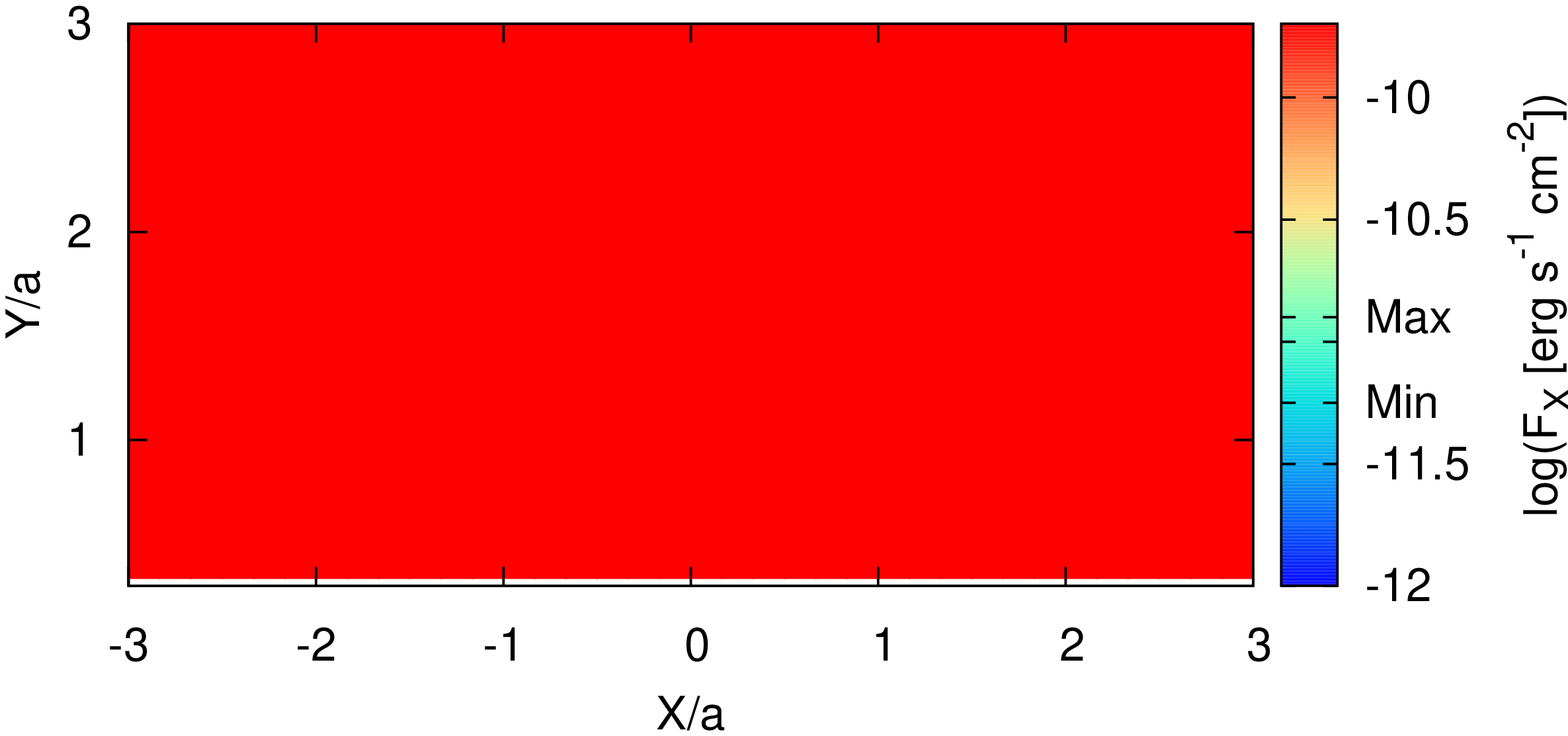}
  \caption[eta=1,delta=0]{As in Fig.~\ref{fig:mev01inj} but showing the integrated energy flux in the 0.3--10 keV energy band.}
  \label{fig:mev01x}

  \centering
  \includegraphics[width=0.2\textwidth, angle=270]{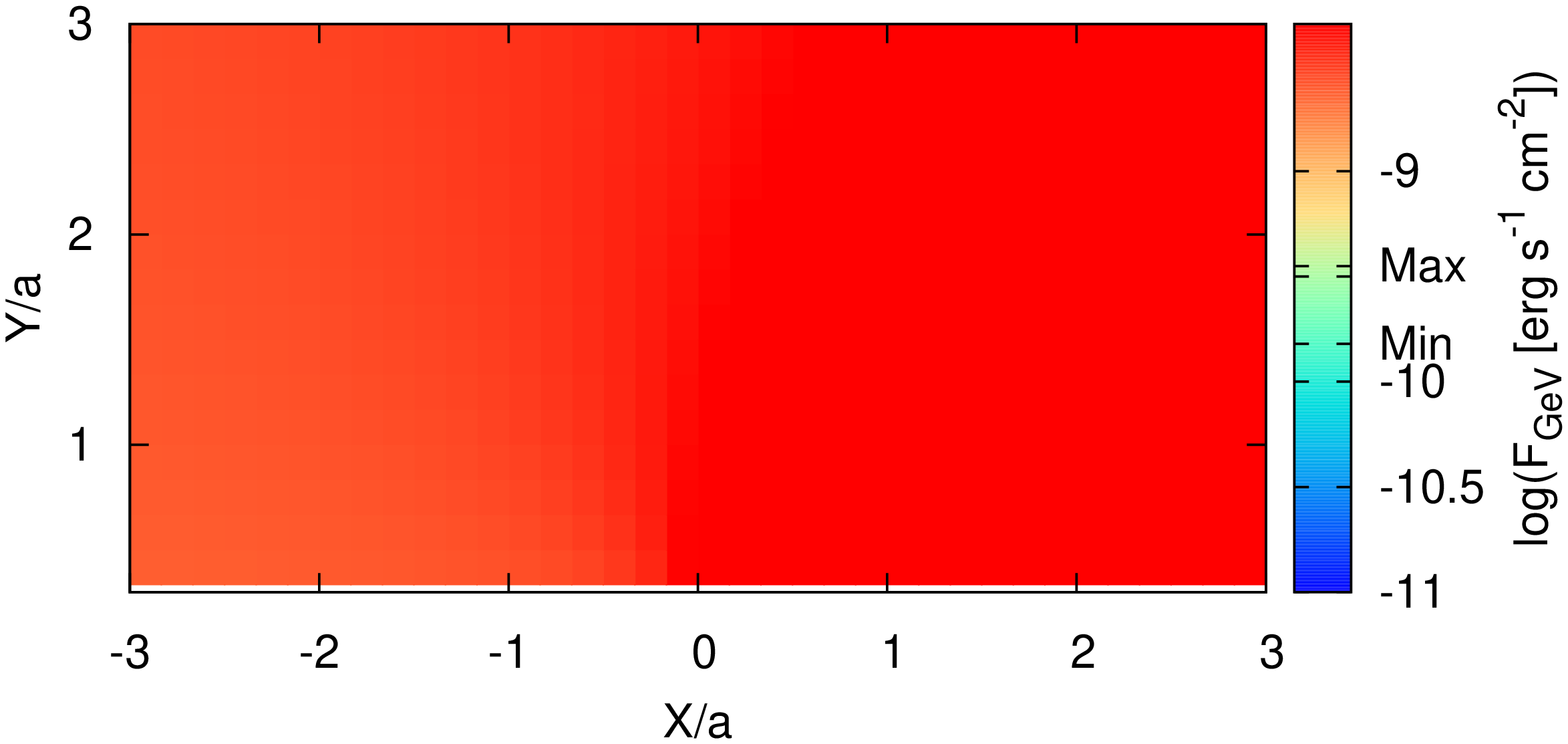}
  \caption[eta=1,delta=0]{As in Fig.~\ref{fig:mev01inj} but showing the integrated energy flux in the 0.1--10 GeV energy band.}
  \label{fig:mev01gev}
  
  \centering
  \includegraphics[width=0.2\textwidth, angle=270]{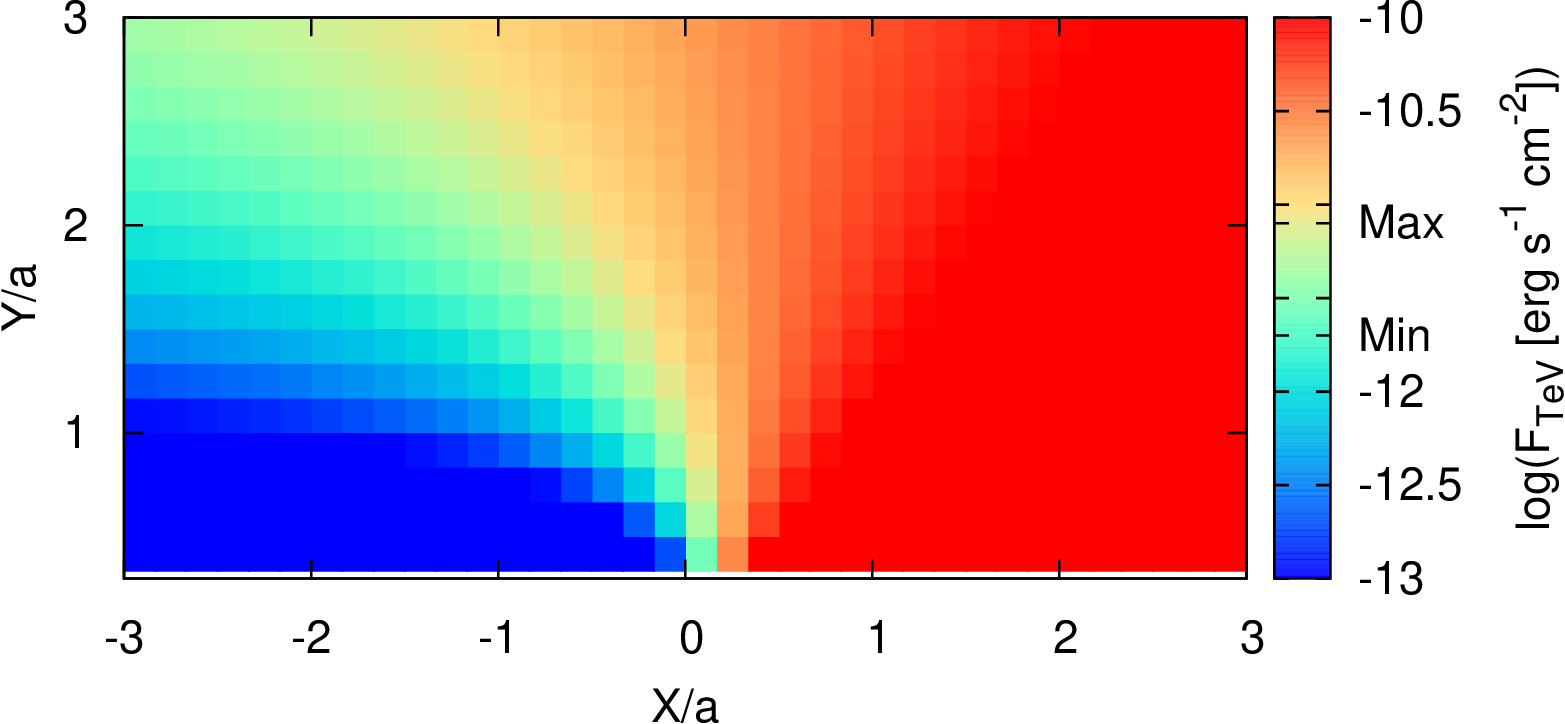}
  \caption[eta=1,delta=0]{As in Fig.~\ref{fig:mev01inj} but showing the integrated energy flux in the 0.1--10 TeV energy band.}
  \label{fig:mev01tev}
  \end{figure}

  In Sec.~\ref{sec:maps}, we presented the maps for intermediate adiabatic losses ($v=10^{9}$ cm s$^{-1}$) 
  and intermediate magnetic fields ($\xi=10^{-2}$, which yields $B$ fields of $\sim$ 10 G close to the massive star to
  $\sim$ 1 G far from it). In this section, we present the maps for the normalization set to reproduce the observed MeV flux for four 
  extreme scenarios, varying between fast/slow adiabatic losses ($v=c$ and $v=10^8$ cm s$^{-1}$, respectively) and high/low 
  magnetic fields ($\xi=1$ -- $B$ between 10--10$^2$ G -- and $\xi=10^{-4}$ -- $B$ between 0.1--1 G --, respectively).
  The results are shown in Figs.~\ref{fig:mev10inj} -- \ref{fig:mev01tev}.


\clearpage
  
\begin{acknowledgements}
We want to thank the referee for his/her useful and constructive comments.
This work is supported by ANPCyT (PICT 2012-00878).
V.B-R. acknowledges financial support from MICINN and European Social Funds through a Ram\'on y Cajal fellowship.
This research has been supported by the Marie Curie Career Integration Grant 321520.
V.B-R. and G.E.R. acknowledges support by the Spanish
Ministerio de Econom\'{\i}a y Competitividad (MINECO)
under grant AYA2013-47447-C3-1-P.
\end{acknowledgements}

\end{document}